\documentclass[12pt,a4paper]{article}
\usepackage{amsmath,caption,xcolor}
\usepackage{listings}
\usepackage{lstautogobble}
\usepackage[top=1.45in, bottom=1.45in, left=1.22in, right=1.22in]{geometry}
\usepackage[pdftex,colorlinks=true]{hyperref}
\hypersetup{citecolor=darkblue,linkcolor=darkblue,urlcolor=darkblue}
\definecolor{darkblue}{rgb}{0,0,.6}

\DeclareCaptionStyle{italic}[justification=centering]{labelfont={bf},textfont={it},labelsep=colon}
\usepackage[nolists,tablesfirst]{endfloat}

\usepackage{graphicx,psfrag,epsf}
\usepackage{enumerate}
\usepackage{natbib}
\usepackage{url} 
\usepackage{booktabs, subfig, bm, paralist,mathpazo,tikz,setspace,todonotes,longtable,microtype,dsfont,rotating} 
\definecolor{darkblue}{rgb}{0,0,.6}

\newcommand{\blind}{0}
\newcommand\possessivecite[1]{\citeauthor{#1}'s \citeyearpar{#1}}

\DeclareMathOperator*{\argmin}{\arg\!\min}
\newsavebox\CBox
\def\textBF#1{\sbox\CBox{#1}\resizebox{\wd\CBox}{\ht\CBox}{\textbf{#1}}}

\graphicspath{{plots/}}

\begin{document}

\def\spacingset#1{\renewcommand{\baselinestretch}%
{#1}\small\normalsize} \spacingset{1}

\if0\blind
{
  \title{\bf Mortality and life expectancy forecasting for a group of populations in developed countries: a multilevel functional data method}
  \author{Han Lin Shang\thanks{Postal address: RSFAS, Level 4, Building 26C, Australian National University, Kingsley Street, Canberra, ACT 2601, Australia; Telephone: +61(2) 612 50535; Fax: +61(2) 612 50087; Email: hanlin.shang@anu.edu.au.}
  \hspace{.2cm}\\
    Research School of Finance, Actuarial Studies and Statistics \\
    Australian National University}
  \maketitle
} \fi

\if1\blind
{
  \bigskip
  \bigskip
  \bigskip
  \begin{center}
    {\LARGE\bf Mortality and life expectancy forecasting for a group of populations in developed countries: a multilevel functional data method}
\end{center}
  \medskip
} \fi

\vspace{1in}

\bigskip

\begin{abstract}

A multilevel functional data method is adapted for forecasting age-specific mortality for two or more populations in developed countries with high-quality vital registration systems. It uses multilevel functional principal component analysis of aggregate and population-specific data to extract the common trend and population-specific residual trend among populations. If the forecasts of population-specific residual trends do not show a long-term trend, then convergence in forecasts may be achieved. This method is first applied to age- and sex-specific data for the United Kingdom, and its forecast accuracy is then further compared with several existing methods, including independent functional data and product-ratio methods, through a multi-country comparison. The proposed method is also demonstrated by age-, sex- and state-specific data in Australia, where the convergence in forecasts can possibly be achieved by sex and state. For forecasting age-specific mortality, the multilevel functional data method is more accurate than the other coherent methods considered. For forecasting female life expectancy at birth, the multilevel functional data method is outperformed by the Bayesian method of \cite{RLG14}. For forecasting male life expectancy at birth, the multilevel functional data method performs better than the Bayesian methods in terms of point forecasts, but less well in terms of interval forecasts. Supplementary materials for this article are available online.

\vspace{.2in}

\end{abstract}

\newpage
\spacingset{1.71}

\section{Introduction}

Many statistical methods have been proposed for forecasting age-specific mortality rates \citep[see][for reviews]{CDE04,Booth06,BT08,GK08,SBH11,BT14}. Of these, a significant milestone in demographic forecasting was the work by \cite{LC92}. They applied a principal component method to age-specific mortality rates and extracted a single time-varying index of the level of mortality rates, from which the forecasts are obtained by a random-walk with drift. The method has since been extended and modified. For example, \citet{RH03} proposed the age-period-cohort Lee-Carter method; \cite{HU07} proposed a functional data model that utilizes nonparametric smoothing and high-order principal components; \cite{GK08} and \cite{WSB+15} considered Bayesian techniques for Lee-Carter model estimation and forecasting; and \cite{LLG13} extended the Lee-Carter method to model the rotation of age patterns for long-term projections.

These works mainly focused on forecasting mortality for a single population, or several populations individually. However, individual forecasts, even when based on similar extrapolative procedures, may imply increasing divergence in mortality rates in the long run, counter to the expected and observed trend toward a global convergence \citep{LL05,Pampel05,Li13}. Thus, joint modeling mortality for two or more populations simultaneously is paramount, as it allows one to model the correlations among two or more populations, distinguish between long-term and short-term effects in the mortality evolution, and explore the additional information contained in the experience of other populations to further improve forecast accuracy. These populations can be grouped by sex, state, ethnic group, socioeconomic status and other attributes. In these cases, it is often desirable to produce coherent forecasts that do not diverge over time (e.g., in demography, \citeauthor{LL05}, \citeyear{LL05}, \citeauthor{BC10}, \citeyear{BC10}, \citeauthor{ARG+11}, \citeyear{ARG+11}, \citeauthor{RLS+12}, \citeyear{RLS+12}, \citeauthor{RCG+13}, \citeyear{RCG+13}, \citeauthor{Li13}, \citeyear{Li13}, \citeauthor{RLG14}, \citeyear{RLG14}, \citeauthor{SLK+15}, \citeyear{SLK+15}; in actuarial science, \citeauthor{JK11}, \citeyear{JK11}, \citeauthor{LH11}, \citeyear{LH11}, \citeauthor{CBD+11}, \citeyear{CBD+11b}, \citeauthor{DCB+11}, \citeyear{DCB+11}).

The definition of coherent in demography varies, but here it means joint modeling of populations, and further that the mortality forecasts do not overlap. In the case of two-sex populations, there may be common features in the groups of data that can first be captured with the common principal components. Further, we can prevent the forecasts of the groups from diverging by requiring the difference in each sex-specific principal component scores to be stationary for different populations $i$ and $j$, so that
\begin{equation*}
\limsup_{t\rightarrow \infty} \text{E}||f_{t,i} - f_{t,j}||<\infty, \quad \text{for all} \ i \ \text{and}  \ j,
\end{equation*}
where $\text{E}||f_{t,i} - f_{t,j}|| = \int_{\mathcal{I}}[f_{t,j}(x) - f_{t,i}(x)]^2dx$ is the $L_2$ norm, $f_t(x)$ represents age-specific mortality for year $t$, and $\mathcal{I}$ denotes a function support range. The problem of jointly forecasting mortality rates for a group of populations has been considered by \cite{Lee00, LL05, Lee06, DDG+06} and \cite{SLK+15} in the context of the Lee-Carter model. These authors proposed the augmented common factor model that extracts a common trend for a group of populations, while acknowledging their individual differences in level, age pattern and short-term trend \citep{LL05}. On the other hand, \cite{HBY13} proposed a functional data model to jointly model the gap between female and male age-specific mortality rates, and \cite{RLG14} proposed a Bayesian method to jointly model the gap between female and male life expectancies at birth.

Based on the work of \cite{LL05}, a general framework is presented by \cite{Lee06} for forecasting life expectancy at birth as the sum of a common trend and the population-specific trend. Coherent forecasting in the framework of \citeauthor{LC92}'s \citeyearpar{LC92} model has recently been extended to the coherent functional data model by \citet*{HBY13}. These authors proposed the product-ratio method, which models the product and ratio functions of the age-specific mortality rates of different populations through a functional principal component decomposition, and forecasts age- and sex-specific mortality coherently by constraining the forecast ratio function via stationary time-series model. The forecasts of product and ratio functions are obtained using the independent functional data method given in \cite{HU07}; the forecast product and ratio functions are then transformed back into the male and female age-specific mortality rates. Illustrated by empirical studies, they found that the product-ratio method generally gives slightly less accurate female mortality forecasts and produces much more accurate male mortality forecasts than the independent functional data method, in which the latter one does not impose a coherent structure.

As an extension of \cite{LL05} and \cite{HBY13}, we consider a multilevel functional data model motivated by the work of \cite{DCC+09}, \cite{CSD09}, \cite{CG10} and \cite{GCC+10}, among many others. The objective of the multilevel functional data method is to model multiple sets of functions that may be correlated among groups. In this paper, we apply this technique to forecast age-specific mortality and life expectancy at birth for a group of populations. We found the multilevel functional data model captures the correlation among populations, models the forecast uncertainty through Bayesian paradigm, and is adequate for use within a probabilistic population modeling framework \citep{RLS+12}. Similar to the work of \cite{LL05,Lee06,DDG+06} and \cite{Li13}, the multilevel functional data model captures the common trend and the population-specific trend. It produces forecasts that are comparable with the ones from the product-ratio method, which themselves are also more accurate than the independent functional data method for male age-specific mortality and life expectancy forecasts. 

The multilevel functional data model is described in Section~\ref{sec:2}. In Section~\ref{sec:relate}, we outline the differences among the multilevel functional data, augmented common factor and independent functional data methods. In Section~\ref{sec:3}, we illustrate the multilevel functional data method by applying it to the age- and sex-specific mortality rates for the United Kingdom (UK). In Section~\ref{sec:4},  we compare the point and interval forecast accuracy among five methods for 32 populations. In Section~\ref{sec:5}, we investigate the performance of the multilevel functional data method with the age-, and sex- and state-specific mortality rates in Australia. In Section~\ref{sec:6}, we provide some concluding remarks, along with some reflections on how the method presented here can be further extended. More information on some theoretical properties of multilevel functional principal component decomposition are deferred to the Supplementary Material A \citep{Shang16}.

\section{A multilevel functional data model}\label{sec:2}

We first present the problem in the context of forecasting male and female age-specific mortality rates, although the method can easily be generalized to any number of populations. Let $y_t^{j}(x_i)$ be the log central mortality rates observed at the beginning of each year for year $t=1,2,\dots,n$ at observed ages $x_1,x_2,\dots,x_p$ where $x$ is a continuous variable, $p$ is the number of ages, and superscript $j$ represents either male or female in the case of two populations. 

Following the functional data framework, we assume there is an underlying continuous and smooth function $f_t^{j}(x)$ that is observed at discrete data points with error. That is
\begin{equation}
y_t^{j}(x_i) = f_t^{j}(x_i) + \delta_t^{j}(x_i)\varepsilon^{j}_{t,i},\label{eq:21}
\end{equation}
where $x_i$ represents the center of each age or age group for $i=1,\dots,p$, $\varepsilon_{t,i}^{j}$ is an independent and identically distributed (iid) standard normal random variable for each age in year $t$, and $\delta_{t}^{j}(x_i)$ measures the variability in mortality at each age in year $t$ for the $j^{\text{th}}$ population. Together, $\delta_t^{j}(x_i)\varepsilon^{j}_{t,i}$ represents the smoothing error. 

Let $m_t^j(x_i) = \exp\left\{y_t^j(x_i)\right\}$ be the observed central mortality rates for age $x_i$ in year $t$ and define $N_t^j(x_i)$ to be the total $j^{\text{th}}$ population of age $x_i$ at 1st January of year $t$. The observed mortality rate approximately follows a binomial distribution with estimated variance
\begin{equation}
\text{Var}\left[m_t^j(x_i)\right] \approx \frac{m_t^j(x_i)\times \left[1-m_t^j(x_i)\right]}{N_t^j(x_i)}.
\end{equation}
Via Taylor's series expansion, the estimated variance associated with the log mortality rate is given by
\begin{equation}
\left(\widehat{\delta}_t^j\right)^2(x_i)\approx \text{Var}\left\{\ln\left[m_t^j(x_i)\right]\right\} = \frac{1-m_t^j(x_i)}{m_t^j(x_i) \times N_t^j(x_i)}. 
\end{equation}
Since $m_t^j(x_i)$ is often quite small, $(\delta_t^j)^2(x_i)$ can be approximated by a Poisson distribution with estimated variance
\begin{equation}
\left(\widehat{\delta}_t^j\right)^2(x_i) \approx \frac{1}{m_t^j(x_i)\times N_t^j(x_i)}.\label{eq:smoo}
\end{equation} 

As suggested by \cite{HU07}, we smooth mortality rates using weighted penalized regression splines with a partial monotonic constraint for ages above 65, where the weights are equal to the inverse variances given in~\eqref{eq:smoo}. The weights are used to model heterogeneity (different variances) in mortality across different ages. Let the weights be the inverse variances $w_t^j(x_i) = 1/\big[(\delta_t^j)^2(x_i)\big]$, the penalized regression spline can be written as:
\begin{equation}
\widehat{f}_t^j(x_i) = \argmin_{\theta_t(x_i)}\sum^M_{i=1}w_t^j(x_i)\Big|y_t^j(x_i) - \theta_t(x_i)\Big|+\alpha\sum^{M-1}_{i=1}\Big|\theta^{'}_t(x_{i+1})-\theta^{'}_t(x_i)\Big|,
\end{equation}
where $i$ represents different ages (grid points) in a total of $M$ grid points, $\alpha$ is a smoothing parameter, and $^{'}$ symbolizes the first derivative of a function. While the $L_1$ loss function and the $L_1$ roughness penalty are employed to obtain robust estimates, the monotonic increasing constraint helps to reduce the noise from estimation of older ages \citep[see also][]{HN99}. In the multilevel functional data model, we first apply~\eqref{eq:21} to smooth multiple sets of curves from different populations that may be correlated. 

The multilevel functional data model can be related to a two-way functional analysis of variance model studied by \cite{MVB+03}, \cite{CF10} and \citet[][Section 5.4]{Zhang14}, it is a special case of the general `functional mixed model' proposed in \cite{MC06}. In the case of two populations, the basic idea is to decompose curves among different populations into an average of total mortality $\mu(x)$, a sex-specific deviation from the averaged total mortality $\eta^{j}(x)$, a common trend across populations $R_t(x)$, a sex-specific residual trend $U_t^{j}(x)$, and measurement error $e_t^{j}(x)$ with finite variance $(\sigma^2)^{j}$. The common and sex-specific residual trends are modeled by projecting them onto the eigenvectors of covariance operators of the aggregate and population-specific centered stochastic processes, respectively. To express our idea, the smoothed mortality rate at year $t$ can be written as:
\begin{equation}
f_t^{j}(x) = \mu(x) + \eta^{j}(x) + R_t(x) + U_t^{j}(x),\qquad x\in \mathcal{I}.\label{eq:2}
\end{equation}
To ensure identifiability, we assume two stochastic processes $R(x)$ and $U^j(x)$ are uncorrelated but we allow correlations among their realizations. 

Because the centered stochastic processes $R(x)$ and $U^j(x)$ are unknown in practice, the population eigenvalues and eigenfunctions can only be approximated through a set of realizations $\bm{R}(x) = \left\{R_1(x),\dots,R_n(x)\right\}$ and $\bm{U}^j(x) = \left\{U_1^j(x),\dots,U_n^j(x)\right\}$. From the covariance function of $\bm{R}(x)$, we can extract a set of functional principal components and their corresponding scores, along with a set of residual functions. Based on the covariance function of residual functions, we can then extract a second set of functional principal components and their associated scores. While the first functional principal component decomposition captures the common trend from total mortality rates, the second functional principal component decomposition captures the sex-specific residual trend.

The sample versions of the aggregate mean function, sex-specific mean function deviation, common trend, and sex-specific residual trend, for a set of dense and regularly spaced functional data, can be estimated by:
\begin{align}
  \widehat{\mu}(x)&=\frac{1}{n}\sum^n_{t=1}f_t^{\text{T}}(x), \label{eq:31}\\
  \widehat{\eta}^{j}(x) &= \widehat{\mu}^{j}(x)-\widehat{\mu}(x),\label{eq:32}\\
  \widehat{R}_t(x) &= \sum^{\infty}_{k=1}\widehat{\beta}_{t,k}\widehat{\phi}_k(x) \approx \sum^K_{k=1}\widehat{\beta}_{t,k}\widehat{\phi}_k(x),\label{eq:3}\\
  \widehat{U}^{j}_t(x) &= \sum^{\infty}_{l=1}\widehat{\gamma}^{j}_{t,l}\widehat{\psi}_l^{j}(x) \approx \sum^L_{l=1}\widehat{\gamma}^{j}_{t,l}\widehat{\psi}_l^{j}(x),\label{eq:4}
\end{align}
where $\{f_1^{\text{T}}(x),\dots,f_n^{\text{T}}(x)\}$ represents a set of smoothed functions for the age-specific total mortality; $\widehat{\mu}(x)$ represents the simple average of the total mortality, whereas $\widehat{\mu}^j(x)$ represents the simple average of females or males; $\{\widehat{\bm{\beta}}_k = (\widehat{\beta}_{1,k},\dots,\widehat{\beta}_{n,k}); k=1,\dots,K\}$ represents the $k^{\text{th}}$ sample principal component scores of $\bm{R}(x)$, $\bm{\Phi}=\left[\widehat{\phi}_1(x),\dots,\widehat{\phi}_{K}(x)\right]$ are the corresponding orthogonal sample eigenfunctions in a square integrable function space. Similarly, $\{\widehat{\bm{\gamma}}_l^j = (\widehat{\gamma}^{j}_{1,l},\dots,\widehat{\gamma}^{j}_{n,l}); l=1,\dots,L\}$ represents the $l^{\text{th}}$ sample principal component scores of $\bm{U}^{j}(x)$, $\bm{\Psi}=\left[\widehat{\psi}_1^{j}(x),\dots,\widehat{\psi}_L^{j}(x)\right]$ are the corresponding orthogonal sample eigenfunctions, $K$, $L$ are truncation lags. As two stochastic processes $R(x)$ and $U^j(x)$ are uncorrelated, $\widehat{\bm{\beta}}_k$ are uncorrelated with $\widehat{\bm{\gamma}}_l^j$. 

Substituting Equations~\eqref{eq:31}--\hspace{-.05in}~\eqref{eq:4} into Equations~\eqref{eq:2}--\hspace{-.05in}~\eqref{eq:21}, we obtain 
\begin{equation*}
  y_t^{j}(x)=\widehat{\mu}(x)+\widehat{\eta}^{j}(x)+\sum^K_{k=1}\widehat{\beta}_{t,k}\widehat{\phi}_k(x)+\sum^L_{l=1}\widehat{\gamma}_{t,l}^{j}\widehat{\psi}_l^{j}(x)+e_t^{j}(x)+\delta^j_t(x)\varepsilon_{t}^{j},
\end{equation*}
where $\widehat{\beta}_{t,k}\sim \text{N}\Big(0,\widehat{\lambda}_k\Big)$, and $\widehat{\lambda}_k$ represents the $k^{\text{th}}$ eigenvalue of empirical covariance operator associated with the common trend; $\widehat{\gamma}_{t,l}^{j}\sim \text{N}\left(0,\widehat{\lambda}_l^{j}\right)$, and $\widehat{\lambda}_l^{j}$ represents the $l^{\text{th}}$ eigenvalue of empirical covariance operator associated with the sex-specific residual trend; and $e_t^{j}(x)\sim N\left(0, (\widehat{\sigma}^2)^j\right)$ represents model errors due to finite truncation. 

Selecting the number of principal components, $K$ and $L$,  is an important practical issue. Four common approaches are cross validation \citep{RS91}, Akaike's information criterion \citep{YMW05}, bootstrap method \citep{HV06}, and explained variance \citep{CG10, Chiou12}. We use a cumulative percentage of total variation to determine $K$ and $L$. The optimal numbers of $K$ and $L$ are determined by: 
\begin{align}
K &= \argmin_{K: K\geq 1}\left\{\sum_{k=1}^K\widehat{\lambda}_k\Big/\sum_{k=1}^{\infty}\widehat{\lambda}_k\mathds{1}{\big\{\widehat{\lambda}_k>0\big\}}\geq P_1\right\},\\
L &= \argmin_{L: L\geq 1}\left\{\sum_{l=1}^L\widehat{\lambda}_l^j\Big/\sum_{l=1}^{\infty}\widehat{\lambda}_l^j\mathds{1}{\big\{\widehat{\lambda}_l^j>0\big\}}\geq P_2\right\}, 
\end{align}
where $\mathds{1}\{\cdot\}$ denotes a binary indicator function. Following \cite{Chiou12}, we chose $P_1=P_2=0.9$.

An important parameter is the proportion of variability explained by aggregate data, which is the variance explained by the within-cluster variability \citep{DCC+09}. A possible measure of within-cluster variability is given by: 
\begin{equation}
\frac{\sum^{\infty}_{k=1}\lambda_k}{\sum^{\infty}_{k=1}\lambda_k+\sum^{\infty}_{l=1}\lambda_l}=\frac{\int_{\mathcal{I}} \text{Var}\left[\bm{R}(x)\right]dx}{\int_{\mathcal{I}}\text{Var}\left[\bm{R}(x)\right]dx + \int_{\mathcal{I}}\text{Var}\left[\bm{U}^{j}(x)\right]dx}.\label{eq:cluster}
\end{equation}
When the common factor can explain the main mode of total variability, the value of within-cluster variability is close to 1. 

For multiple populations, the other important parameter is the total variability for a population, given by
\begin{equation}
\frac{1}{n}\sum^n_{t=1}[f_t(x) - \bar{f}(x)][f_t(w) - \bar{f}(w)], \qquad x, w\in \mathcal{I}.\label{eq:total_var}
\end{equation}
This allows us to identify the population with larger variability. 

Conditioning on the estimated principal components $\bm{\Phi}$, $\bm{\Psi}$ and continuous functions $\bm{y}^{j}=\Big[y_1^{j}(x),\dots,y_n^{j}(x)\Big]$, the $h$-step-ahead point forecasts of $y_{n+h}^{j}(x)$ are given by:
\begin{align*}
  \widehat{y}_{n+h|n}^{j}(x)&=\text{E}\left[y_{n+h}(x)\middle|\mu(x), \eta(x), \bm{\Phi},\bm{\Psi},\bm{y}^{j}\right]\\
  &=\widehat{\mu}(x)+\widehat{\eta}^{j}(x)+\sum^K_{k=1}\widehat{\beta}_{n+h|n,k}\widehat{\phi}_k(x)+\sum^L_{l=1}\widehat{\gamma}^{j}_{n+h|n,l}\widehat{\psi}_l^{j}(x),
\end{align*}
where $\widehat{\beta}_{n+h|n,k}$ and $\widehat{\gamma}_{n+h|n,l}^{j}$ are the forecast principal component scores, obtained from a univariate time-series forecasting method, such as the random walk with drift (rwf) or autoregressive integrated moving average (ARIMA)$(p,d,q)$ model. The automatic algorithm of \cite{HK08} is able to choose the optimal orders $p, q$ and $d$ automatically. $d$ is selected based on successive Kwiatkowski-Phillips-Schmidt-Shin (KPSS) unit-root test \citep{KPSS92}. KPSS tests are used for testing the null hypothesis that an observable time series is stationary around a deterministic trend. We first test the original time series for a unit root; if the test result is significant, then we test the differenced time series for a unit root. The procedure continues until we obtain our first insignificant result. Having identified $d$, the orders of $p$ and $q$ are selected based on the Akaike information criterion \citep{Akaike74} with a correction for finite sample sizes. The maximum likelihood method can then be used to estimate these parameters. It is noteworthy that a multivariate time-series method, such as vector autoregressive model, can also be used to model and forecast stationary principal component scores \citep[see for example,][]{ANH15}.

\cite{HBY13} used the autoregressive fractionally integrated moving average (ARFIMA) in the product-ratio method (see Section~\ref{sec:pr}), which allows non-integer values for the difference parameter, to forecast the principal component scores. For any two populations, convergent forecasts are obtained when $\left\{\widehat{\gamma}_{n+h|n,l}^{\text{F}}-\widehat{\gamma}_{n+h|n,l}^{\text{M}}\right\}$ is stationary for each $l$.

As pointed out by \cite{LL05}, if $\left\{\widehat{\gamma}_{n+h|n,l}^{\text{F}}-\widehat{\gamma}_{n+h|n,l}^{\text{M}}; l=1,\dots,L\right\}$ has a trending long-term mean, the Li and Lee method fails to achieve convergence. As an extension of the Li and Lee method, the proposed method may also fail to achieve convergence. However, if the common mean function and common trend capture the long-term effect, the Li-Lee and multilevel functional data methods produce convergent forecasts, as the forecasts of residual trends would be flat.

To quantify forecast uncertainty, the interval forecasts of $y_{n+h}^{j}(x)$ can be obtained through a Bayesian paradigm equipped with Markov chain Monte Carlo (MCMC) for estimating all variance parameters and drawing samples from the posterior of principal component scores. Given errors are assumed to be normally distributed, a hierarchical regression model is able to capture fixed and random effects \citep[see for example][Chapter 11.1]{RCG+13, Hoff09}. With a set of MCMC outputs, the forecasts of future sample path are given by:
\begin{align}
  \widehat{y}_{n+h|n}^{b,j}(x) = \ & \text{E}\left[y_{n+h}(x)\middle|\mu(x), \eta(x), \bm{\Phi},\bm{\Psi},\bm{y}^{j}\right]\notag\\
  = \ & \widehat{f}_{n+h}^{b,j}(x) + \widehat{\delta}_{n+h}^{b,j}(x)\varepsilon_{n+h}^{b,j}, \notag\\
  = \ & \widehat{\mu}(x)+\widehat{\eta}^{j}(x)+\sum^K_{k=1}\widehat{\beta}_{n+h|n,k}^b\widehat{\phi}_k(x)+\sum^L_{l=1}\widehat{\gamma}_{n+h|n,l}^{b,j}\widehat{\psi}_l^{j}(x)+ \label{eq:interval} \\
  &\widehat{e}_{n+h}^{b,j}(x)+\widehat{\delta}_{n+h}^{b,j}(x)\varepsilon_{n+h}^{b,j},\notag 
\end{align}
for $b=1,\dots,B$. We first simulate $\left\{\widehat{\beta}^b_{1,k},\dots,\widehat{\beta}^b_{n,k}\right\}$ drawn from its full conditional density, and then obtain $\widehat{\beta}_{n+h|n,k}^b$ using a univariate time-series forecasting method for each simulated sample; similarly, we first simulate $\left\{\widehat{\gamma}^{b,j}_{1,l},\dots,\widehat{\gamma}^{b,j}_{n,l}\right\}$ drawn from its full conditional density, and then obtain $\widehat{\gamma}_{n+h|n,l}^{b,j}$ for each simulated sample; $\left(\widehat{\sigma}^2\right)^{b,j}$ is drawn from its full conditional density. The derivation of full conditional densities is given in the Supplement B \citep{Shang16}, while some WinBUGS computation code is presented in the Supplement C \citep{Shang16}. As we pre-smooth the functional data, we must add the smoothing error $\widehat{\delta}_{n+h}^{b,j}(x)\varepsilon_{n+h}^{b,j}$, where $\widehat{\delta}_{n+h}^{b,j}(x)$ is simulated from its posterior and $\varepsilon_{n+h}^{b,j}$ is drawn from $N(0,1)$. 

The total number of MCMC draws is 20,000 iterations, the first 10,000 iterations are used for the burn-in, whereas the remaining 10,000 iterations are recorded. Among these recorded draws, we keep every $10^{\text{th}}$ draw in order to reduce autocorrelation. The prediction interval is constructed from the percentiles of the bootstrapped mortality forecasts. The point and interval forecasts of life expectancy are obtained from the forecast age-specific mortality rates using the life table method \citep[see for example,][]{PHG01}. In this paper, we focus on forecasting life expectancy at birth, described simply as life expectancy hereafter.

\section{Relationship to two existing coherent methods}\label{sec:relate}

\subsection{Relationship to the augmented common factor method}\label{sec:aug}

The multilevel functional data method can be viewed as a generalization of the augmented common factor method of \cite{LL05}. They proposed the following model for the two-sex case, which can be expressed using a functional data model notation:
\begin{equation*}
  y_t^j(x_i) = \widehat{\mu}^j(x_i)+\widehat{\beta}_t\widehat{\phi}(x_i)+\widehat{\gamma}_{t}^j\widehat{\psi}^j(x_i)+e_t^j(x_i),
\end{equation*}
where $x_i$ represents a discrete age or age group, $\widehat{\mu}^j(x_i)$ is the age- and sex-specific mean, $(\widehat{\beta}_1,\dots,\widehat{\beta}_n)$ is the mortality index of the common factor, which can be forecast by random walk with drift; $\widehat{\phi}(x_i)$ is the first estimated principal component of the common factor of \possessivecite{LC92} model (based on log mortality), and it measures the sensitivity of the log total mortality to changes in $\{\beta_1,\dots,\beta_n\}$ over time; $\widehat{\gamma}_t^j$ is the time component of the additional factor, and it can be forecast by an autoregressive (AR) process of order 1; $\widehat{\psi}^j(x_i)$ is the first estimated principal component of the residual matrix that is specific to males or females; and $e_t^j(x_i)$ is the error term. $\widehat{\beta}_t\widehat{\phi}(x_i)$ specifies the long-term trend in mortality change and random fluctuations that are common for all populations, whereas $\widehat{\gamma}_t^j\widehat{\psi}^j(x_i)$ describes the short-term changes that are specific only for $j^{\text{th}}$ population. The augmented common factor model takes into account the mortality trends in all populations by applying the Lee-Carter method twice, subject to identifiability constraints $\sum_{i=1}^p\widehat{\phi}(x_i)=1$ and $\sum^n_{t=1}\widehat{\beta}_t=0$. The eventual constant ratio between the age-specific mortality rates will thus be adjusted to the short term according to the population-specific deviations from the common pattern and trend \citep*{JVK13}. If the $|\widehat{\gamma}_{n+h|n}^{\text{F}}-\widehat{\gamma}_{n+h|n}^{\text{M}}|$ values become constant, this model leads to non-divergent forecasts in the long run but not necessarily in the short term in the case of two populations \citep{LL05}.

There are two main differences between the proposed multilevel functional data method and \possessivecite{LL05} method. First, \possessivecite{LL05} method uses a single principal component to capture the largest amount of variation. In contrast, the multilevel functional data method includes the option of incorporating more than just one component by selecting the number of components based on the cumulative percentage of total variation in the data \citep{CG10,Chiou12}. An examination of the residual contour plots can help to reveal the existence of any systematic patterns not being accounted for. In such cases, the additional principal components capture patterns in the data that may not necessarily be explained by the first principal component. As noted by \cite{HBY13}, the use of multiple principal components does not introduce additional model complexity because the scores are uncorrelated and components are orthogonal by construction. In a similar vein, \cite{BMS02} considered up to three components in total when analyzing data of both sexes combined, and found that clustering in the residuals was diminished after the addition of extra components. \cite{DDG+06} modeled five countries' data simultaneously with a number of components, and \cite{Li13} modeled Australian female and male mortality and life expectancy jointly using more than one component. 

The second main difference between the proposed multilevel functional data method and that of \cite{LL05} is that the latter restricted the univariate time-series forecasting method to be random-walk with drift for $\widehat{\beta}_t$ and AR(1) for $\widehat{\gamma}_{t}^j$. These choices for the univariate time-series forecasting method may not necessarily be optimal for a given time series. In contrast, we implemented the \textit{auto.arima} algorithm of \cite{HK08}, which selects the optimal order of ARIMA process based on the corrected Akaike information criterion.

\subsection{Relationship to the product-ratio method}\label{sec:pr}

Let us again consider modeling mortality in the two-sex case. The product-ratio method begins by obtaining the product and ratio functions of all series. The product function can be seen as the sum of all series in the log scale, whereas the ratio function can be seen as the differences among series in the log scale. It first applies an independent functional data method to forecast the future realizations of product and ratio functions, then transforms the forecasts of product and ratio functions back to the original male and female age-specific mortality rates. The convergent forecasts are achieved through the ARFIMA modeling of the ratio function, which implicitly prevents it from diverging in a long-run. This constraint ultimately results in a better forecast accuracy than the independent functional data method for males, but worse forecast accuracy for females. A possible explanation is that the product-ratio method improves the goodness of fit for males at the cost of reduced goodness of fit for females. 

The prediction intervals of mortality are constructed based on the normality assumption in \cite{HBY13}, although it is possible to use a bootstrap method \citep[see for example,][]{HS09}. In contrast, in the multilevel functional data method, the prediction intervals of mortality were constructed based on Bayesian paradigm. The validity of Bayesian paradigm for principal component scores has been given in \citet[][supplement A]{DCC+09}. For a small sample size, a Bayesian sampling technique is known to produce more accurate interval forecast accuracy than the one based on the normality assumption \citep[see][p.174 for details]{Chernick08}. 

\section{Application to UK age- and sex-specific mortality}\label{sec:3}

\begin{figure}[!htbp]
\centering
\includegraphics[width=8.3cm]{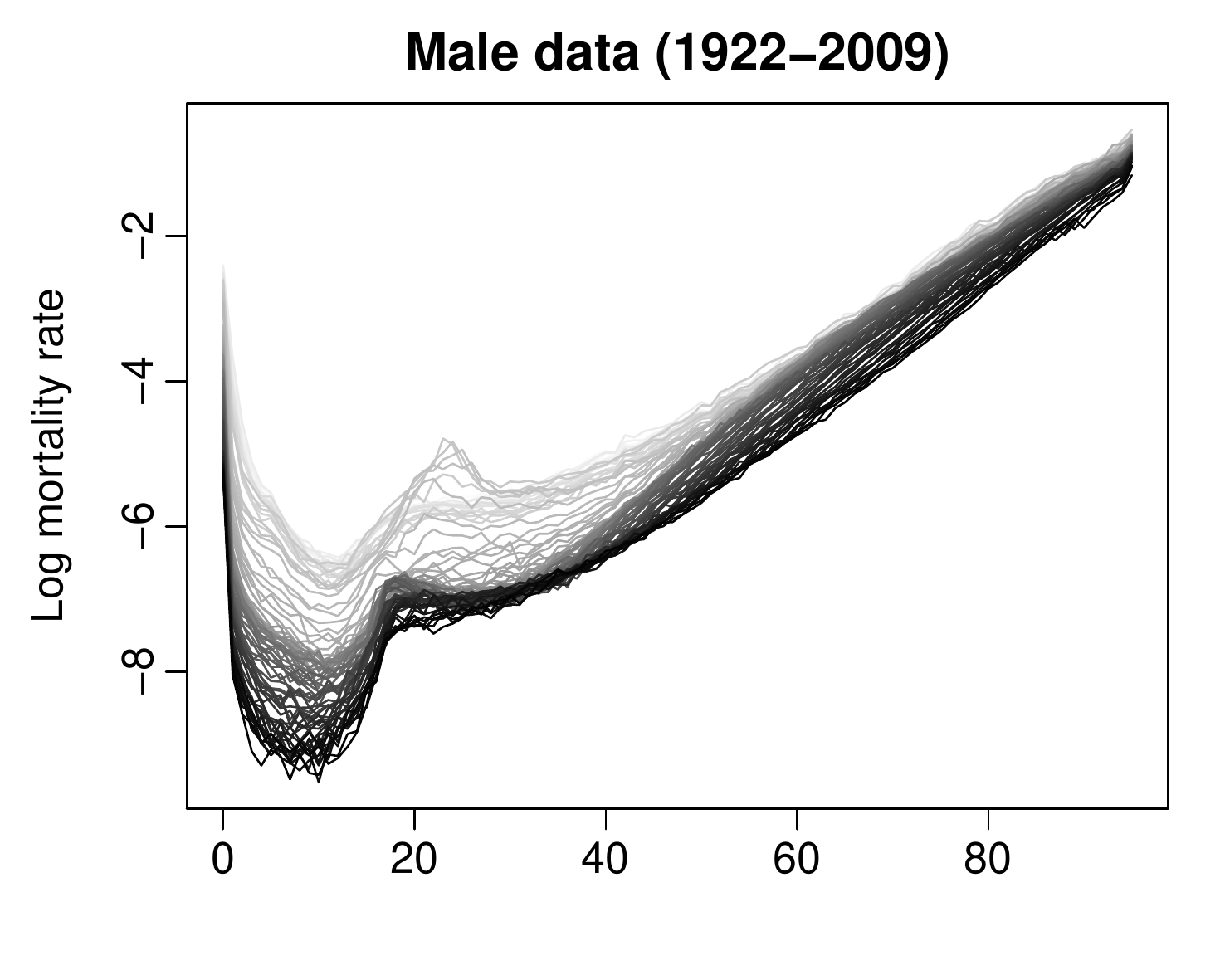} 
\quad
\includegraphics[width=8.3cm]{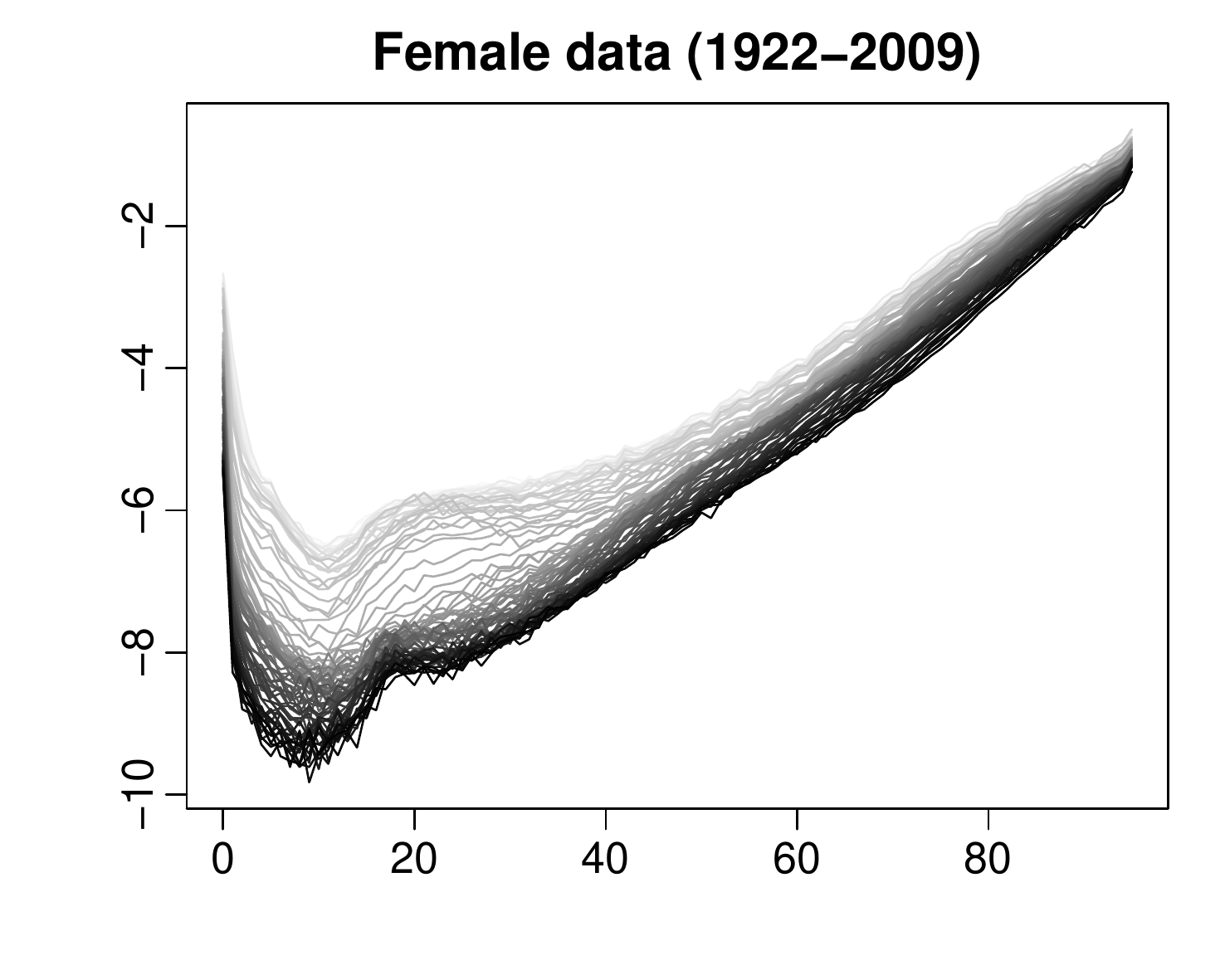}
\\
\includegraphics[width=8.3cm]{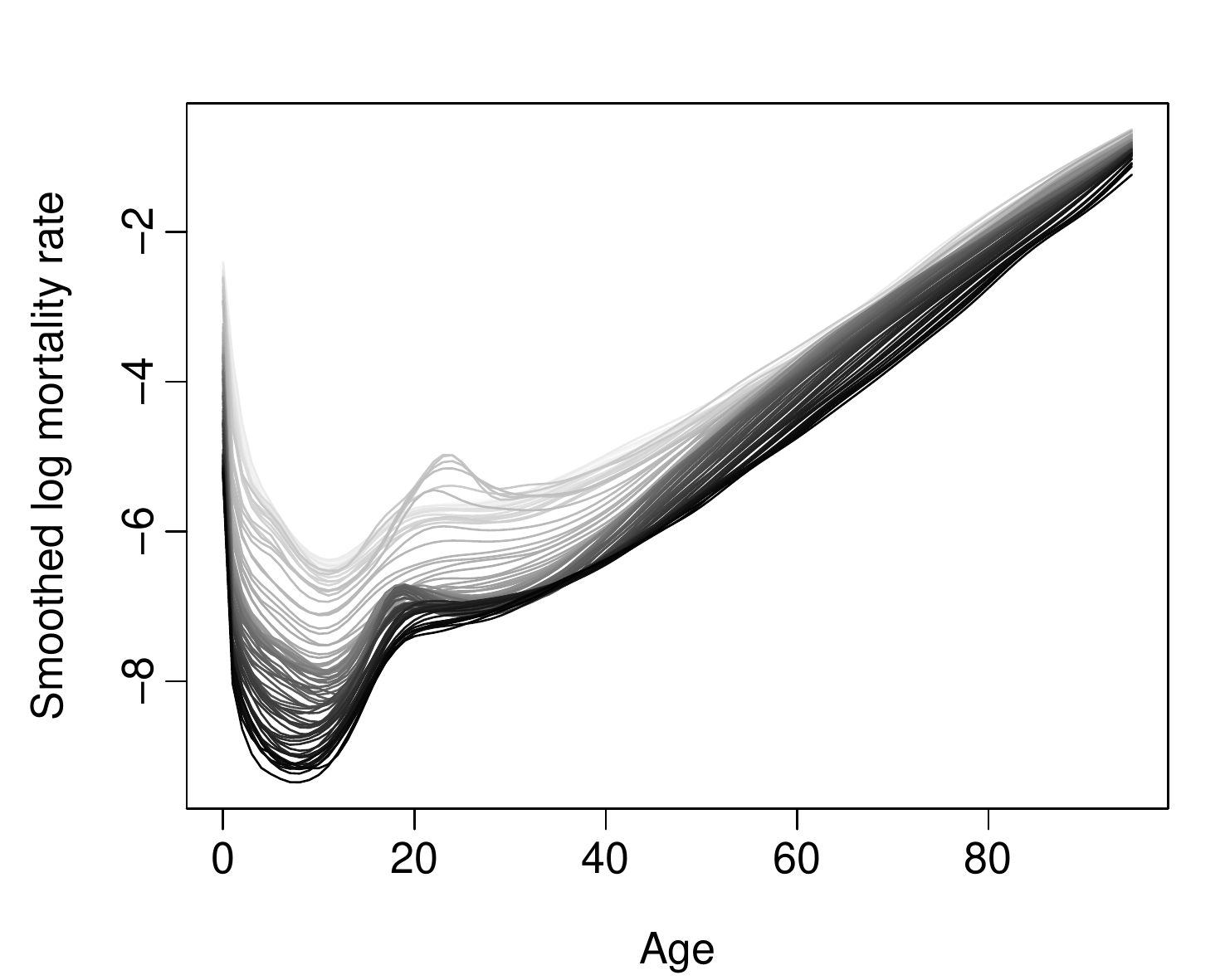}
\quad
\includegraphics[width=8.3cm]{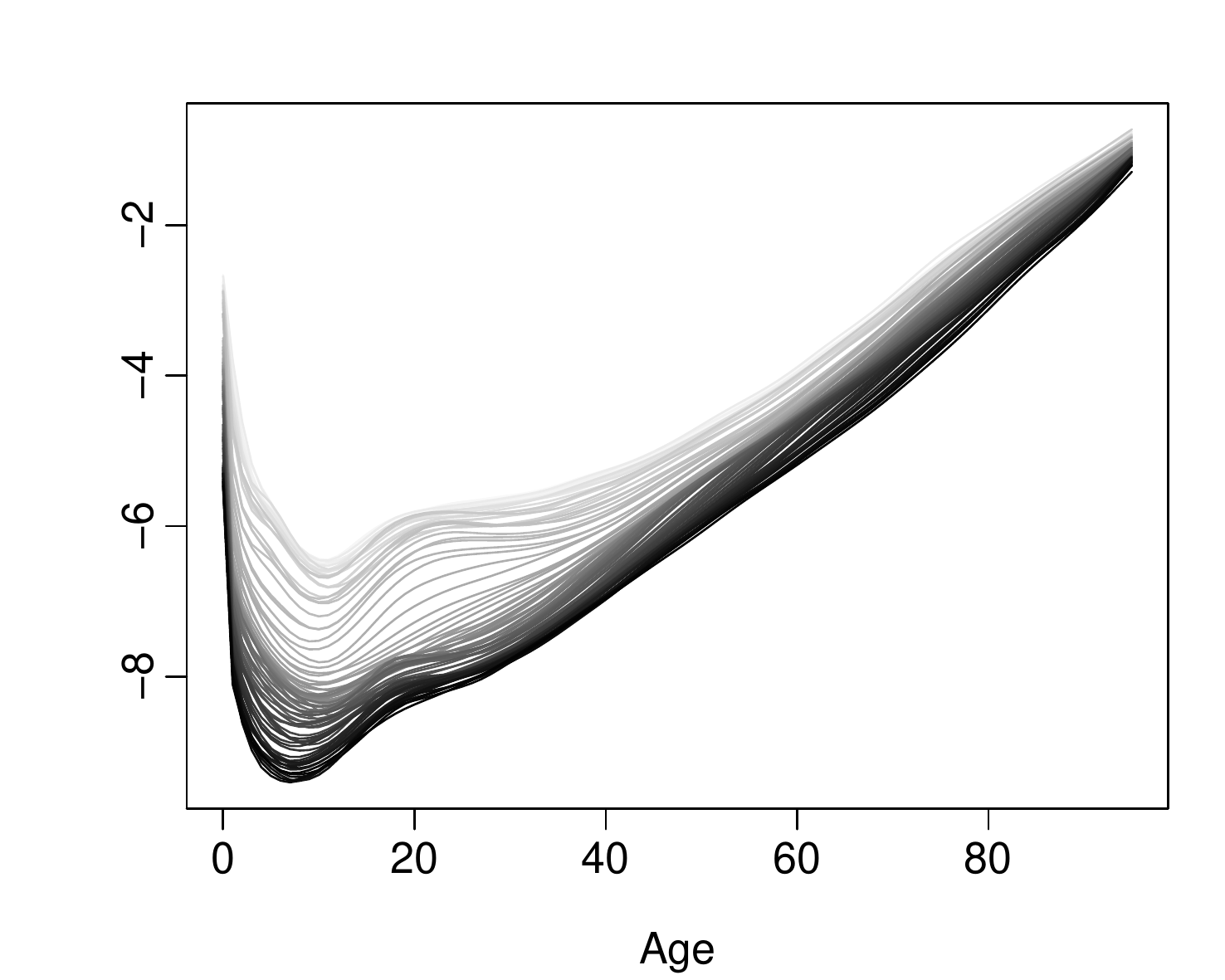}
\caption{Observed and smoothed age-specific male and female log mortality rates in the UK. Data from the distant past are shown in light gray, and the most recent data are shown in dark gray.}\label{fig:1}
\end{figure}

Age- and sex-specific raw mortality data for the UK between 1922 and 2009 are available from the \cite{HMD13}. For each sex in a given calendar year, the mortality rates obtained by the ratio between ``number of deaths" and ``exposure to risk", are arranged in a matrix for age and calendar year. By analyzing the changes in mortality as a function of both age $x$ and year $t$, it can be seen that mortality rates have shown a gradual decline over time. To provide an idea of this evolution, we present the functional time-series plot for male and female log mortality rates in Figure~\ref{fig:1}. Mortality rates dip from their early childhood high, climb in the teen years, stabilize in the early 20s, and then steadily increase with age. We further notice that for both males and females, mortality rates are declining over time, especially in the younger and older ages. Despite the higher male mortality rates in comparison to females, the difference becomes smaller and smaller over years at the older ages.

\begin{figure}[!htbp]
  \centering
  \includegraphics[width=\textwidth]{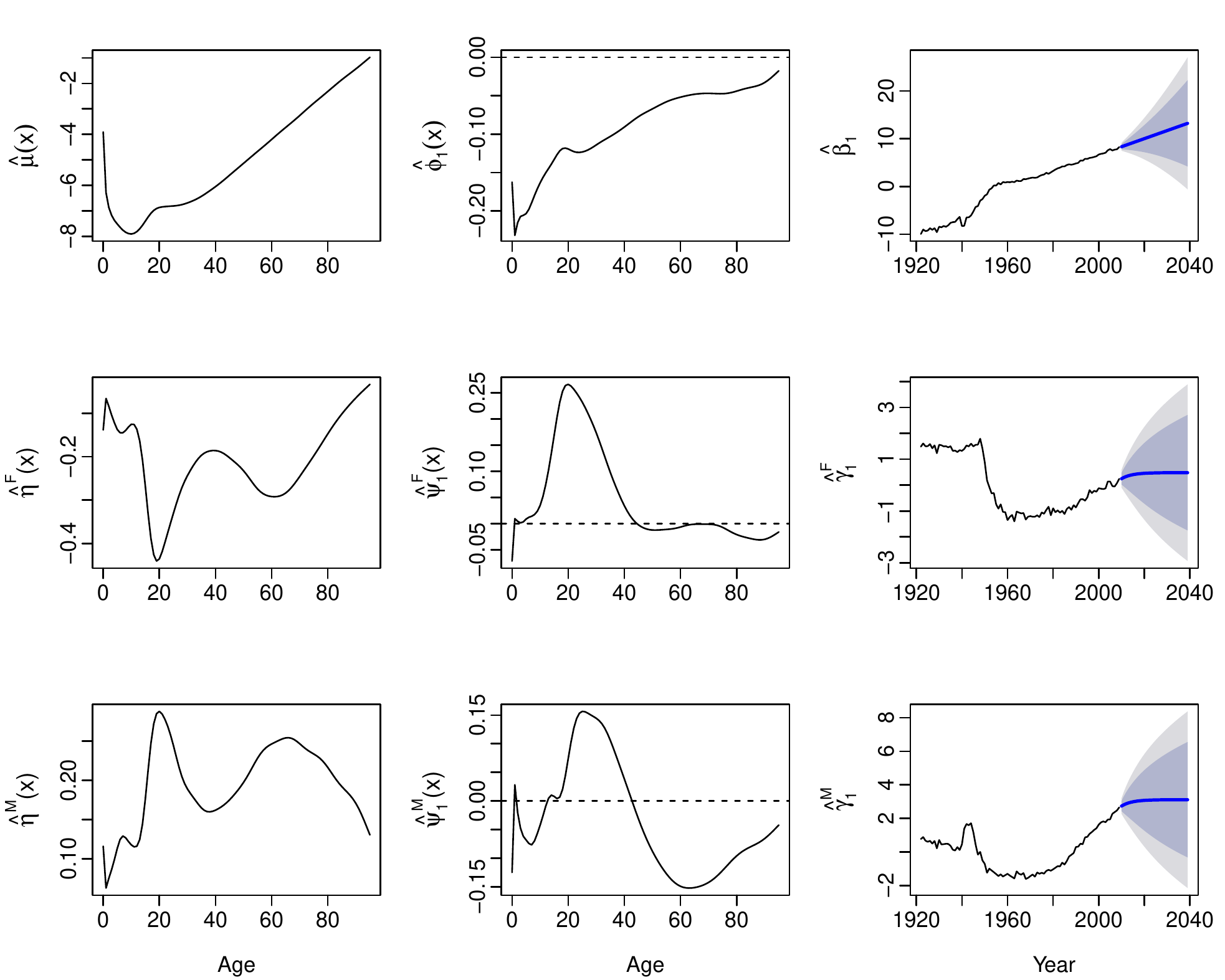}
  \caption{Estimated common mean function, first common functional principal component, and associated scores for UK total mortality (top); estimated mean function deviation for females, first functional principal component, and associated scores for UK female mortality (middle); estimated mean function deviation for males, first functional principal component, and associated scores for UK male mortality (bottom). The dark and light gray regions show the 80\% and 95\% prediction intervals, respectively.}\label{fig:2}
\end{figure}

In the top panel of Figure~\ref{fig:2}, we display the estimated common mean function $\widehat{\mu}(x)$, first estimated common principal component $\widehat{\phi}_1(x)$ and corresponding principal component scores $\left\{\widehat{\beta}_{1,1},\dots,\widehat{\beta}_{n,1}\right\}$ along with 30-years-ahead forecasts. The first common functional principal component captures more than 98\% of the total variation in the age-specific total mortality. In the middle panel of Figure~\ref{fig:2}, we display the estimated mean function deviance of females from the overall mean function $\widehat{\eta}^{\text{F}}(x)$, first functional principal component for females $\widehat{\psi}_1^{\text{F}}(x)$ and corresponding principal component scores $\left\{\widehat{\gamma}_{1,1}^{\text{F}},\dots,\widehat{\gamma}_{n,1}^{\text{F}}\right\}$ with 30-years-ahead forecasts. In the bottom panel of Figure~\ref{fig:2}, we display the estimated mean function deviance of males from the overall mean function $\widehat{\eta}^{\text{M}}(x)$, first functional principal component for males $\widehat{\psi}_1^{\text{M}}(x)$ and corresponding principal component scores $\left\{\widehat{\gamma}_{1,1}^{\text{M}},\dots,\widehat{\gamma}_{n,1}^{\text{M}}\right\}$ with 30-years-ahead forecasts. In this data set, the first three functional principal components explain at least 90\% of the remaining 10\% total variations for both females and males. Due to limited space, we present only the first functional principal component, which captures more than 64\% and 50\% of the remaining 10\% total variations for females and males, respectively. Based on~\eqref{eq:cluster}, the proportion of variability explained by the total mortality is 94\% for females and 95\% for males, respectively.

From Figure~\ref{fig:2}, it is apparent that the basis functions are modeling different movements in mortality rates: $\widehat{\phi}_1(x)$ primarily models mortality changes in children and adults, $\widehat{\psi}_1^{\text{F}}(x)$ models mortality changes between late-teens and 40, and $\widehat{\psi}_1^{\text{M}}(x)$ models the differences between young adults and those over 60. From the forecast common principal component scores, the mortality changes in children and adults are likely to continue in the future with increasing forecast uncertainty. From the forecasts of sex-specific principal component scores, there are no clear trends associated with each sub-population, as the forecasts would be flat. Thus, it is likely to achieve convergent forecasts between female and male sub-populations.

\begin{figure}[!htbp]
\centering
\includegraphics[width=7.5cm]{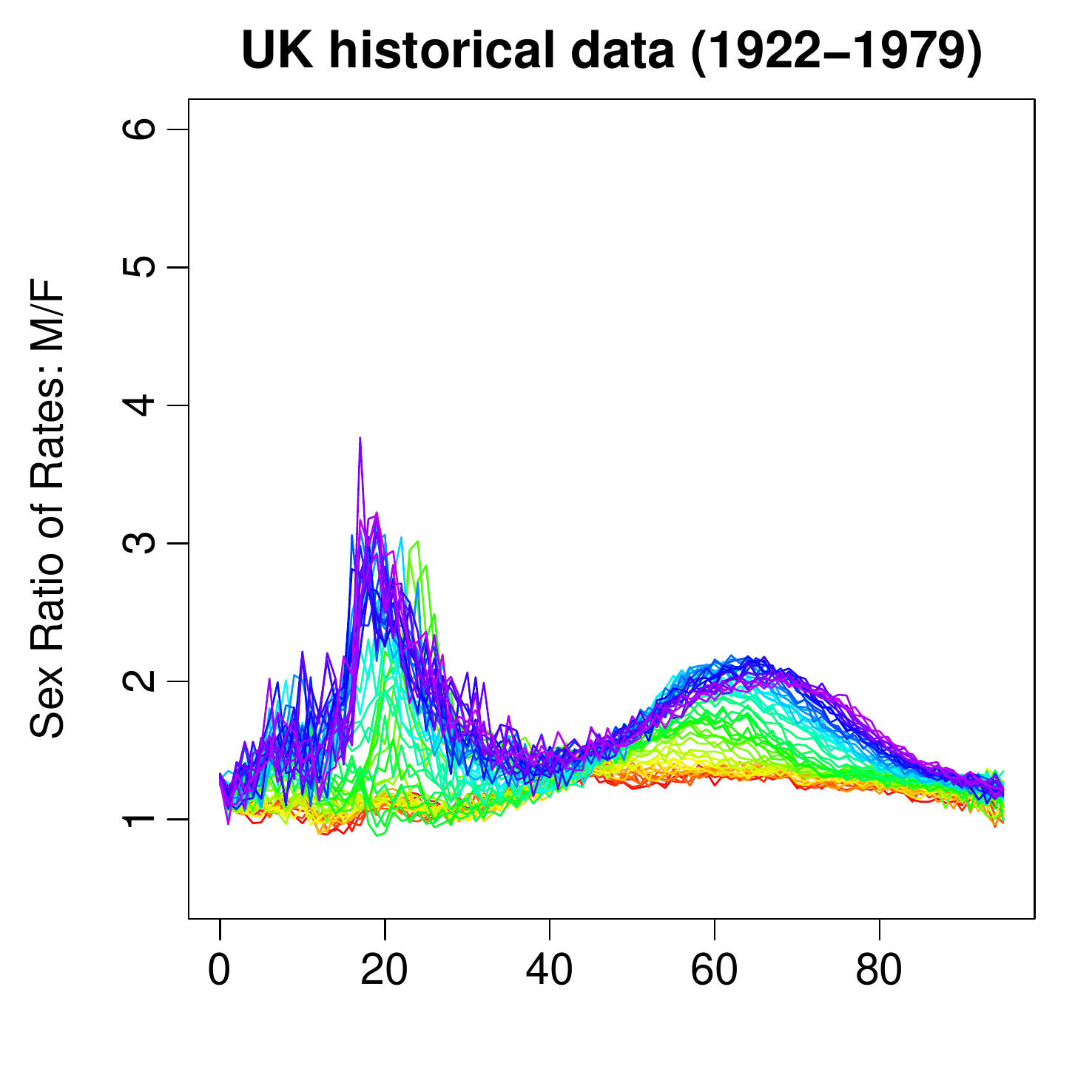}
\quad
\includegraphics[width=7.5cm]{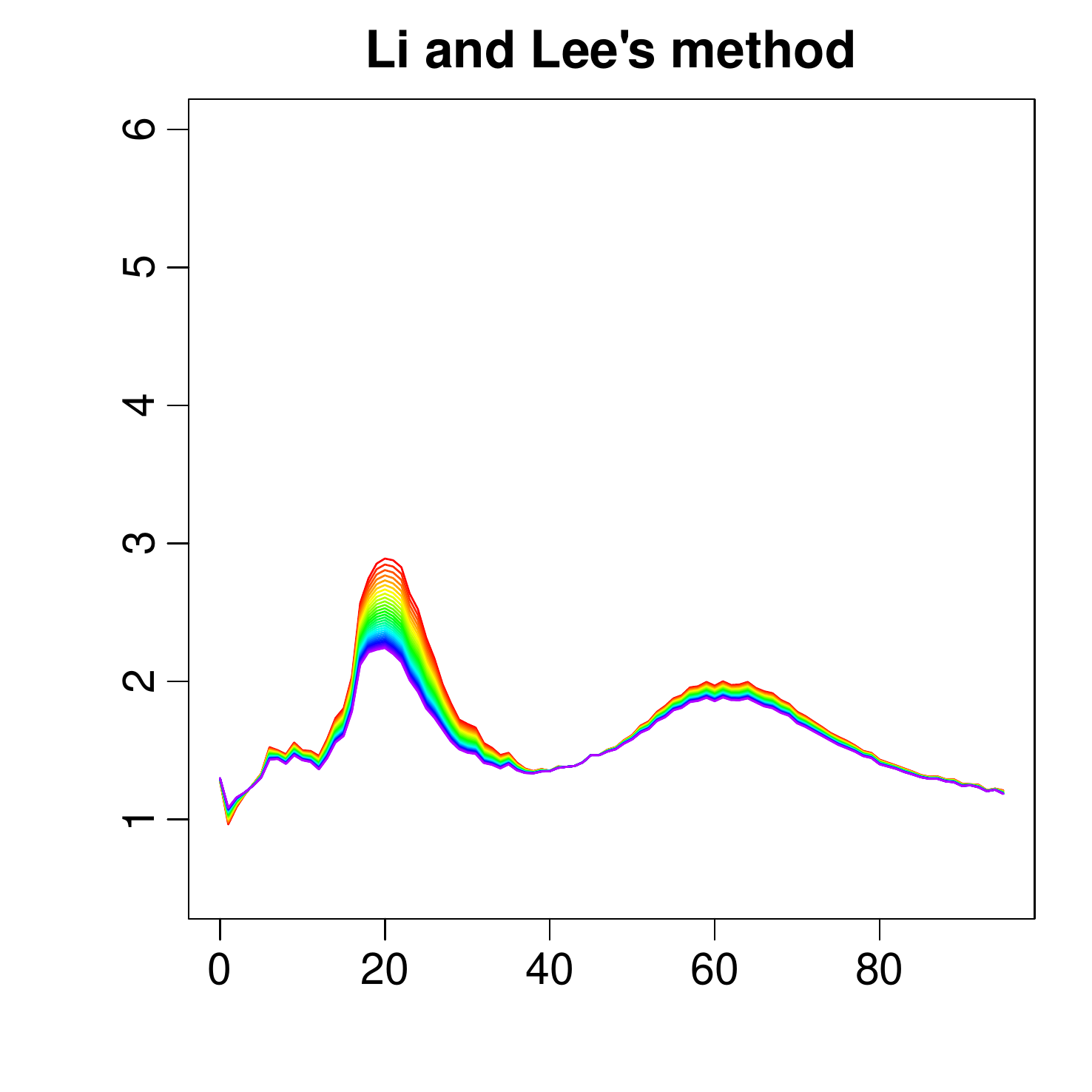}
\\
\includegraphics[width=7.5cm]{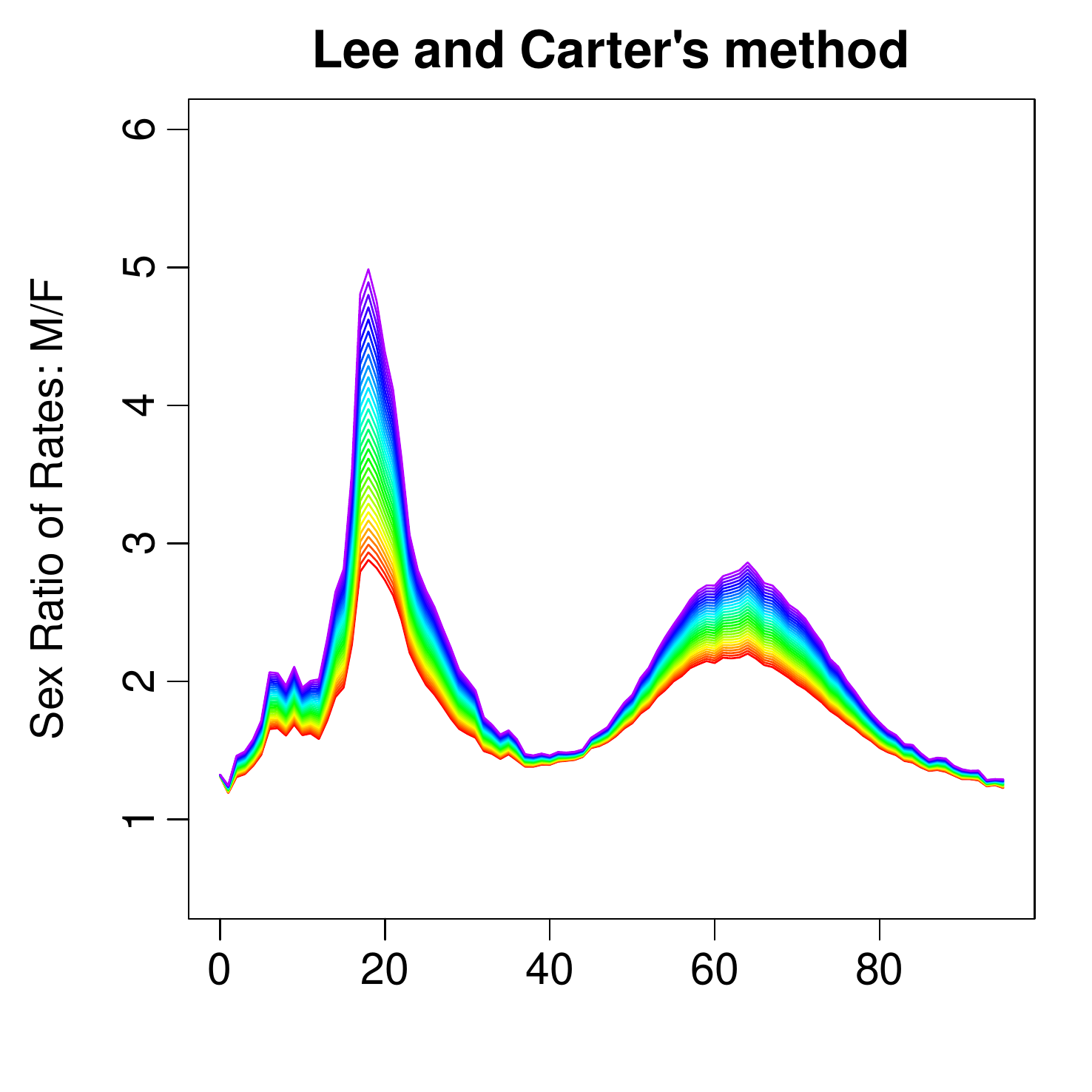}
\quad
\includegraphics[width=7.5cm]{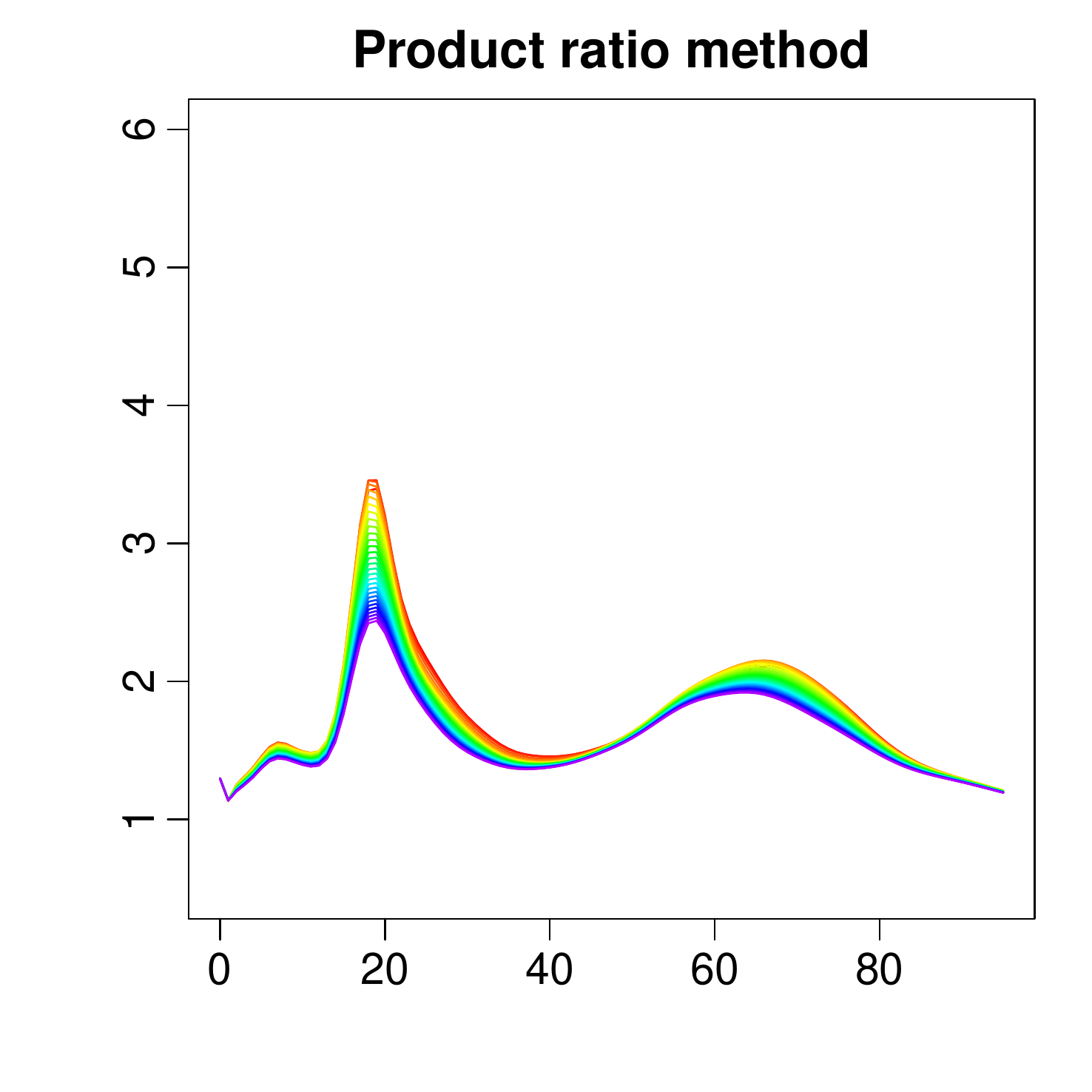}
\\
\includegraphics[width=7.5cm]{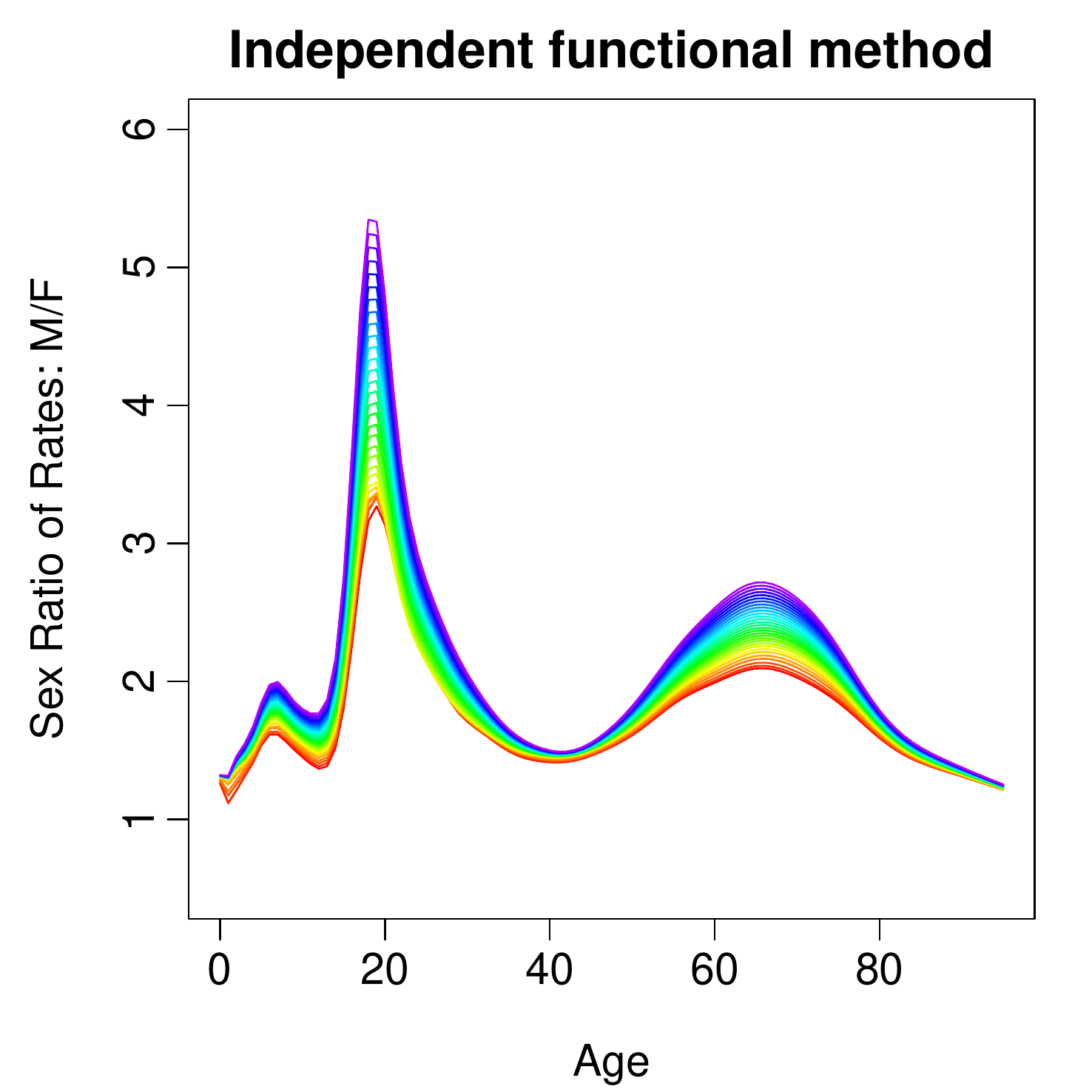}
\quad
\includegraphics[width=7.5cm]{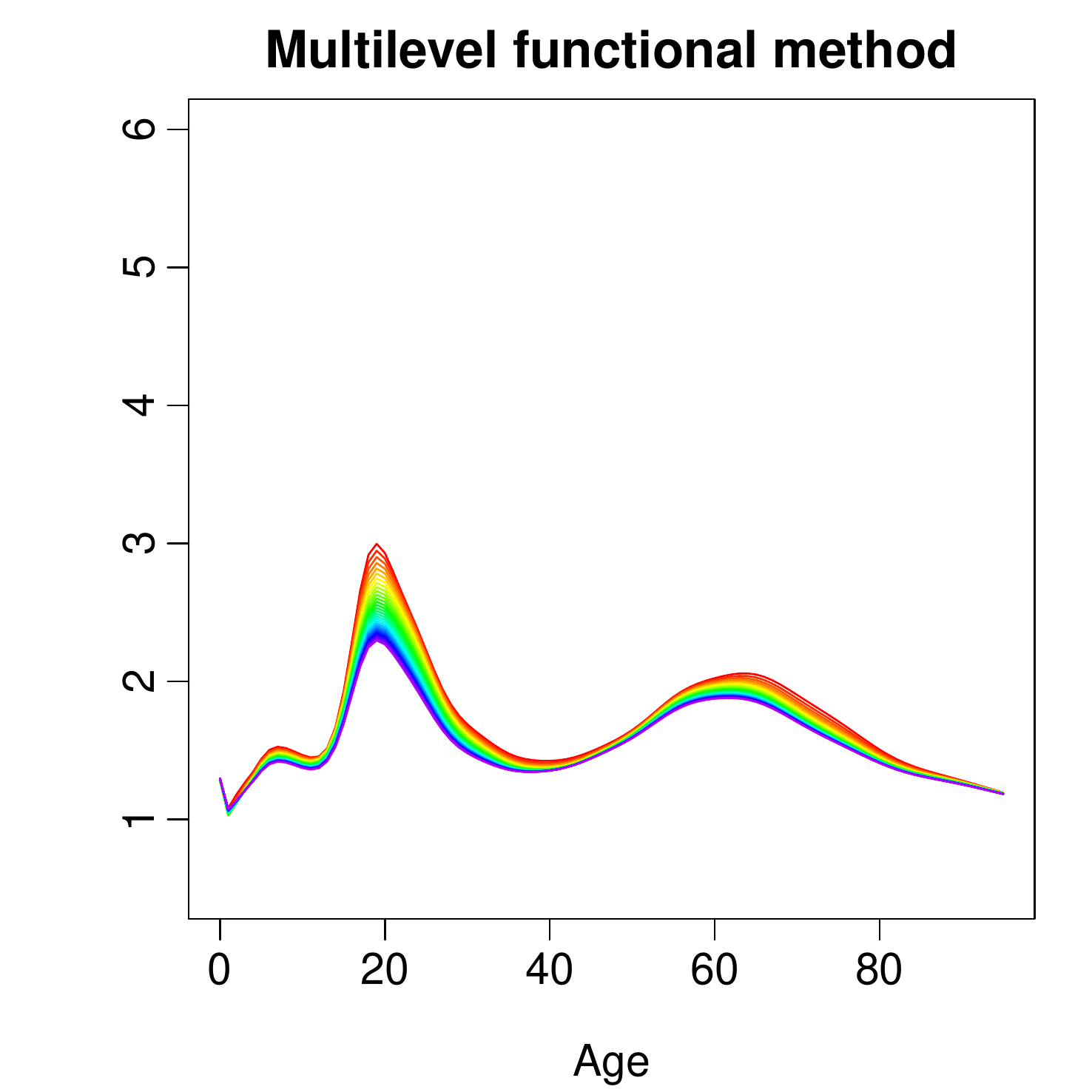}
\caption{30-years-ahead forecasts of mortality sex ratios from 1980 to 2009 in the UK data using Lee and Carter's method, Li and Lee's method, the independent functional data method, the product-ratio method, and the multilevel functional data method (rwf). The forecast curves are plotted using a rainbow color palette; the most recent forecast curves are shown in red, whereas the long-term forecast curves are shown in purple.}\label{fig:3}
\end{figure}

In the first column of Figure~\ref{fig:3}, we plot the historical mortality sex ratios (Male/Female) from 1922 to 1979, alongside the 30-years-ahead forecasts of mortality sex ratios from 1980 to 2009 by the non-coherent forecasting methods, namely Lee and Carter's method and the independent functional data method. In the second column, we show the 30-years-ahead forecasts of mortality sex ratios from 1980 to 2009, using coherent forecasting methods, including Li and Lee's method, and the product-ratio and multilevel functional data methods. We found that all the coherent forecasting methods exhibit a quite similar pattern, with much smaller sex ratios than the non-coherent forecasting methods. Our results confirm the expected trend toward convergence, where the gap in mortality forecasts between males and females gradually converges to a constant for each age. The convergent forecasts demonstrate biological characteristics, for example, the mortality of females has been lower than that of males, it would be counter-intuitive if forecasts of the recent convergence of mortality which has been observed in many developed countries leads to the opposite situation. Our results further reflect the importance of joint modeling, which has already been adopted for the official mortality projection in New Zealand \citep{WD14}.

\section{Multi-country comparison}\label{sec:4}

While joint modeling mortality for multiple populations offers the advantage of avoiding possible undesirable divergence in the forecasts, little is known about whether these methods can improve forecast accuracy at various lengths of forecast horizon. In order to investigate the forecast accuracy of the multilevel functional data method, we consider 15 other developed countries for which data are also available in the \cite{HMD13}. These raw mortality rates are shown in Table~\ref{tab:multi}, along with their respective data periods, within-cluster variability in~\eqref{eq:cluster} and total variance in~\eqref{eq:total_var}. The selected countries are all developed countries with relatively long data series commencing at or before 1950. It was desirable to have a long available data period, in order to obtain consistent sample estimators \citep*{BJR08}. Including the UK data, 32 sex-specific populations were obtained for all analyses. Note that the age groups are single years of age from 0 to 94 and then a single age group for 95 and above, in order to avoid the excessive fluctuations at older ages.

\begin{table}[!htbp]
\centering
\setlength{\tabcolsep}{7.2pt}
\caption{Data period and within-cluster variability for each country.}\label{tab:multi}
\begin{tabular}{@{}lcccc@{}}\toprule
Country 				& Data period 	& \multicolumn{2}{c}{Within-cluster variability} & Variance ratio \\
					&			& Female & Male & Female vs Male \\\toprule
Australia				& 1921 : 2011	& 0.91	& 0.92 & 1 : 1.18		\\
Austria				& 1947 : 2010	& 0.92	& 0.94 & 1 : 1.24		\\
Belgium				& 1841 : 2012	& 0.95	& 0.96 & 1 : 1.13		\\			
Canada   				& 1921 : 2009 	& 0.91 	& 0.94 & 1 : 1.17   	\\
Denmark 				& 1835 : 2011 	& 0.95 	& 0.96 & 1 : 1.11  	\\
France				& 1816 : 2012	& 0.95	& 0.94 & 1 : 1.14		\\
Finland   				& 1878 : 2009 	& 0.93 	& 0.93 & 1 : 1.24 	\\
Italy	        				& 1872 : 2009 	& 0.95 	& 0.94 & 1 : 1.14 	\\
Japan				& 1947 : 2012	& 0.94	& 0.97 & 1 : 1.18		\\
Netherlands			& 1850 : 2009	& 0.97	& 0.97 & 1 : 1.10	 	\\
Norway    				& 1846 : 2009 	& 0.94 	& 0.96 & 1 : 1.16	\\
Spain				& 1908 : 2009	& 0.95	& 0.96 & 1 : 1.19		\\
Sweden 				& 1751 : 2011 	& 0.96 	& 0.96 & 1 : 1.11	\\	
Switzerland 			& 1876 : 2011 	& 0.95 	& 0.97 & 1 : 1.16	\\	
United Kingdom		& 1922 : 2009	& 0.94 	& 0.94 & 1 : 1.16	\\
United States of America 	& 1933 : 2010 	& 0.92 	& 0.94 & 1 : 1.20	\\\bottomrule
\end{tabular}
\end{table}

\subsection{Forecast accuracy evaluation}

\subsubsection{Evaluation of point forecast accuracy}\label{sec:evalpoint}

We split our age- and sex-specific data into a training sample (including data from years $1$ to $(n-30)$) and a testing sample (including data from years $(n-29)$ to $n$), where $n$ represents the total number of years in the data. The length of the fitting period differs by country (see Table~\ref{tab:multi}). We implement a rolling origin approach, following \cite{HBY13} and \cite{SBH11}. A rolling origin analysis of a time-series model is commonly used to assess model and parameter stabilities over time. A common technique to assess the constancy of a model's parameter is to compute parameter estimates and their forecasts over a rolling origin of a fixed size through the sample \citep[see][Chapter 9 for more details]{ZW06}. The advantage of the rolling origin approach is that it allows us to assess the point and interval forecast accuracy among methods for different forecast horizons. With the initial training sample, we produce one- to 30-year-ahead forecasts, and determine the forecast errors by comparing the forecasts with actual out-of-sample data. As the training sample increases by one year, we produce one- to 29-year-ahead forecasts and calculate the forecast errors. This process continues until the training sample covers all available data. We compare these forecasts with the holdout samples to determine the out-of-sample point forecast accuracy.

To measure overall point forecast accuracy and bias, we use the root mean squared forecast error (RMSFE), mean absolute forecast error (MAFE), and mean forecast error (MFE), averaged across ages and forecasting years. Averaged over 16 countries, they are defined as:
\begin{align*}
\text{RMSFE}(h) &= \frac{1}{16}\sum^{16}_{c=1}\sqrt{\frac{1}{(31-h)\times p}\sum^{n}_{k=n-30+h}\sum^p_{i=1}\left[m_k^{c}(x_i)-\widehat{m}_k^c(x_i)\right]^2},\\
\text{MAFE}(h) &= \frac{1}{16}\sum^{16}_{c=1}\frac{1}{(31-h)\times p}\sum^{n}_{k=n-30+h}\sum^p_{i=1}\left|m_k^c(x_i)-\widehat{m}_k^c(x_i)\right|,\\
\text{MFE}(h) &= \frac{1}{16}\sum^{16}_{c=1}\frac{1}{(31-h)\times p}\sum^{n}_{k=n-30+h}\sum^p_{i=1}\left[m_k^c(x_i)-\widehat{m}_k^c(x_i)\right],
\end{align*}
where $m_k^c(x_i)$ denotes mortality rate at year $k$ in the forecasting period for age $x_i$ in country $c$, and $\widehat{m}_k^c(x_i)$ denotes the point forecast. The ordering of the 16 countries are given in Table~\ref{tab:multi}. The RMSFE and MAFE are the average of squared and absolute errors and they measure forecast precision regardless of sign. The MFE is the average of errors and it measures bias. 

\subsubsection{Evaluation of interval forecast accuracy}

To assess interval forecast accuracy, we use the interval score of \cite{GR07} \citep[see also][]{GK14}. For each year in the forecasting period, one-year-ahead to 30-year-ahead prediction intervals were calculated at the $(1-\alpha)\times 100\%$ nominal coverage probability. We consider the common case of symmetric $(1-\alpha)\times 100\%$ prediction interval, with lower and upper bounds that are predictive quantiles at $\alpha/2$ and $1-\alpha/2$, denoted by $m_k(x_l)$ and $m_k(x_u)$ for a given year $k$. As defined by \cite{GR07}, a scoring rule for the interval forecast of mortality at age $x_i$ is:
\begin{align*}
S_{\alpha}\left[m_k(x_l), m_k(x_u); m_k(x_i)\right] = \left[m_k(x_u)-m_k(x_l)\right]+&\frac{2}{\alpha}[m_k(x_l)-m_k(x_i)]\mathds{1}\{m_k(x_i)<m_k(x_l)\}+\\
&\frac{2}{\alpha}\left[m_k(x_i)-m_k(x_u)\right]\mathds{1}\{m_k(x_i)>m_k(x_u)\},
\end{align*}
where $\alpha$ denotes the level of significance, customarily $\alpha=0.2$. The interval score rewards for a narrow prediction interval, if and only if the true observation lies within the prediction interval. The optimal score is achieved when $m_k(x_i)$ lies between $m_k(x_l)$ and $m_k(x_u)$, and the distance between $m_k(x_l)$ and $m_k(x_u)$ is minimal. 

From different ages, countries and years in the forecasting period, the mean interval score averaged across 16 countries is defined by:
\begin{align*}
\bar{S}_{\alpha}(h) &= \frac{1}{16\times (31-h)\times p}\sum^{16}_{c=1}\sum_{k=n-30+h}^{n}\sum_{i=1}^{p}S_{\alpha,k}^{c}[m_k(x_l),m_k(x_u);m_k(x_i)].
\end{align*}

\subsection{Multi-country comparison of point forecast accuracy}

Based on the averaged MAFE and RMSFE across 30 horizons shown in Table~\ref{tab:multicount}, the Lee-Carter method performs overall the worst among the methods considered. \cite{LM01} and \cite{LLG13} stated that mortality at older ages has been declining more quickly (on a log scale) than at younger ages, which contradicts the stationarity assumption of mortality improvement in the Lee-Carter method. Thus, it has been systematically under-predicting improvements in life expectancy over time. This confirms the fact that progress in life expectancy has been and continues to rise \citep[see also][]{OV02}.

\begin{table}[!htbp]
\centering
\tabcolsep 0.12in
\caption{Point forecast accuracy of mortality and life expectancy for females and males by method, as measured by the averaged MAFE, RMSFE, and MFE. For mortality, the forecast errors were multiplied by 100 in order to keep two decimal places. The minimal forecast errors are underlined for females and males, whereas the minimal overall forecast error is highlighted in bold. FDM represents functional data model.}\label{tab:multicount}
\begin{tabular}{@{}lccc|ccc|ccc@{}}\toprule
Method  	 		& \multicolumn{3}{c|}{MAFE} & \multicolumn{3}{c|}{RMSFE} & \multicolumn{3}{c}{MFE} \\ 
				& F & M   & $\frac{\text{F+M}}{2}$	 & F & M   &$\frac{\text{F+M}}{2}$	 & F & M  & $\frac{\text{F+M}}{2}$\\\midrule
  \multicolumn{6}{l}{\hspace{-.1in}{\underline{Mortality ($\times 100$)}}} \\
Lee-Carter			& 0.76 			& 0.89 			& 0.83 			& 1.68 				& 1.74 			& 1.71 			& -0.74 			& -0.85 			& -0.80 \\ 
Li-Lee				& 0.84 			& 0.65 			& 0.75 			& 1.76 				& 1.36 			& 1.56			& -0.83 			& -0.57 			& -0.70 \\ 
Independent FDM		& \underline{0.42}	& 0.69 			& 0.56		& \underline{1.00} 		& 1.33 			& \textBF{1.17}		& \underline{-0.28} 	& -0.60 			& -0.44 \\ 
Product-ratio			& 0.60 			& \underline{0.58} 	& 0.59			& 1.32 				& \underline{1.22} 	& 1.27			& -0.51 			& \underline{-0.44} 	& -0.48 \\ 
Multilevel FDM (arima) 	& 0.49 			& 0.60 			& \textBF{0.55}		& 1.13 				& \underline{1.22} 			& 1.18			& -0.36 			& -0.47 	& \textBF{-0.42} \\ 
Multilevel FDM (rwf) 		& 0.72 			& 0.60 			& 0.66			& 1.54 				& 1.24 			& 1.39			& -0.68 			& -0.50 			& -0.59 \\ 
	  \\
 \multicolumn{6}{l}{\hspace{-.1in}{\underline{e(0)}}} \\ 			
Lee-Carter 			& 2.33 			& 3.04 			& 2.69 	 	 & 2.36 			& 3.10 			& 2.73		& 2.26 			& 2.97 			& 2.62 \\ 
Li-Lee 				& 3.00 			& 1.92 			& 2.46	 	 & 3.03 			& 2.00 			& 2.52 		& 3.00 			& 1.73 	& 2.37 \\ 
Independent FDM		& \underline{1.53} 	& 3.06			& 2.30 		 & \underline{1.62} 	& 3.11 			& 2.37 		& \underline{1.24}	& 3.05 			& 2.15 \\ 
Product-ratio			& 2.19 			& 1.91			& 2.05		 & 2.26 			& 2.02 			& 2.14		& 1.95 			& 1.76 			& 1.86 \\ 
Multilevel FDM (arima) 	& 1.65 			& 2.19			& \textBF{1.92}	 & 1.73 			& 2.28 			& \textBF{2.00} 	& 1.30 			& 2.13 			& \textBF{1.72} \\ 
Multilevel FDM (rwf)		& 2.57 			& \underline{1.84} 	& 2.21		 & 2.61	 		& \underline{1.90} 	& 2.26		& 2.53 			& \underline{1.66} 			& 2.10 \\ \bottomrule
\end{tabular}
\end{table}	

The functional data methods use the automatic ARIMA algorithm for selecting the optimal difference operator $d$, for which the mortality improvement will then be stationary. Generally, the functional data methods give more accurate forecasts than the Lee-Carter and Li-Lee methods. The independent functional data method performs consistently the best for forecasting female mortality, followed by the multilevel functional data (arima) and product-ratio methods. The superiority of the independent functional data method over the coherent forecasting methods is manifested by a population with small variabilities over age and time, such as in female mortality. In terms of male and overall forecast errors, the product-ratio and multilevel functional data methods perform similarly: they both produce more accurate forecasts than those from the independent functional data method.

From the averaged MFE across 30 horizons, the coherent forecasting methods produce less-biased forecasts than the non-coherent forecasting methods for males. The independent functional data method gives the least-biased forecasts of female mortality. For male mortality, the product-ratio method and multilevel functional data method (arima) perform about the same in terms of bias, and they both produce less-biased forecasts than the ones from the independent functional data method. 

With the forecast age-specific mortality, we can also forecast life expectancy \citep[see][for details]{PHG01}. Based on the averaged MAFE, RMSFE, and MFE across 30 horizons, we again found that the functional data methods generally give smaller overall forecast errors and bias across two sexes, in comparison to the Lee-Carter and Li-Lee methods. The independent functional data method performs the best for forecasting female life expectancy, followed by the multilevel functional data (arima) and product ratio methods. For male data, the multilevel functional data method (rwf) gives the most accurate point forecasts. The product-ratio and multilevel functional data methods both produce more accurate point forecasts than the ones from the independent functional data method. Of the two approaches,  the multilevel functional data method (arima) performs the best based on simple averaging of the forecast errors over two sub-populations.

To achieve optimal point forecast accuracy and bias, the independent functional data method should be used for forecasting female mortality and life expectancy, whereas the product-ratio or multilevel functional data method (rwf) should be implemented for forecasting male mortality and male life expectancy, respectively. Based on the simple average of two sub-populations, the multilevel functional data method (arima) generally performs the overall best in all. With respect to the automatic ARIMA and random-walk with drift (rwf), the automatic ARIMA method is recommended to forecast principal component scores in the multilevel functional data method for age-specific female mortality and life expectancy. In contrast, the rwf method is suitable to forecast principal component scores for age-specific male mortality and life expectancy.

\subsection{Multi-country comparison of interval forecast accuracy}

The prediction intervals for age-specific mortality are obtained from~\eqref{eq:interval}, and the prediction intervals for life expectancy are obtained from the percentiles of simulated life expectancies. The simulation method takes the nonlinear relationship between age-specific mortality and life expectancy into account, thus giving an asymmetric prediction interval \citep{HBY13}. Based on the averaged mean interval scores shown in Table~\ref{tab:score_multicount}, the independent functional data method produces the most accurate forecasts for female mortality, followed by the multilevel functional data (arima) method. For male mortality, the multilevel functional data model (rwf) performs the best, followed by the Li-Lee method. Averaged across both sexes, the multilevel functional data method (arima) performs the best. For forecasting female life expectancy, the multilevel functional data method (arima) produces the most accurate interval forecasts, followed by the independent functional data method. For forecasting male life expectancy, the multilevel functional data method (rwf) gives the best interval forecast accuracy. Averaged across both sexes, the multilevel functional data method (arima) performs the best.
			
\begin{table}[htbp]
\centering
\tabcolsep 0.25in
\caption{Interval forecast accuracy of mortality and life expectancy for females and males by method, as measured by the averaged mean interval score. For mortality, the mean interval scores were multiplied by 100 in order to keep two decimal places.}\label{tab:score_multicount}
\begin{tabular}{@{}lcccrrr@{}}
  \toprule
 	Method & \multicolumn{3}{c}{Mortality ($\times 100$)} & \multicolumn{3}{c}{e(0)} \\ 
	& F & M  & $\frac{\text{F+M}}{2}$	 & F & M   &$\frac{\text{F+M}}{2}$ \\\midrule
  Lee-Carter  			& 6.14 			& 7.25 			&	6.70			& 11.41 			& 55.54 			& 33.48	\\   
  Li-Lee  				& 4.51 			& 3.01			&	3.76			& 19.61 			& 9.04 			& 14.33	\\  
  Independent FDM 		& \underline{2.05} 	& 3.66 			&	2.86			& 8.09 			& 17.93 			& 13.01	\\  
  Product-ratio 			& 3.17 			& 3.64 			&	3.41			& 12.93 			& 8.46 			& 10.70	\\  
  Multilevel FDM (arima) 	& 2.45 			& 3.04 			&	\textBF{2.75}	& \underline{7.76} 	& 10.49			& \textBF{9.13}	\\
  Multilevel FDM (rwf) 	& 3.99 			& \underline{2.92}	&	3.46			& 14.95 			& \underline{7.66} 	& 11.31	\\  
     \bottomrule
\end{tabular}
\end{table}

Apart from the mean forecast errors and mean interval scores, we also consider the maximum absolute forecast error, maximum root squared forecast error, and maximum interval score, for measuring the extreme point and interval errors across different ages and years in the forecasting period. Their results in the multi-country comparison are included in the supplement D \citep{Shang16}. 

\subsection{Comparison between the functional data models and a Bayesian method}

\cite{RLG14} proposed a Bayesian hierarchical model for joint probabilistic projection of male and female life expectancies that ensures coherence between them by projecting the gap between female life expectancy and male life expectancy. This method starts with probabilistic projection of life expectancy for females obtained from a Bayesian hierarchical model, then models the gap in life expectancy between females and males. The probabilistic projection of life expectancy for males can be obtained by combining the former two quantities. Computationally, this method is implemented in the \textit{bayesLife} package \citep{SR15} in R \citep{Team13}. In Tables~\ref{tab:bay_1} and~\ref{tab:bay_2}, we compare the forecast accuracy between the multilevel functional data and Bayesian methods for forecasting life expectancy.

\begin{table}[!htbp] \centering 
\tabcolsep 0.07in
\caption{Point and interval forecast accuracy between the multilevel functional data method and Bayesian method for forecasting female life expectancy at birth (e(0)). Using the data until 1979, we forecast the e(0) for years 1984, 1989, 1994, 1999, 2004 and 2009.}\label{tab:bay_1}
\begin{tabular}{@{}lrrrrrr|rrrrrr@{}} 
\toprule
& \multicolumn{6}{c}{Multilevel functional data method} & \multicolumn{6}{c}{Bayesian method} \\
Country & 1984 & 1989 & 1994 & 1999 & 2004 & 2009 & 1984 & 1989 & 1994 & 1999 & 2004 & 2009 \\ 
\hline \\[-1.8ex] 
\multicolumn{13}{l}{\hspace{-.09in} \underline{MAFE}}  \\
AUS & $0.54$ & $1.84$ & $2.22$ & $2.81$ & $3.51$ & $4.55$ & $0.98$ & $0.78$ & $1.49$ & $2.02$ & $2.51$ & $2.74$ \\ 
AUT & $0.71$ & $1.46$ & $1.74$ & $2.30$ & $2.96$ & $3.13$ & $0.78$ & $1.30$ & $1.43$ & $1.84$ & $2.35$ & $2.43$ \\ 
BEL & $1.63$ & $2.40$ & $3.07$ & $3.56$ & $4.17$ & $4.39$ & $0.94$ & $1.15$ & $1.39$ & $1.53$ & $1.79$ & $1.66$ \\ 
CAN & $0.20$ & $1.01$ & $1.85$ & $2.41$ & $2.78$ & $3.02$ & $0.74$ & $0.40$ & $0.03$ & $0.17$ & $0.10$ & $0.11$ \\ 
DEN & $0.20$ & $0.04$ & $0.06$ & $0.40$ & $0.99$ & $1.91$ & $0.58$ & $1.29$ & $1.83$ & $1.78$ & $1.56$ & $1.02$ \\ 
FRA & $1.78$ & $2.81$ & $3.65$ & $3.89$ & $4.87$ & $5.10$ & $0.74$ & $1.09$ & $1.50$ & $1.25$ & $1.92$ & $1.74$ \\ 
FIN & $1.66$ & $1.65$ & $2.60$ & $3.13$ & $4.03$ & $4.55$ & $0.61$ & $0.40$ & $0.10$ & $0.14$ & $0.27$ & $0.33$ \\ 
ITA & $1.79$ & $2.59$ & $2.86$ & $3.40$ & $4.43$ & $4.33$ & $0.78$ & $1.09$ & $0.99$ & $1.24$ & $1.99$ & $1.65$ \\ 
JPN & $0.53$ & $1.25$ & $1.62$ & $1.97$ & $2.95$ & $3.25$ & $0.94$ & $1.24$ & $1.29$ & $1.38$ & $2.18$ & $2.30$ \\ 
NET & $1.41$ & $1.52$ & $1.58$ & $1.35$ & $1.96$ & $2.80$ & $0.43$ & $0.06$ & $0.36$ & $0.84$ & $0.48$ & $0.07$ \\ 
NOR & $0.99$ & $0.74$ & $1.47$ & $1.62$ & $2.51$ & $2.98$ & $0.21$ & $0.44$ & $0.11$ & $0.24$ & $0.34$ & $0.43$ \\ 
SPA & $1.42$ & $1.79$ & $2.21$ & $2.05$ & $2.55$ & $2.96$ & $1.27$ & $1.01$ & $1.26$ & $1.05$ & $1.42$ & $1.74$ \\ 
SWE & $1.40$ & $1.76$ & $2.33$ & $2.59$ & $3.12$ & $3.56$ & $0.60$ & $0.39$ & $0.37$ & $0.09$ & $0.11$ & $0.09$ \\ 
SWI & $1.26$ & $1.82$ & $2.23$ & $2.64$ & $3.28$ & $3.59$ & $0.41$ & $0.21$ & $0.05$ & $0.13$ & $0.00$ & $0.10$ \\ 
UK & $0.74$ & $0.60$ & $1.20$ & $1.10$ & $1.86$ & $2.50$ & $0.74$ & $0.48$ & $0.98$ & $0.80$ & $1.46$ & $2.00$ \\ 
USA & $1.02$ & $2.03$ & $2.88$ & $3.84$ & $4.31$ & $4.53$ & $0.21$ & $0.26$ & $0.61$ & $1.10$ & $1.01$ & $0.80$ \\ 
\hline \\[-1.8ex]
Mean & 1.08 & 1.58 & 2.10 & 2.44 & 3.14 & 3.57 & \textBF{0.68} & \textBF{0.72} & \textBF{0.86} & \textBF{0.98} & \textBF{1.22} & \textBF{1.20}  \\\midrule
\multicolumn{13}{l}{\hspace{-.042in}\underline{Mean interval score}} \\
AUS & $1.83$ & $3.13$ & $4.81$ & $7.29$ & $9.49$ & $13.13$ & $2.06$ & $2.78$ & $3.48$ & $4.22$ & $5.52$ & $5.30$ \\
AUT & $2.92$ & $5.24$ & $8.92$ & $13.94$ & $20.97$ & $27.48$ & $2.10$ & $3.28$ & $4.17$ & $5.02$ & $5.75$ & $6.42$ \\ 
BEL & $5.51$ & $10.75$ & $17.32$ & $19.93$ & $26.41$ & $25.64$ & $2.12$ & $3.12$ & $4.01$ & $4.65$ & $5.38$ & $6.03$ \\ 
CAN & $1.80$ & $2.59$ & $3.42$ & $6.10$ & $6.07$ & $7.09$ & $1.96$ & $2.94$ & $3.72$ & $4.50$ & $5.02$ & $5.78$ \\ 
DEN & $3.34$ & $4.18$ & $5.03$ & $6.40$ & $11.54$ & $19.91$ & $1.97$ & $2.96$ & $3.74$ & $4.47$ & $5.16$ & $5.75$ \\
FRA & $6.31$ & $14.65$ & $21.48$ & $23.77$ & $32.17$ & $35.97$ & $2.05$ & $3.21$ & $4.11$ & $4.79$ & $5.43$ & $6.04$ \\
FIN & $8.90$ & $4.16$ & $11.13$ & $14.58$ & $23.71$ & $26.42$ & $2.25$ & $3.47$ & $4.66$ & $5.61$ & $6.45$ & $7.23$ \\
ITA & $4.03$ & $9.51$ & $11.21$ & $17.16$ & $27.69$ & $24.23$ & $2.16$ & $3.28$ & $4.11$ & $4.96$ & $5.68$ & $6.28$ \\ 
JPN & $2.08$ & $3.88$ & $5.57$ & $6.73$ & $8.17$ & $9.30$ & $2.17$ & $3.35$ & $4.24$ & $4.90$ & $5.58$ & $6.25$ \\
NET & $3.83$ & $4.93$ & $5.85$ & $6.22$ & $6.75$ & $7.05$ & $1.80$ & $2.56$ & $3.28$ & $3.86$ & $4.30$ & $4.65$ \\
NOR & $2.75$ & $2.51$ & $5.60$ & $6.36$ & $15.46$ & $19.65$ & $1.85$ & $2.53$ & $3.15$ & $3.68$ & $4.16$ & $4.69$ \\
SPA & $3.62$ & $5.88$ & $10.81$ & $8.35$ & $17.02$ & $20.55$ & $4.48$ & $3.11$ & $3.87$ & $4.59$ & $5.33$ & $5.80$ \\ 
SWE & $3.22$ & $4.13$ & $7.05$ & $9.15$ & $14.60$ & $15.85$ & $1.86$ & $2.78$ & $3.36$ & $4.07$ & $4.71$ & $5.29$ \\
SWI & $2.62$ & $3.90$ & $7.51$ & $9.20$ & $14.75$ & $17.56$ & $1.96$ & $3.06$ & $4.06$ & $5.06$ & $5.94$ & $6.68$ \\ 
UK & $4.15$ & $2.60$ & $7.85$ & $7.93$ & $14.89$ & $21.47$ & $2.00$ & $2.90$ & $3.67$ & $4.15$ & $4.71$ & $5.32$ \\ 
USA & $1.81$ & $2.45$ & $3.02$ & $3.41$ & $3.64$ & $4.04$ & $2.06$ & $3.01$ & $3.76$ & $4.57$ & $5.18$ & $5.71$ \\\midrule
Mean & 3.67 & 5.28 & 8.54 & 10.41 & 15.83 & 18.46 & \textBF{2.18} & \textBF{3.02} & \textBF{3.84} & \textBF{4.57} & \textBF{5.27} & \textBF{5.83}  \\\bottomrule
\end{tabular} 
\end{table}

\begin{table}[!htbp] \centering 
\tabcolsep 0.07in
\caption{Point and interval forecast accuracy between the multilevel functional data method and Bayesian method for forecasting male life expectancy at birth (e(0)). Using the data until 1979, we forecast the e(0) for years 1984, 1989, 1994, 1999, 2004 and 2009.}\label{tab:bay_2}
\begin{tabular}{@{}lrrrrrr|rrrrrr@{}} 
\toprule
& \multicolumn{6}{c}{Multilevel functional data method} & \multicolumn{6}{c}{Bayesian method} \\
Country & 1984 & 1989 & 1994 & 1999 & 2004 & 2009 & 1984 & 1989 & 1994 & 1999 & 2004 & 2009 \\ 
\hline \\[-1.8ex] 
\multicolumn{13}{l}{\hspace{-.085in} \underline{MAFE}}  \\
AUS & $1.32$ & $1.70$ & $2.97$ & $4.01$ & $5.28$ & $5.97$ & $1.61$ & $1.90$ & $3.08$ & $4.19$ & $5.44$ & $6.07$ \\ 
AUT & $0.09$ & $0.69$ & $0.96$ & $1.79$ & $2.66$ & $2.98$ & $0.72$ & $1.73$ & $2.13$ & $2.96$ & $3.89$ & $4.22$ \\ 
BEL & $1.04$ & $1.85$ & $2.28$ & $2.72$ & $3.82$ & $4.52$ & $0.81$ & $1.46$ & $1.83$ & $2.06$ & $2.97$ & $3.53$ \\ 
CAN & $1.13$ & $1.32$ & $1.69$ & $2.36$ & $3.32$ & $4.09$ & $1.56$ & $1.62$ & $1.87$ & $2.43$ & $3.31$ & $3.99$ \\ 
DEN & $0.00$ & $0.23$ & $0.07$ & $1.02$ & $1.51$ & $2.77$ & $0.35$ & $0.67$ & $0.43$ & $0.39$ & $0.76$ & $1.89$ \\ 
FRA & $0.39$ & $0.99$ & $1.66$ & $2.50$ & $3.84$ & $4.57$ & $0.57$ & $0.92$ & $1.21$ & $1.65$ & $2.59$ & $2.93$ \\ 
FIN & $1.27$ & $0.81$ & $2.06$ & $2.42$ & $3.46$ & $4.20$ & $1.42$ & $0.86$ & $1.95$ & $2.03$ & $2.82$ & $3.21$ \\ 
ITA & $0.70$ & $1.08$ & $1.09$ & $1.81$ & $3.10$ & $3.53$ & $1.00$ & $1.53$ & $1.60$ & $2.43$ & $3.74$ & $4.20$ \\ 
JPN & $0.26$ & $0.23$ & $0.85$ & $1.43$ & $1.02$ & $1.02$ & $0.34$ & $0.27$ & $0.34$ & $0.84$ & $0.44$ & $0.32$ \\ 
NET & $0.74$ & $0.92$ & $0.73$ & $0.58$ & $0.43$ & $1.62$ & $0.72$ & $1.13$ & $1.79$ & $2.27$ & $3.52$ & $4.90$ \\ 
NOR & $0.01$ & $0.70$ & $0.49$ & $0.65$ & $2.06$ & $2.75$ & $0.26$ & $0.17$ & $1.18$ & $1.35$ & $2.72$ & $3.38$ \\ 
SPA & $0.67$ & $0.20$ & $0.19$ & $0.17$ & $0.58$ & $1.48$ & $0.90$ & $0.17$ & $0.17$ & $0.31$ & $1.09$ & $2.02$ \\ 
SWE & $0.16$ & $0.25$ & $0.95$ & $1.48$ & $2.38$ & $3.03$ & $0.86$ & $1.09$ & $1.71$ & $2.15$ & $2.86$ & $3.24$ \\ 
SWI & $1.02$ & $0.89$ & $1.14$ & $2.00$ & $2.97$ & $3.45$ & $0.60$ & $0.43$ & $0.70$ & $1.59$ & $2.49$ & $2.99$ \\ 
UK & $1.03$ & $1.20$ & $1.94$ & $2.23$ & $3.43$ & $4.30$ & $1.10$ & $1.25$ & $2.00$ & $2.29$ & $3.44$ & $4.26$ \\ 
USA & $0.13$ & $0.30$ & $0.39$ & $0.33$ & $0.59$ & $0.94$ & $0.99$ & $0.73$ & $0.83$ & $1.71$ & $2.08$ & $2.57$ \\\midrule 
Mean & $\textBF{0.62}$ & $\textBF{0.83}$ & $\textBF{1.22}$ & $\textBF{1.72}$ & $\textBF{2.53}$ & $\textBF{3.20}$ & $0.86$ & $1.00$ & $1.43$ & $1.92$ & $2.76$ & $3.36$ \\\hline \\[-1.8ex] 
\multicolumn{13}{l}{\hspace{-.04in}\underline{Mean interval score}} \\
AUS & $5.87$ & $6.07$ & $17.17$ & $27.12$ & $24.19$ & $17.86$ & $6.58$ & $4.56$ & $14.18$ & $22.25$ & $31.65$ & $35.93$ \\
AUT & $1.83$ & $2.51$ & $2.95$ & $7.67$ & $14.82$ & $16.00$ & $2.62$ & $3.83$ & $4.83$ & $5.78$ & $9.44$ & $8.90$ \\ 
BEL & $2.47$ & $5.21$ & $6.31$ & $6.79$ & $9.78$ & $16.47$ & $2.52$ & $3.76$ & $4.87$ & $5.84$ & $6.71$ & $7.54$ \\ 
CAN & $1.67$ & $2.11$ & $2.58$ & $6.36$ & $15.08$ & $21.74$ & $5.58$ & $3.47$ & $4.31$ & $5.27$ & $8.89$ & $12.18$ \\ 
DEN & $1.84$ & $2.28$ & $2.63$ & $2.84$ & $2.74$ & $5.76$ & $2.34$ & $3.48$ & $4.17$ & $4.91$ & $5.54$ & $6.13$ \\ 
FRA & $4.75$ & $6.53$ & $7.86$ & $8.93$ & $10.70$ & $11.78$ & $2.57$ & $3.97$ & $5.23$ & $6.35$ & $7.40$ & $8.53$ \\ 
FIN & $3.86$ & $5.34$ & $6.37$ & $7.02$ & $16.47$ & $21.92$ & $3.14$ & $4.22$ & $5.41$ & $6.62$ & $7.59$ & $8.66$ \\ 
ITA & $3.94$ & $4.99$ & $5.92$ & $6.44$ & $7.63$ & $7.89$ & $2.59$ & $3.88$ & $4.89$ & $5.77$ & $8.81$ & $9.50$ \\
JPN & $1.61$ & $1.83$ & $2.24$ & $2.35$ & $2.24$ & $2.59$ & $2.91$ & $4.55$ & $6.09$ & $7.32$ & $8.51$ & $9.61$ \\  
NET & $4.30$ & $5.17$ & $6.54$ & $6.95$ & $8.19$ & $8.40$ & $2.26$ & $3.31$ & $4.08$ & $4.74$ & $14.80$ & $26.17$ \\ 
NOR & $2.24$ & $2.97$ & $3.84$ & $4.23$ & $4.73$ & $5.10$ & $2.31$ & $3.29$ & $4.00$ & $4.61$ & $6.35$ & $10.30$ \\ 
SPA & $4.04$ & $5.14$ & $5.82$ & $6.76$ & $6.64$ & $6.80$ & $2.61$ & $3.83$ & $4.72$ & $5.61$ & $6.39$ & $7.23$ \\ 
SWE & $3.19$ & $3.80$ & $8.15$ & $8.94$ & $10.25$ & $11.53$ & $2.27$ & $3.29$ & $4.09$ & $4.71$ & $7.45$ & $9.94$ \\
SWI & $1.90$ & $2.37$ & $2.71$ & $7.93$ & $8.81$ & $9.86$ & $2.43$ & $3.60$ & $4.54$ & $5.49$ & $6.26$ & $6.84$ \\
UK & $1.57$ & $2.16$ & $5.44$ & $6.63$ & $17.44$ & $25.34$ & $2.46$ & $3.56$ & $4.37$ & $5.22$ & $10.36$ & $15.70$ \\ 
USA & $1.45$ & $1.87$ & $2.44$ & $2.64$ & $2.87$ & $3.24$ & $2.53$ & $3.74$ & $4.78$ & $5.68$ & $6.51$ & $7.30$ \\\midrule
Mean & \textBF{2.91} &  \textBF{3.77} &  5.56 & 7.47 & 10.16 & 12.02 & 2.98  & \textBF{3.77} &  \textBF{5.29} &  \textBF{6.64} &  \textBF{9.54} & \textBF{11.90} \\\bottomrule
\end{tabular} 
\end{table} 


For females, the Bayesian method is recommended. For males, the multilevel functional data method is preferable, in terms of point forecast accuracy. In terms of interval forecast accuracy, the Bayesian method is slightly advantageous for long-term forecasts. We found that the Bayesian (a simpler and direct) method outperforms the multilevel functional data method for long-term projection of life expectancy. The Bayesian method shows a superior interval forecast accuracy for two reasons: 
\begin{inparaenum}
\item[1)] the Bayesian method uses the historical life expectancy data to produce forecasts, whereas the multilevel functional data method uses the historical age-specific mortality to produce these age-specific mortality rate forecasts, which are then combined non-linearly to give life expectancy forecasts. Oftentimes, the direct forecasting method outperforms the indirect forecasting method.
\item[2)] the Bayesian method uses the prior information to assist its forecasts, in particular at longer forecast horizon.
\end{inparaenum}
By contrast, the multilevel functional data method is a time-series extrapolation, which works reasonably well in the short time. However, it does not work well for long term. Given that different changes are at play at different phases of a mortality transition, the age components of change in the past are not necessarily informative of longer-term future.

\section{Application to Australian age- and sex- and state-specific mortality}\label{sec:5}

First, we consider the age- and state-wise total mortality rates from 1950 to 2003 in Australia, available in the \textit{addb} package of \cite{Hyndman10} in R \citep{Team13}. This data set contains mortality rates for six states of Australia: Victoria (VIC), New South Wales (NSW), Queensland (QLD), South Australia (SA), Western Australia (WA), and Tasmania (TAS). The Australian Capital Territory and the Northern Territory are excluded from the analysis due to many missing values in the available data.

\begin{figure}[!htbp]
  \centering
  \includegraphics[width=16.5cm]{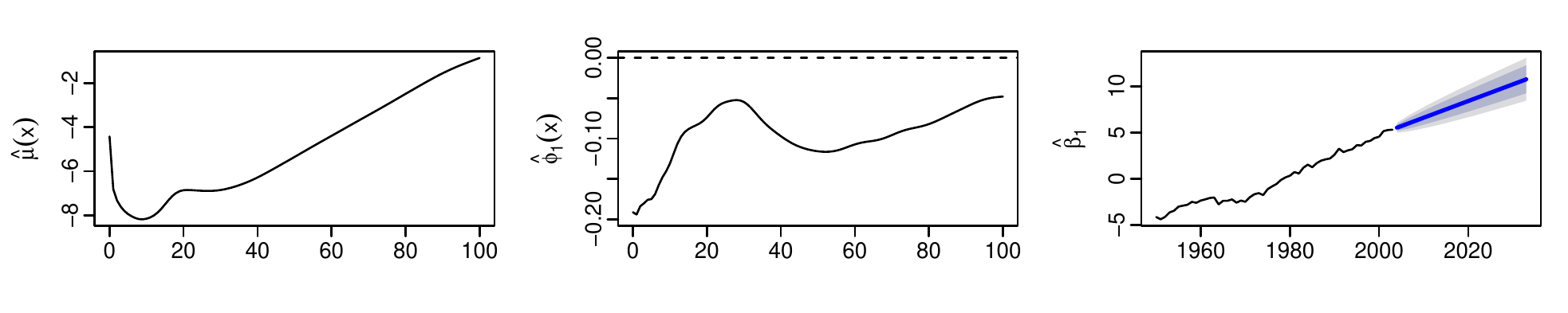}
  \includegraphics[width=16.5cm]{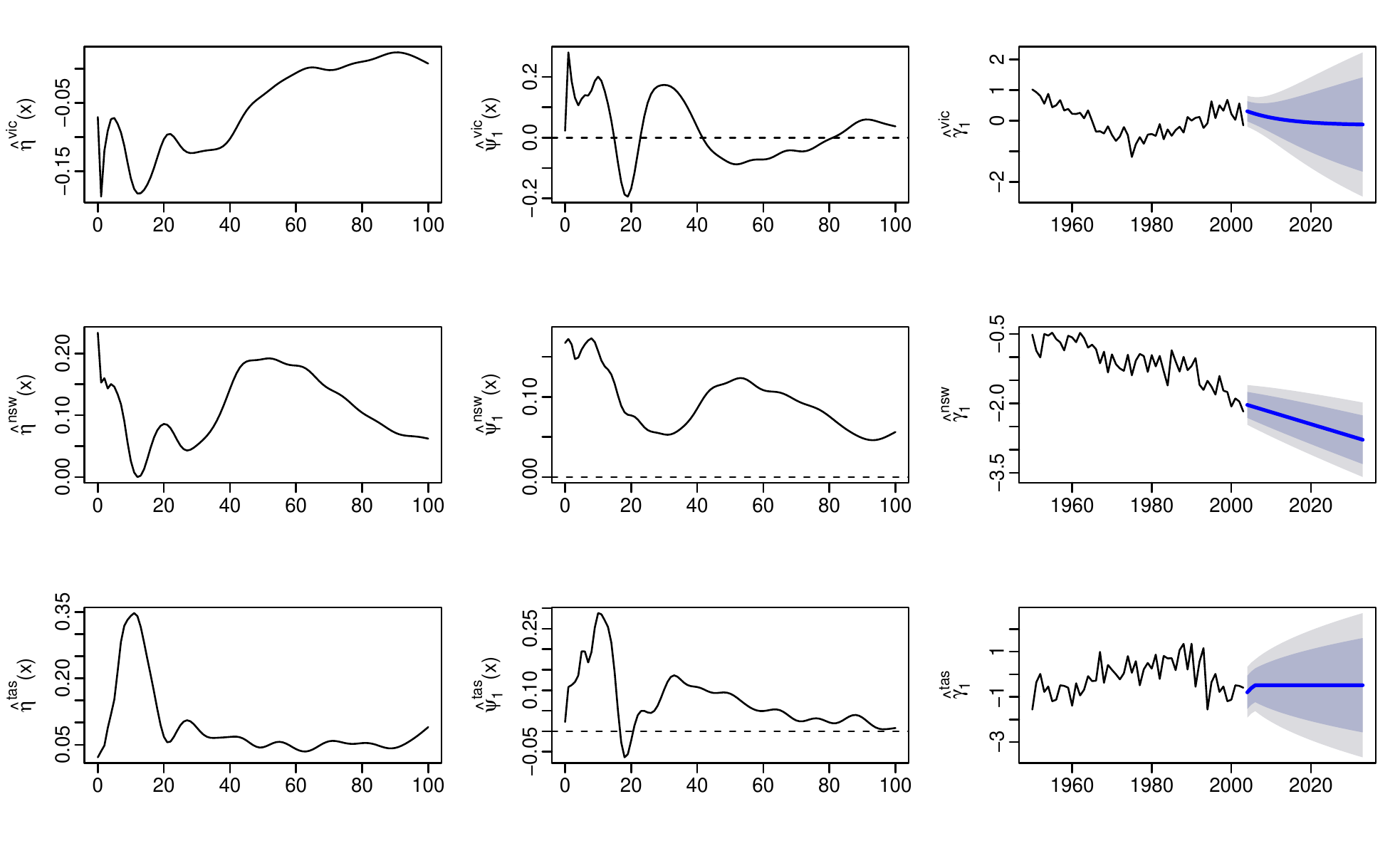}
  \includegraphics[width=16.5cm]{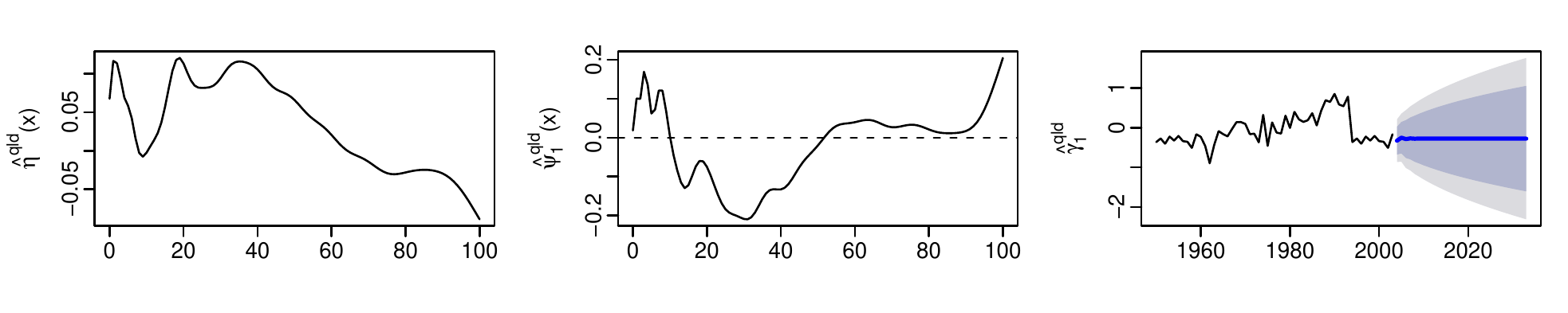}
  \includegraphics[width=16.5cm]{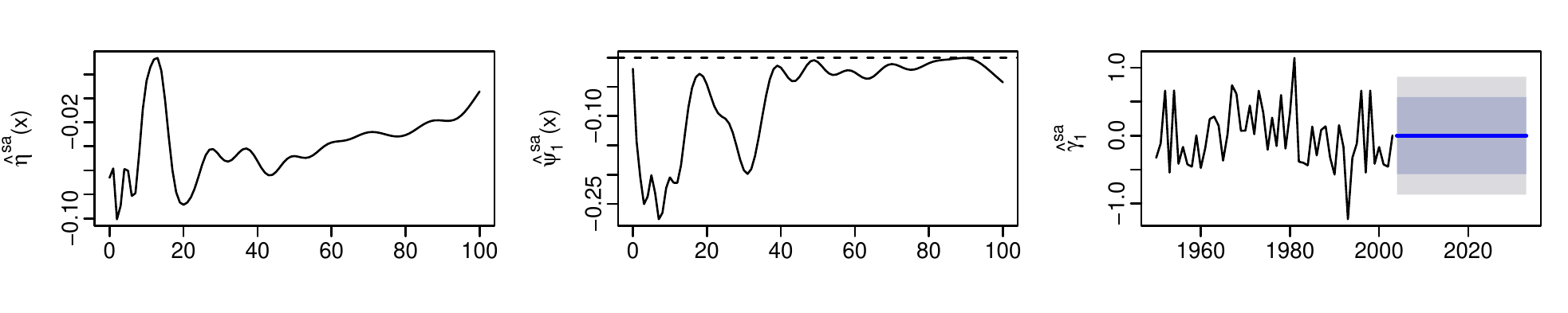}
  \includegraphics[width=16.5cm]{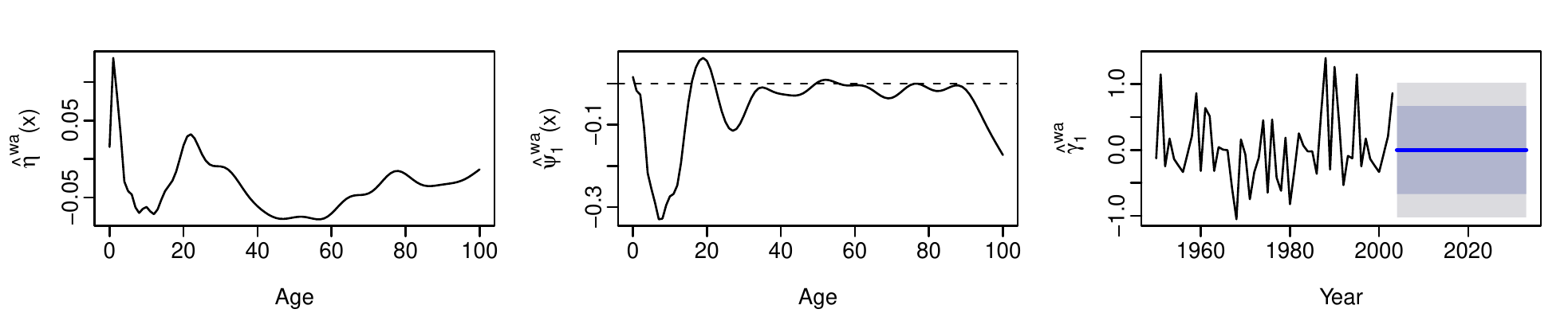}
  \caption{The first common functional principal component and its associated scores for the aggregate mortality data (top), followed by the first functional principal component and associated scores for the state-wise total age-specific mortality rates in VIC, NSW, TAS, QLD, SA and WA, respectively. The dark and light gray regions show the 80\% and 95\% prediction intervals.}\label{fig:20}
\end{figure}

In Figure~\ref{fig:20}, we show the estimated overall mean function $\widehat{\mu}(x)$, first common functional principal component $\widehat{\phi}_1(x)$ and corresponding scores $\left\{\widehat{\beta}_{1,1},\dots,\widehat{\beta}_{n,1}\right\}$ with 30-years-ahead forecasts. The first common functional principal component accounts for at least 90\% of total variation in the total mortality. The retained number of functional principal components for each state is the one that explains at least 90\% of the remaining 10\% total variations in the data. Due to limited space, we present only the first principal components for the six states, which explain 27\%, 68\%, 26\%, 22\%, 22\%, and 28\% of the remaining 10\% total variations for VIC, NSW, TAS, QLD, SA, WA, respectively. Based on~\eqref{eq:cluster}, the proportion of variability explained by the aggregate data (the simple average of total mortality across states) is $71\%, 71\%, 33\%, 63\%, 50\%$, and $50\%$ for VIC, NSW, TAS, QLD, SA, WA, respectively.  

In Figure~\ref{fig:20}, we also show the estimated mean function deviation, first state-specific functional principal component $\widehat{\psi}_1^s(x)$ and principal component scores $\{\widehat{\gamma}^s_{1,1},\dots,\widehat{\gamma}^s_{n,1}\}$ with 30-years-ahead forecasts, where $s$ denotes a state. The convergence in forecasts is likely to be achieved by the multilevel functional data method, because the forecasts of principal component scores for each state do not show a long-term trend, with the exception of NSW. From a statistical perspective, this may be because the NSW has the largest proportion of variability that can not be explained by the aggregate data. From a social perspective, NSW is the state that attracts the most migrants in Australia (\url{http://www.abs.gov.au/ausstats/abs@.nsf/mf/3412.0}). 

\begin{figure}[!htbp]
\centering
{\includegraphics[width=5.5cm]{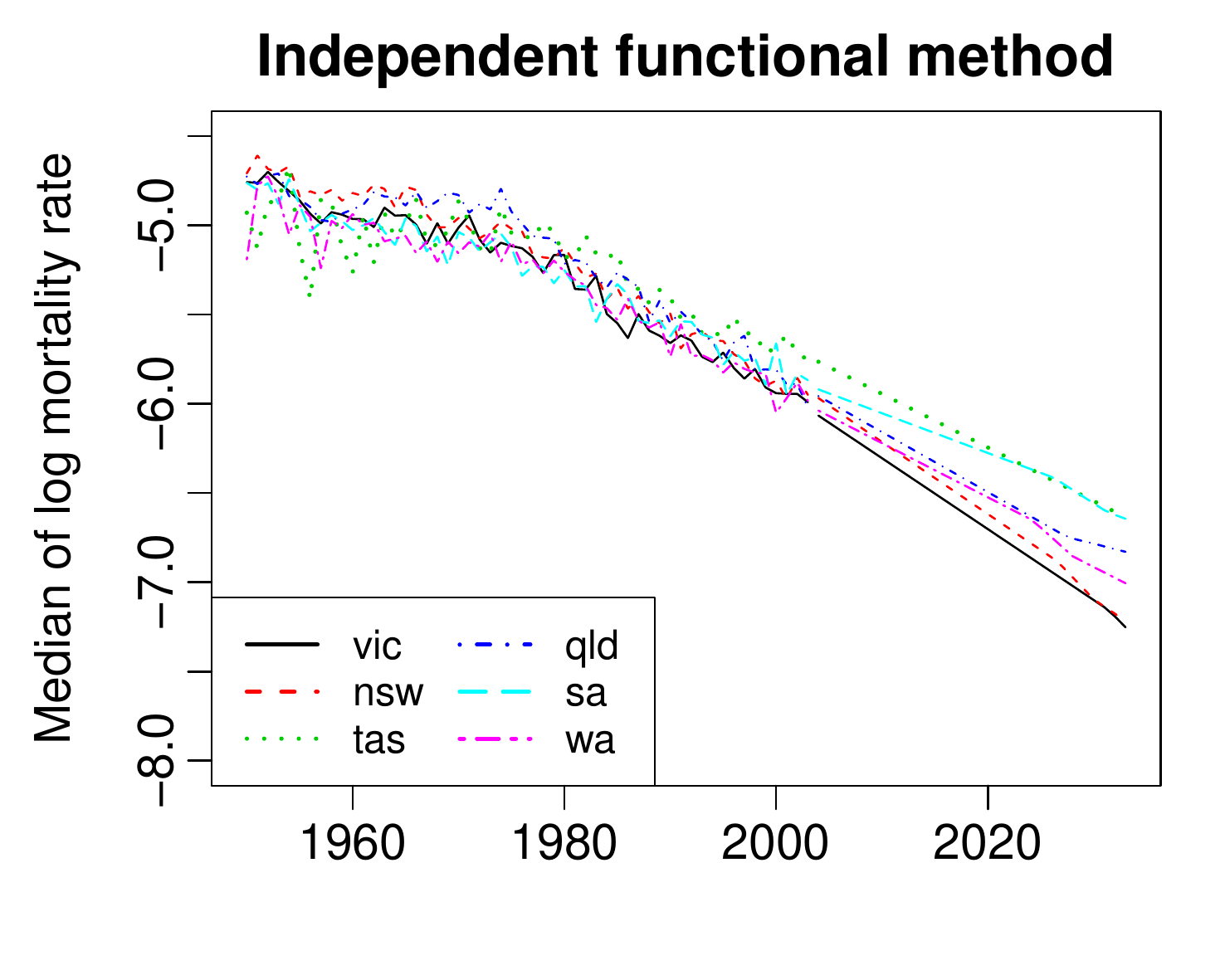}}
{\includegraphics[width=5.5cm]{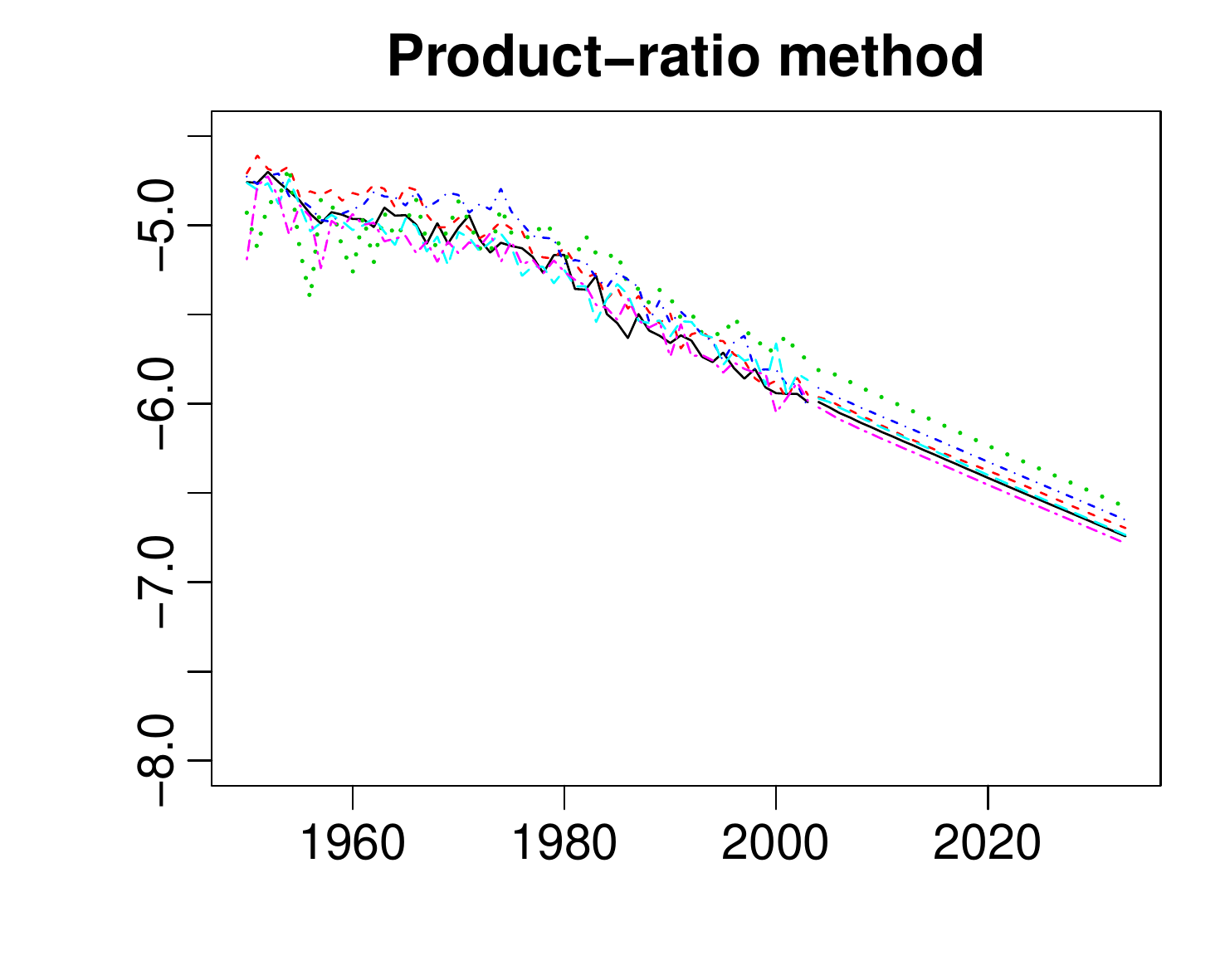}}
{\includegraphics[width=5.5cm]{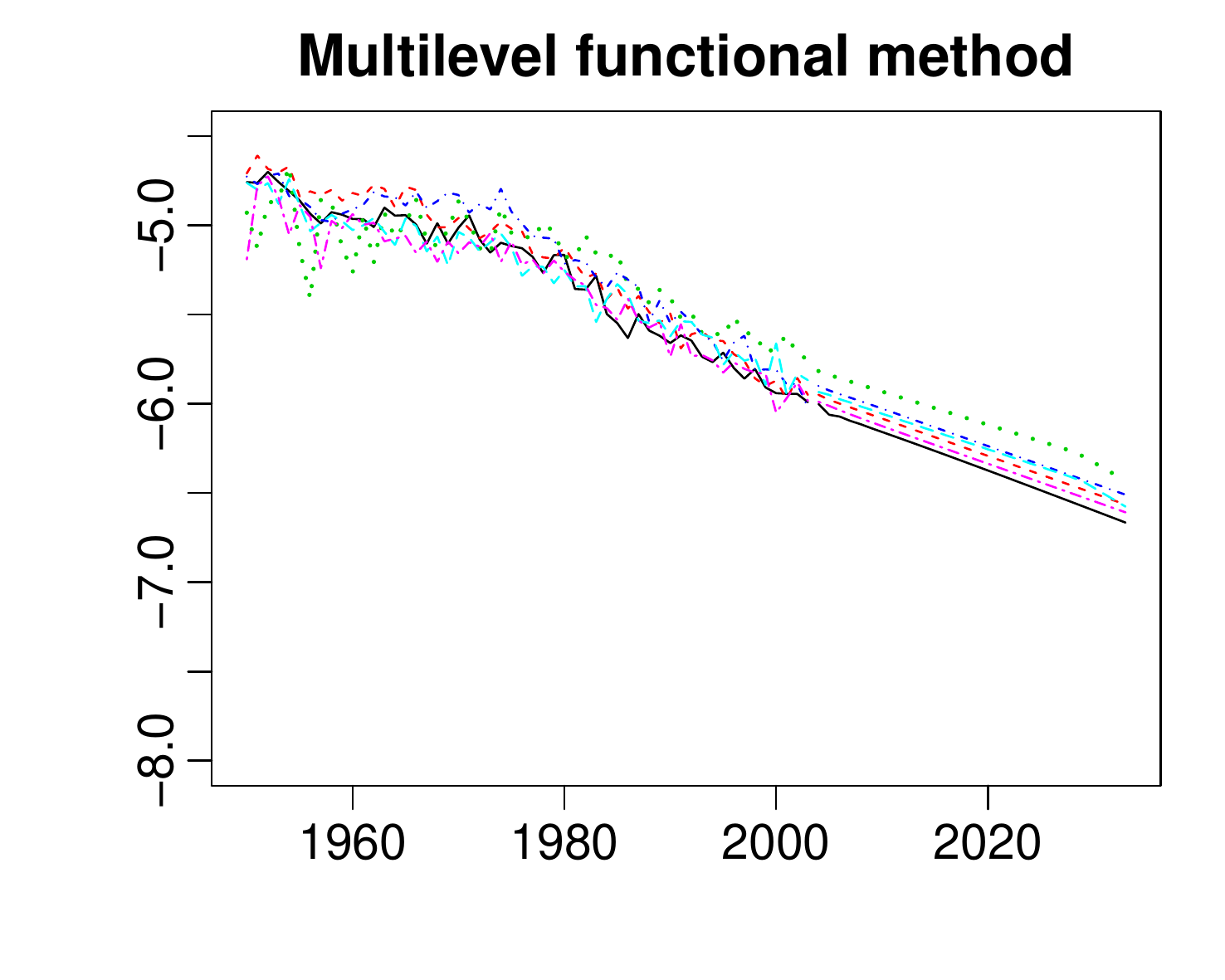}}
\\
{\includegraphics[width=5.5cm]{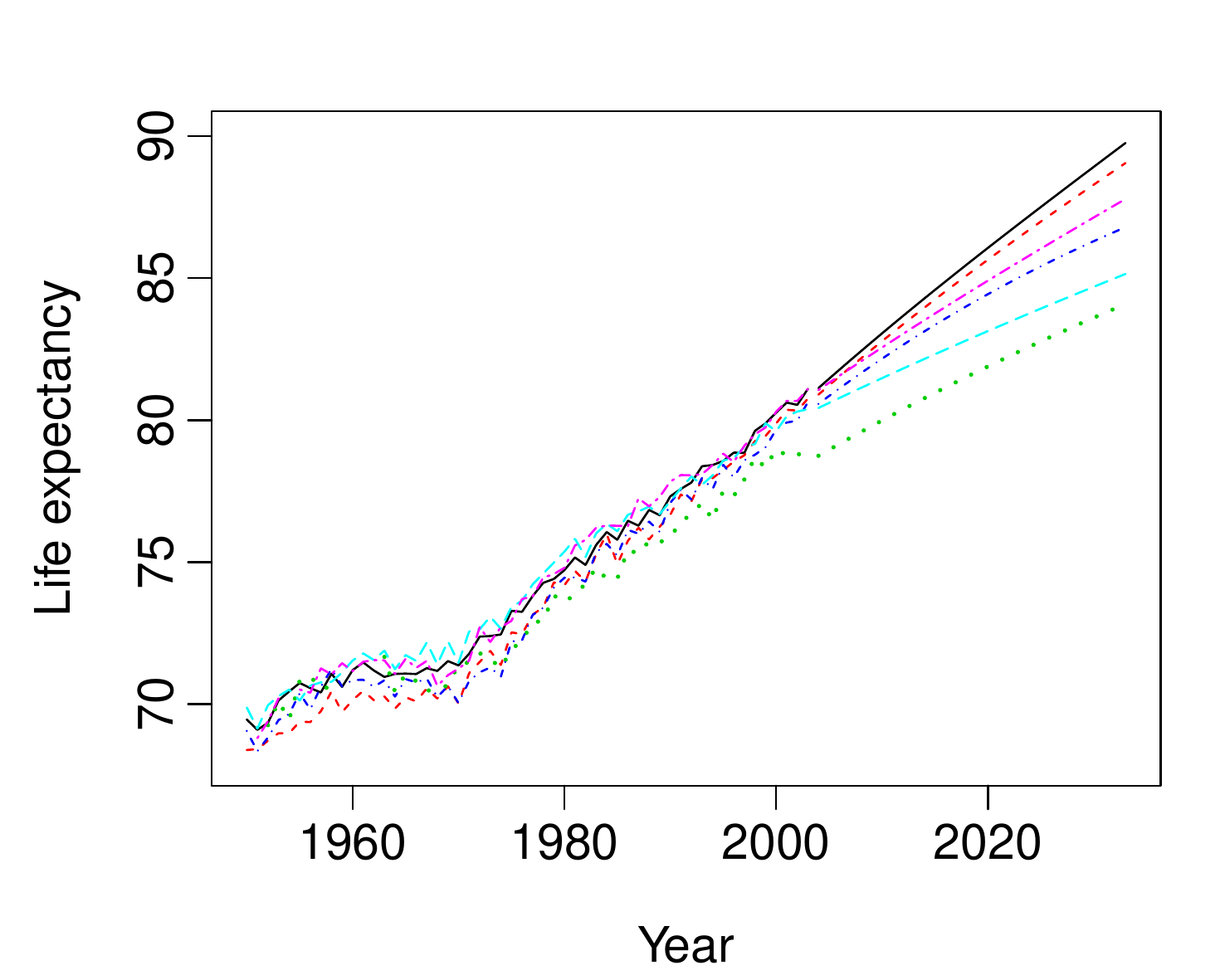}}
{\includegraphics[width=5.5cm]{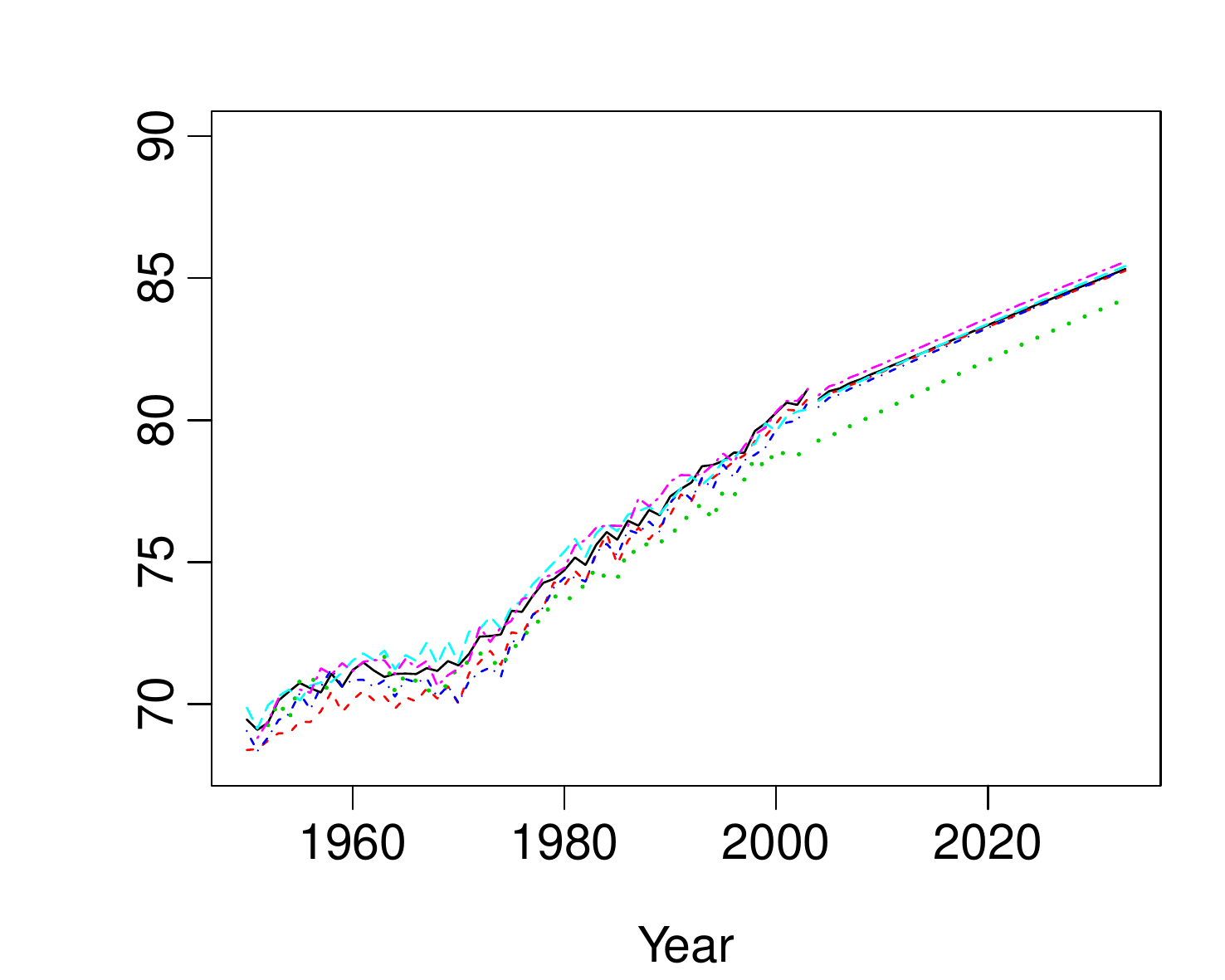}}
{\includegraphics[width=5.5cm]{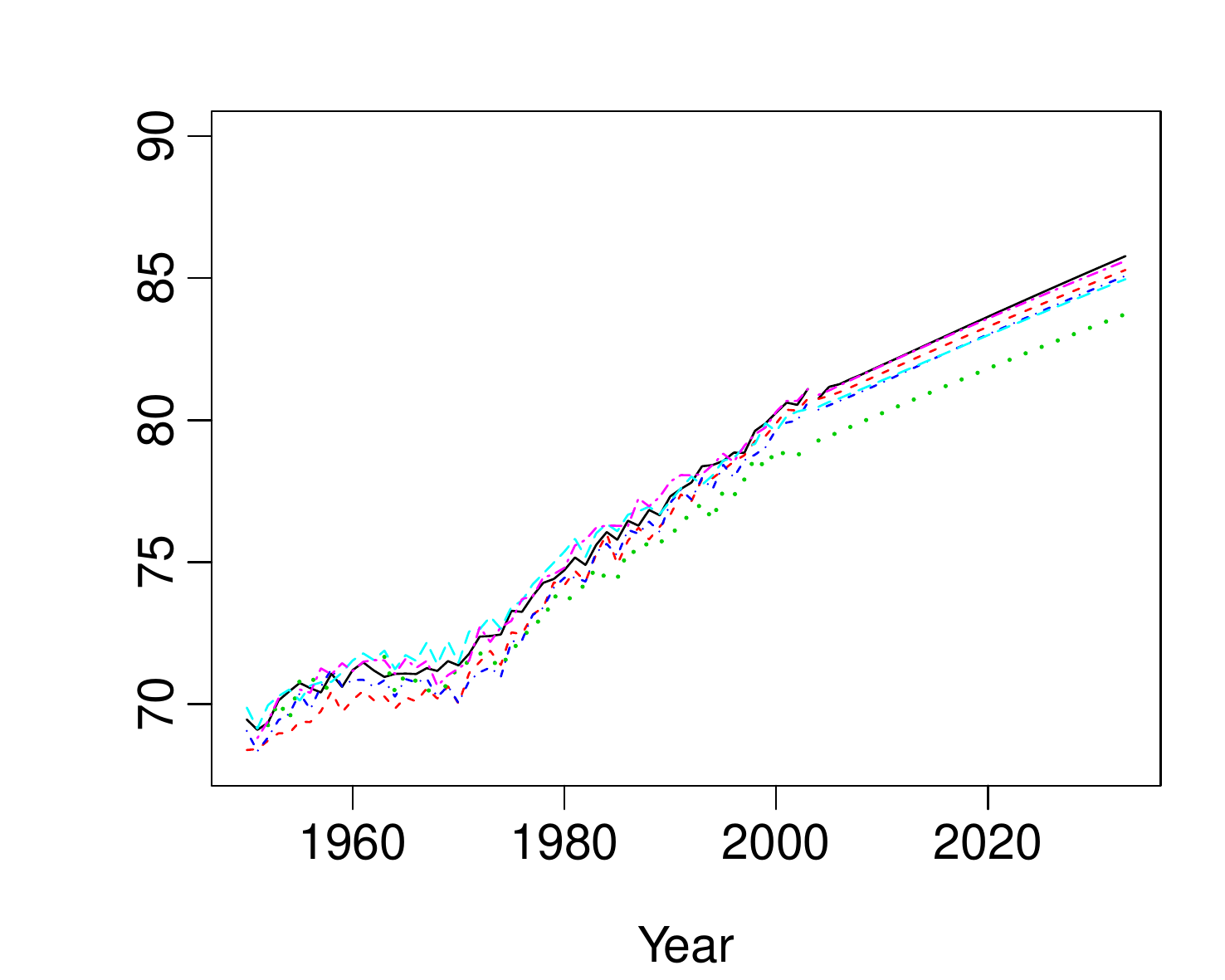}}
\caption{Based on historical mortality rates (1950--2003), we forecast future mortality rates and life expectancy from 2004 to 2033, for the independent functional data, product-ratio, and multilevel functional data methods.}\label{fig:22}
\end{figure}

Figure~\ref{fig:22} shows 30-years-ahead forecasts of median log mortality rates and life expectancy from 2004 to 2033 for all states, for the independent functional data, product-ratio and multilevel functional data methods. We focus on these three methods in this application, because they generally outperform the Lee-Carter and Li-Lee methods as demonstrated in Section~\ref{sec:4}. For the independent functional data method, the gap in mortality and life expectancy forecasts among states diverges. In contrast, the product-ratio and multilevel functional data methods are quite similar, and the gaps between female and male age-specific mortality and life expectancy converge, respectively. 

\subsection{Comparisons of point and interval forecast accuracy}\label{sec:6.1}

Table~\ref{tab:state_point_accuracy} displays the point and interval forecast accuracy for both age- and state-specific total mortality rates and life expectancy at each forecast horizon. As measured by the averaged MAFE, RMSFE, MFE and averaged mean interval score across 30 horizons, the independent functional data method performs the worst, whereas the multilevel functional data method (rwf) performs the best, for forecasting age- and state-specific total mortality and life expectancy. As the product-ratio and multilevel functional data methods perform similarly, it is paramount to incorporate correlation among sub-populations in forecasting, as this allows us to search for characteristics within and among series. 

\begin{table}[!htbp]
\begin{small}
\centering
 \setlength\tabcolsep{9.5pt}
\caption{Point and interval forecast accuracy of mortality and life expectancy (e(0)) across different states by method and forecast horizon, as measured by the averaged MAFE, RMSFE, MFE, and averaged mean interval score. The minimal forecast errors are underlined for each state, whereas the minimal overall forecast error is highlighted in bold.}\label{tab:state_point_accuracy}
\begin{tabular}{@{}llrrrrrrr@{}}
  \toprule
	& & VIC & NSW & QLD & TAS & SA & WA & Mean \\
  \hline
Mortality & \underline{\text{MAFE}}  \\
($\times 100$)	& Independent FDM 	 	& 0.61 & 0.63 & 0.77 & 0.96 & 0.70 & 0.70 & 0.73 \\ 
& Product-ratio 			& 0.56 & 0.55 & 0.45 & 0.53 & 0.47 & 0.53 & 0.51 \\ 
& Multilevel FDM (arima) 	& 0.53 & 0.51 & 0.47 & 0.53 & 0.46 & 0.52 & 0.51 \\ 
& Multilevel FDM (rwf) 	& \underline{0.47} & \underline{0.47} & \underline{0.41} & \underline{0.49} & \underline{0.41} & \underline{0.46} & \textBF{0.45} \\ 
\\
& \underline{\text{RMSFE}}  \\
&  Independent FDM 	 & 1.36 & 1.42 & 1.69 & 1.96 & 1.48 & 1.53 & 1.57 \\ 
&  Product-ratio 		 & 1.08 & 1.04 & 0.87 & 1.26 & 0.97 & 1.06 & 1.05 \\ 
&  Multilevel FDM (arima) 	 & 1.03 & 0.97 & 0.95 & 1.23 & 0.96 & 1.05 & 1.03 \\ 
&  Multilevel FDM (rwf) 	 & \underline{0.91} & \underline{0.88} & \underline{0.82} & \underline{1.18} & \underline{0.86} & \underline{0.93} & \textBF{0.93} \\ 
\\
& \underline{\text{MFE}} \\
&  Independent FDM 	 & \underline{-0.31} & \underline{-0.16} & -0.41 & -0.86 & -0.48 & -0.40 & -0.43 \\ 
&  Product-ratio 		 & -0.52 & -0.49 & -0.32 & -0.25 & -0.35 & -0.43 & -0.39 \\ 
&  Multilevel FDM (arima) 	 & -0.48 & -0.43 & -0.32 & -0.25 & -0.33 & -0.42 & -0.37 \\ 
&  Multilevel FDM (rwf) 	 & -0.42 & -0.39 & \underline{-0.20} & \underline{-0.14} & \underline{-0.26} & \underline{-0.33} & \textBF{-0.29} \\ 
  \\
& \underline{\text{Mean interval score}} \\
 & Independent FDM 	& 4.00 & 3.55 & 5.42 & 4.95 & 5.01 & 4.52 & 4.58 \\
 & Product-ratio			& 2.85 & 2.78 & 2.75 & 2.44 & 2.43 & 2.69 & 2.66 \\
 & Multilevel FDM (arima) & 2.47 & 2.14 & 2.42 & 1.81 & 1.85 & 2.50 & 2.20 \\
 & Multilevel FDM (rwf) 	& \underline{2.10} & \underline{2.06} & \underline{2.01} & \underline{1.55} & \underline{1.58} & \underline{2.04} & \textBF{1.89} \\
\\   
e(0) & \underline{\text{MAFE}} \\
& Independent FDM  	& 2.34 & 2.75 & 3.19 & 4.63 & 3.06 & 3.08 & 3.17 \\ 
& Product-ratio	 		& 3.07 & 3.30 & 2.83 & 2.08 & 2.46 & 2.93 & 2.78 \\ 
& Multilevel FDM (arima)  & 2.96 & 3.05  & 2.81 & 2.39 & 2.39 & 2.88 & 2.75 \\ 
& Multilevel FDM (rwf)    	& \underline{2.79} & \underline{3.01} & \underline{2.49} & \underline{1.76} & \underline{2.17} & \underline{2.64} & \textBF{2.48} \\ 
\\
& \underline{\text{RMSFE}}  \\
&  Independent FDM  	& 2.92 & \underline{3.05} & 3.75 & 4.67 & 3.35 & 3.56 & 3.55 \\ 
&   Product-ratio 		& 3.14 & 3.38 & 2.94 & 2.20 & 2.61 & 3.03 & 2.88 \\ 
&   Multilevel FDM (arima)& 3.04 & 3.16 & 2.95 & 2.53 & 2.53 & 2.99 & 2.87 \\ 
&   Multilevel FDM (rwf)  	& \underline{2.86} & 3.10 & \underline{2.60} & \underline{1.89} & \underline{2.32} & \underline{2.75} & \textBF{2.59} \\ 
\\
& \underline{\text{MFE}}  \\
&   Independent FDM  	& \underline{2.26} & \underline{1.75} & 2.62 & 4.63 & 2.79 & \underline{2.53} & 2.76 \\ 
&   Product-ratio 		& 3.07 & 3.29 & 2.81 & 2.05 & 2.45 & 2.93 & 2.77 \\ 
&   Multilevel FDM (arima)& 2.95 & 3.03 & 2.79 & 2.37 & 2.37 & 2.87 & 2.73 \\ 
&   Multilevel FDM (rwf)    & 2.78 & 3.00 & \underline{2.47} & \underline{1.69} & \underline{2.16} & 2.64 & \textBF{2.46} \\
\\
& \underline{\text{Mean interval score}} \\
& Independent FDM 		&  21.04 & 25.05 & 30.46 & 24.20  	& 19.85 & 16.34 & 22.82 \\
& Product-ratio 			& 22.70 & 24.66 & 13.53 & 19.95 	& 17.10 & 21.14 & 19.85 \\
& Multilevel FDM (arima) 	& 20.79 & 20.64 & 15.04 & 18.44 	& 15.79 & 19.59 & 18.38 \\
& Multilevel FDM (rwf) 	& \underline{17.09} & \underline{18.81} & \underline{9.41} & \underline{14.26} 	& \underline{12.27} & \underline{15.79} & \textBF{14.60} \\
  \bottomrule  
\end{tabular}
\end{small}
\end{table}

\subsection{Application to Australian age-, sex- and state-specific mortality}

\begin{figure}[!htbp]
\centering
\begin{tikzpicture}
\tikzstyle{every node}=[circle,draw,inner sep=0.5pt]
\tikzstyle[level distance=.1cm]
\tikzstyle[sibling distance=.05cm]
\tikzstyle{level 3}=[sibling distance=6mm,font=\tiny]
\tikzstyle{level 2}=[sibling distance=10mm,font=\footnotesize]
\tikzstyle{level 1}=[sibling distance=21mm,font=\small]
\node{Total}
child {node {VIC}
	child {node {Female}}
	child {node {Male}}
	}
child {node {NSW}
	child {node {Female}}
	child {node {Male}}
	}
child {node {QLD}
	child {node {Female}}
	child {node {Male}}
	}
child {node {TAS}
	child {node {Female}}
	child {node {Male}}
	}
child {node {SA}
	child {node {Female}}
	child {node {Male}}
}
child {node {WA}
	child {node {Female}}
	child {node {Male}}
};       
\end{tikzpicture}
\caption{A two-level hierarchical tree diagram.}\label{fig:hier_2}
\end{figure}
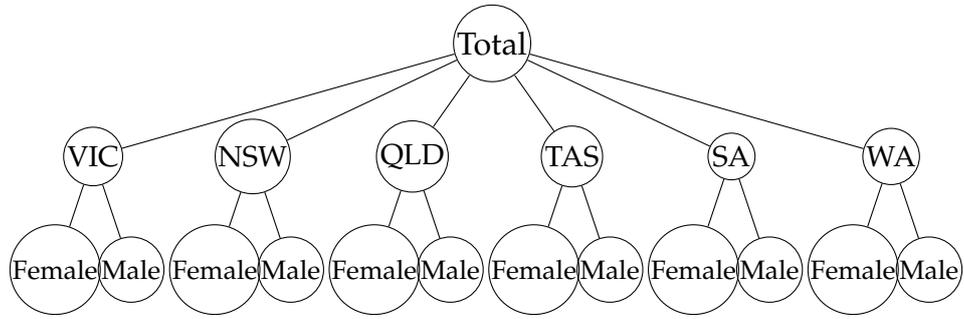

We extend the multilevel functional data method to two or more sub-populations in a hierarchy. This is related to hierarchical/grouped time series \citep[see, for example,][]{HAA+11}. A grouped structure is depicted in the two-level hierarchical diagram, presented in Figure~\ref{fig:hier_2}. 

Following a bottom-up hierarchical structure, we first extract a common trend from the total mortality within each state. For the $j^{\text{th}}$ population in state $s$, the multilevel functional data model can be written as:
\begin{equation}
f_t^{j,s}(x) = \mu^{j,s}(x) + R_t^{s}(x) + U_t^{j, s}(x), \label{eq:mult1}
\end{equation}
where $f_t^{j,s}(x)$ represents the female or male mortality in state $s$ at year $t$; $\mu^{j,s}(x)$ is the mean function of female or male mortality in state $s$; $R_t^{s}(x)$ captures the common trend across two populations for a state; and $U_t^{j, s}(x)$ captures the sex-specific residual trend for a state. Based on~\eqref{eq:cluster}, the proportion of variability explained by the total mortality in each state is 65\%, 69\%, 25\%, 53\%, 43\%, and 37\% for females, and 59\%, 59\%, 22\%, 54\%, 41\%, and 38\% for males.

We can also extract the common trend from the averaged mortality across all states for females and males. For the $j^{\text{th}}$ population in state $s$, the multilevel functional data model can be written as:
\begin{equation}
f_t^{j,s}(x) = \mu^{j,s}(x) + S_t^{j}(x)+W_t^{j,s}(x),\label{eq:mult2}
\end{equation}
where $S_t^{j}(x)$ captures the common trend across six populations; and $W_t^{j,s}(x)$ captures the state-specific residual trend. By combining~\eqref{eq:mult1} and~\eqref{eq:mult2}, we obtain
\begin{align}
f_t^{j,s}(x) &= \mu^{j,s}(x) + \frac{R_t^{s}(x) + U_t^{j, s}(x) + S_t^{j}(x)+W_t^{j,s}(x)}{2}.
\end{align}

\begin{table}[!htbp]
\begin{small}
\centering
 \setlength\tabcolsep{12pt}
\caption{Point forecast errors ($\times 100$) of mortality across states and sexes by method, as measured by the averaged MAFE, RMSFE, and MFE. The minimal forecast errors are underlined for each state and each sex, whereas the minimal overall forecast error is highlighted in bold.}\label{tab:state_point_mort}
\begin{tabular}{@{}llrrrrrrr@{}}\toprule
Sex & Method &  VIC & NSW & QLD & TAS & SA & WA & Mean \\\midrule
& \multicolumn{6}{l}{\underline{\text{MAFE}}} & \\
F 	&  Independent FDM		& 0.46 & 0.41 & 0.90 & 0.56 & 0.59 & 0.76 & 0.61 \\ 
	&  Product-ratio 		& 0.58 & 0.56 & \underline{0.47} & 0.60 & 0.51 & 0.50 & 0.54 \\ 
	&  Multilevel FDM (arima) & 0.39 & \underline{0.37} & 0.48 & 0.35 & \underline{0.35} & 0.36 & 0.38 \\ 
	&  Multilevel FDM (rwf) 	& \underline{0.38} & \underline{0.37} & \underline{0.47} & \underline{0.32} & \underline{0.35} & \underline{0.35} & \textBF{0.37} \\
  \\ 
 M 	& Independent FDM 		& 0.90 & 0.85 & 1.31 & 1.12 & 1.03 & 1.20 & 1.07 \\ 
	&  Product-ratio 		& \underline{0.75} & \underline{0.71} & \underline{0.59} & 0.83 & \underline{0.67} & \underline{0.83} & \textBF{0.73} \\ 
 	& Multilevel FDM (arima) 	& 0.98 & 0.94 & 1.13 & 0.85 & 0.88 & 1.08 & 0.98 \\ 
 	& Multilevel FDM (rwf) 	& 0.91 & 0.86 & 0.93 & \underline{0.73} & 0.79 & 0.98 & 0.87 \\ 
  \\
$\frac{\text{F+M}}{2}$ & Independent FDM 	& 0.68 & 0.63 & 1.11 & 0.84 & 0.81 & 0.98 & 0.84 \\ 
				 & Product-ratio 		& 0.66 & 0.63 & \underline{0.53} & 0.72 & 0.59 & \underline{0.66} & 0.63 \\ 
				 & Multilevel FDM (arima) & 0.69 & 0.66 & 0.80 & 0.60 & 0.62 & 0.72 & 0.68 \\ 
				 & Multilevel FDM (rwf)  	& \underline{0.65} & \underline{0.62} & 0.70 & \underline{0.53} & \underline{0.57} & \underline{0.66} & \textBF{0.62} \\ 
  \\
 &  \multicolumn{6}{l}{\underline{\text{RMSFE}}} \\
  F & Independent FDM 	& 1.20 & 0.99 & 2.02 & 1.34 & 1.35 & 1.63 & 1.42 \\ 
  & Product-ratio 		& 1.19 & 1.14 & \underline{0.99} & 1.48 & 1.12 & 1.08 & 1.17 \\ 
 & Multilevel FDM (arima) & 0.85 & 0.79 & 1.28 & 0.82 & \underline{0.81} & 0.86 & 0.90 \\ 
 & Multilevel FDM (rwf) 	& \underline{0.81} & \underline{0.78} & 1.26 & \underline{0.73} & 0.82 & \underline{0.81} & \textBF{0.87} \\ 	
  \\
  M & Independent FDM 	& 1.90 & 1.66 & 2.91 & 2.59 & 2.09 & 2.53 & 2.28 \\ 
  & Product-ratio 		& \underline{1.58} & \underline{1.41} & \underline{1.26} & 2.22 & \underline{1.51} & \underline{1.98} & \textBF{1.66} \\ 
 & Multilevel FDM (arima) & 1.94 & 1.77 & 2.58 & 1.70 & 1.83 & 2.36 & 2.03 \\ 
 & Multilevel FDM (rwf)  	& 1.77 & 1.58 & 2.30 & \underline{1.51} & 1.63 & 2.12 & 1.82 \\ 
  \\
$\frac{\text{F+M}}{2}$ & Independent FDM & 1.55 & 1.33 & 2.46 & 1.97 & 1.72 & 2.08 & 1.85 \\ 
 & Product-ratio 					& 1.39 & 1.28 & \underline{1.12} & 1.85 & 1.32 & 1.53 & 1.41 \\ 
&  Multilevel FDM (arima) 				& 1.40 & 1.28 & 1.93 & 1.26 & 1.32 & 1.61 & 1.46 \\ 
 & Multilevel FDM (rwf)  				& \underline{1.29} & \underline{1.18} & 1.78 & \underline{1.12} & \underline{1.23} & \underline{1.46} & \textBF{1.35} \\
  \\
  & \multicolumn{6}{l}{\underline{\text{MFE}}}\\
 F &  Independent FDM  	& -0.16 & -0.09 & -0.77 & -0.23 & -0.50 & -0.60 & -0.39 \\ 
 & Product-ratio  		& -0.55 & -0.51 & -0.37 & -0.38 & -0.42 & -0.41 & -0.44 \\ 
 & Multilevel FDM (arima) & \underline{-0.34} & \underline{-0.30} & \underline{-0.15} & -0.21 & \underline{-0.21} & -0.23 & -0.24 \\ 
 & Multilevel FDM (rwf) 	& \underline{-0.34} & -0.32 & \underline{-0.15} & \underline{-0.16} & -0.22 & \underline{-0.20} & \textBF{-0.23} \\ 
  \\
 M & Independent FDM  	& -0.66 & -0.71 & -1.07 & -0.79 & -0.73 & -0.98 & -0.82 \\ 
 & Product-ratio  		& \underline{-0.65} & \underline{-0.61} & \underline{-0.36} & \underline{-0.24} & \underline{-0.41} & \underline{-0.66} & \textBF{-0.49} \\ 
&  Multilevel FDM (arima) & -0.87 & -0.82 & -0.69 & -0.65 & -0.62 & -0.91 & -0.76 \\ 
 & Multilevel FDM (rwf) 	& -0.83 & -0.77 & \underline{-0.36} & -0.48 & -0.58 & -0.84 & -0.64 \\ 
  \\
$\frac{\text{F+M}}{2}$ & Independent FDM & \underline{-0.41} & \underline{-0.40} & -0.92 & -0.51 & -0.62 & -0.79 & -0.60 \\ 
 &  Product-ratio 					& -0.60 & -0.56 & -0.37 & \underline{-0.31} & -0.42 & -0.54 & -0.46 \\ 
 &  Multilevel FDM (arima) 			& -0.60 & -0.56 & -0.42 & -0.43 & -0.42 & -0.57 & -0.50 \\ 
 &  Multilevel FDM (rwf) 				& -0.59 & -0.54 & \underline{-0.26} & -0.32 & \underline{-0.40} & \underline{-0.52} & \textBF{-0.43}  
	\\\bottomrule		
\end{tabular}
\end{small}
\end{table}

\begin{table}[!htbp]
\begin{small}
\centering
 \setlength\tabcolsep{12pt}
\caption{Point forecast accuracy of life expectancy across states and sexes by method, as measured by the averaged MAFE, RMSFE, and MFE. The minimal forecast errors are underlined for each state and each sex, whereas the minimal overall forecast error is highlighted in bold.}\label{tab:state_point_e0}
\begin{tabular}{@{}llrrrrrrr@{}}\toprule
Sex & Method &  VIC & NSW & QLD & TAS & SA & WA & Mean \\\midrule
&\multicolumn{6}{l}{\underline{\text{MAFE}}} & \\
F 	&  Independent FDM 	& \underline{1.92} & \underline{1.94} & 4.48 & 2.49 & 2.91 & 3.87 & 2.93 \\ 
 	& Product-ratio 			& 2.94 & 3.07 & 2.67 & 2.26 & 2.42 & 2.68 & 2.67 \\ 
 	&  Multilevel FDM (arima) & 1.97 & 2.05 & 1.62 & 1.82 & \underline{1.48} & \underline{1.76} & \textBF{1.78} \\ 
 	&  Multilevel FDM (rwf) 	& 2.08 & 2.26 & \underline{1.32} & \underline{1.76} & 1.57 & \underline{1.76} & \textBF{1.78} \\ 
  \\
 M 	& Independent FDM  	& 3.44 & 3.65 & 5.51 & 4.24 & 4.47 & 4.80 & 4.35 \\ 
  	& Product-ratio 			& \underline{3.18} & \underline{3.44} & 2.93 & \underline{2.24} & \underline{2.53} & \underline{3.07} & \textBF{2.90} \\ 
  	& Multilevel FDM (arima) 	& 3.91 & 4.08 & 4.09 & 3.85 & 3.33 & 3.95 & 3.87 \\ 
 	 & Multilevel FDM (rwf) 	& 3.95 & 4.20 & \underline{2.84} & 3.63 & 3.29 & 3.87 & 3.63 \\ 
  \\
 $\frac{\text{F+M}}{2}$ 	& Independent FDM  	& \underline{2.68} & \underline{2.79} & 5.00 & 3.36 & 3.69 & 4.33 & 3.64 \\ 
 					& Product-ratio 			& 3.06 & 3.26 & 2.80 & \underline{2.25} & 2.48 & 2.87 & 2.78 \\ 
					& Multilevel FDM (arima) 	& 2.94 & 3.06 & 2.86 & 2.83 & \underline{2.40} & 2.85 & 2.83 \\ 
					& Multilevel FDM (rwf) 	& 3.02 & 3.23 & \underline{2.08} & 2.69 & 2.43 & \underline{2.81} & \textBF{2.71} \\ 
  \\
& \multicolumn{6}{l}{\underline{\text{RMSFE}}} & \\
 F 	& Independent FDM  	& 2.45 & 2.18 & 4.55 & 3.02 & 3.23 & 4.11 & 3.26 \\ 
  	& Product-ratio 			& 3.03 & 3.20 & 2.83 & 2.42 & 2.61 & 2.82 & 2.82 \\ 
 	& Multilevel FDM (arima) 	& \underline{2.09} & \underline{2.17} & 1.77 & 2.01 & \underline{1.66} & 1.92 & \textBF{1.93} \\ 
 	& Multilevel FDM (rwf) 	& 2.18 & 2.36 & \underline{1.51} & \underline{1.91} & 1.75 & \underline{1.91} & \textBF{1.93} \\ 
  \\
 M &  Independent FDM  		& 3.71 & 3.86 & 5.55 & 4.58 & 4.66 & 5.05 & 4.57 \\ 
  & Product-ratio 			& \underline{3.23} & \underline{3.49} & 3.00 & \underline{2.32} & \underline{2.61} & \underline{3.14} & \textBF{2.96} \\ 
 &  Multilevel FDM (arima) 	& 4.06 & 4.25 & 4.25 & 4.04 & 3.54 & 4.14 & 4.05 \\ 
  & Multilevel FDM (rwf) 		& 4.00 & 4.25 & \underline{2.92} & 3.70 & 3.39 & 3.94 & 3.70 \\ 
  \\
$\frac{\text{F+M}}{2}$ 	&  Independent FDM  	& 3.08 & \underline{3.02} & 5.05 & 3.80 & 3.94 & 4.58 & 3.91 \\ 
  					& Product-ratio 			& 3.13 & 3.35 & 2.91 & \underline{2.37} & 2.61 & 2.98 & 2.89 \\ 
				  	& Multilevel FDM (arima) 	& \underline{3.07} & 3.21 & 3.01 & 3.02 & 2.60 & 3.03 & 2.99 \\ 
				 	&  Multilevel FDM (rwf) 	& 3.09 & 3.30 & \underline{2.21} & 2.81 & \underline{2.57} & \underline{2.92} & \textBF{2.82} \\
  \\
&  \multicolumn{6}{l}{\underline{\text{MFE}}} & \\
  F & Independent FDM  	& \underline{0.98} & \underline{1.00} & 4.48 & \underline{1.48} & 2.90 & 3.27 & 2.35 \\ 
  & Product-ratio 		& 2.93 & 3.06 & 2.66 & 2.25 & 2.41 & 2.68 & 2.67 \\ 
 &  Multilevel FDM (arima)& 1.97 & 2.03 & 1.61 & 1.80 & \underline{1.45} & \underline{1.76} & \textBF{1.77} \\ 
 &  Multilevel FDM (rwf) 	& 2.08 & 2.25 & \underline{1.26} & 1.75 & 1.54 & \underline{1.76} & \textBF{1.77} \\ 
  \\
  M & Independent FDM  	& 3.43 & 3.62 & 5.51 & 3.95 & 4.47 & 4.71 & 4.28 \\ 
  & Product-ratio 		& \underline{3.17} & \underline{3.44} & 2.91 & \underline{2.23} & \underline{2.51} & \underline{3.07} & \textBF{2.89} \\ 
 &  Multilevel FDM (arima)& 3.91 & 4.06 & 4.09 & 3.82 & 3.28 & 3.94 & 3.85 \\ 
 &  Multilevel FDM (rwf) 	& 3.95 & 4.19 & \underline{2.81} & 3.62 & 3.29 & 3.87 & 3.62 \\ 
  \\
 $\frac{\text{F+M}}{2}$ &  Independent FDM  	& \underline{2.21} & \underline{2.31} & 5.00 & 2.72 & 3.69 & 3.99 & 3.32 \\ 
 &  Product-ratio 						& 3.05 & 3.25 & 2.79 & \underline{2.24} & 2.46 & 2.87 & 2.78 \\ 
 &  Multilevel FDM (arima) 				& 2.94 & 3.05 & 2.85 & 2.81 & \underline{2.37} & 2.85 & 2.81 \\ 
 &  Multilevel FDM (rwf) 					& 3.02 & 3.22 & \underline{2.03} & 2.68 & 2.42 & \underline{2.81} & \textBF{2.70} 
	\\\bottomrule		
\end{tabular}
\end{small}
\end{table}

\begin{table}[!htbp]
\centering
 \setlength\tabcolsep{8.5pt}
\caption{Interval forecast accuracy of mortality and life expectancy across states and sexes by method, as measured by the averaged mean interval score. The minimal forecast errors are underlined for each state and each sex, whereas the minimal overall forecast error is highlighted in bold.}\label{tab:state_interval}
\begin{tabular}{@{}llrrrrrrr@{}}\toprule
Sex & Method &  VIC & NSW & QLD & TAS & SA & WA & Mean \\\midrule
&\multicolumn{6}{l}{\underline{\text{Mortality} ($\times 100$)}}  & \\
F &  Independent FDM 		& 3.12 & 2.28 & 4.93 & 3.57 & 3.46 & 4.44 & 3.63 \\ 
   &  Product-ratio 			& 2.76 & 2.64 & 3.11 & 2.30 & 2.43 & 2.64 & 2.65 \\ 
   &  Multilevel FDM (arima) 	& 1.83 & 1.74 & 2.41 & 1.66 & \underline{1.70} & \underline{1.71} & 1.84 \\ 
   &  Multilevel FDM (rwf) 		& \underline{1.78} & \underline{1.73} & \underline{2.36} & \underline{1.54} & 1.72 & \underline{1.71} & \textBF{1.81} \\
   \\
M & Independent FDM 		& 6.00 & 5.10 & 7.50 & 7.37 & 6.79 & 7.35 & 6.68 \\
    & Product-ratio 			& \underline{3.63} & \underline{3.52} & \underline{4.10} & \underline{3.12} & \underline{3.46} & \underline{3.84} & \textBF{3.61} \\   
    &  Multilevel FDM (arima) 	& 6.71 & 6.62 & 6.57 & 5.57 & 5.63 & 6.99 & 6.35 \\
    &  Multilevel FDM (rwf) 		& 4.61 & 4.50 & 4.68 & 3.81 & 4.07 & 4.81 & 4.41 \\
\\
$\frac{\text{F}+\text{M}}{2}$ 	& Independent FDM 		& 4.56 & 3.69 & 6.22 & 5.47 & 5.12 & 5.90 &  5.16	\\
		  				& Product-ratio			& \underline{3.20} & \underline{3.08} & 3.60 & 2.71 & 2.94 & \underline{3.24} & 3.13	\\
						& Multilevel FDM (arima)	& 4.27 & 4.18 & 4.49 & 3.62 & 3.66 & 4.35 & 4.10 \\
						& Multilevel FDM (rwf)	& \underline{3.20} & 3.11 & \underline{3.52} & \underline{2.68} & \underline{2.90} & 3.26 & \textBF{3.11} \\
\\			
&\multicolumn{6}{l}{\underline{\text{e(0)}}} & \\
F & Independent FDM 		& \underline{7.76} 	& 13.31 	& 33.49 	& 13.91 	& 8.09 	& 11.75 	& 14.72 \\
   & Product-ratio			& 20.09 	& 21.50 	& 14.10 	& 17.84 	& 15.49 	& 17.70 	& 17.79 \\
   & Multilevel FDM (arima)	& 9.43 	& 9.74 	& 6.98 	& 8.37 	& 6.46 	& 7.37 	& 8.06  \\
   & Multilevel FDM (rwf)		& 8.07 	& \underline{8.93} 	& \underline{5.29} 	& \underline{6.51} 	& \underline{5.88} 	& \underline{6.50} 	& \textBF{6.86} \\
\\
M & Independent FDM		& 33.67 & 35.66 & 49.16 & 37.06 & 34.50 & 29.49 & 36.59  \\    
    & Product-ratio			& \underline{22.01} & \underline{24.30} & \underline{11.71} & \underline{18.38} & \underline{15.97} & \underline{20.81} & \textBF{18.86} \\
    & Multilevel FDM (arima)	& 32.51 & 33.57 & 29.37 & 31.44 & 27.92 & 32.57 & 31.23 \\
    & Multilevel FDM (rwf)		& 26.07 & 28.34 & 16.58 & 22.55 & 20.28 & 25.37 & 23.20 \\
\\
$\frac{\text{F}+\text{M}}{2}$ & Independent FDM 		& 20.72 & 24.49 & 41.32 & 25.48 & 21.30 & 20.62 & 25.65 \\
					   & Product-ratio	  		& 21.05 & 22.90 & 12.90 & 18.11 & 15.73  & 19.25 & 18.32 \\
					   & Multilevel FDM (arima) 	& 20.97 & 21.66 & 18.17 & 19.90 & 17.19 & 19.97 & 19.64 \\
					   & Multilevel FDM (rwf)		& \underline{17.07} & \underline{18.63} & \underline{10.94} & \underline{14.53} & \underline{13.08} & \underline{15.94} &  \textBF{15.03} \\			
\bottomrule
\end{tabular}
\end{table}

Tables~\ref{tab:state_point_mort},~\ref{tab:state_point_e0} and~\ref{tab:state_interval} show the point and interval forecast accuracy among different functional data methods. As measured by the averaged MAFE, RMSFE, MFE and averaged mean interval score across 30 horizons, the multilevel functional data method (rwf) gives the smallest errors for forecasting female mortality rate and life expectancy, as well as the smallest overall errors, whereas the product-ratio method produces the most accurate forecasts for male mortality rate and life expectancy. 

Apart from the expected error loss function, we also consider the maximum point and interval forecast error criteria. Their results are also included in the supplement D \citep{Shang16}. 

\section{Conclusion}\label{sec:6}

In this paper, we adapt the multilevel functional data model to forecast age-specific mortality and life expectancy for a group of populations. We highlight the relationships among the adapted multilevel functional data, augmented common factor method and product-ratio method. 

As demonstrated by the empirical studies consisting of two populations, we found that the independent functional data method gives the best forecast accuracy for females, whereas the multilevel functional data and product-ratio methods produce more accurate forecasts for males. Based on their averaged forecast errors, the multilevel functional data method (arima) should be used in the case of two sub-populations, in particular for females. 

In the case of more than two populations, it is evident that the multilevel functional data and product-ratio methods consistently outperform the  independent functional data method. The multilevel functional data method (rwf) gives the most accurate mortality and life expectancy forecasts for age- and state-specific total mortality. When we further disaggregated the age- and state-specific total mortality by sex, we found that the multilevel functional data method (rwf) should be used for forecasting female mortality and life expectancy, whereas the product-ratio method should be applied for forecasting male mortality and life expectancy. 

The superiority of the product-ratio and multilevel functional data methods over the independent functional data method is manifested by a population with large variability over age and year. For example, the male data generally show greater variability over age and year than do the female data; as a result the product-ratio and multilevel functional data methods perform better in terms of forecast accuracy than the independent functional data method. Because the product-ratio and multilevel functional data methods produce better forecast accuracy than the independent functional data method overall, this may lead to their use by government agencies and statistical bureaus involved in short-term demographic forecasting. For long-term forecast horizons, any time-series extrapolation methods, including the proposed one, may not be accurate as the underlying model may no longer be optimal. Given that different changes are at play in different phases of a mortality transition, the age components of change in the past are not necessarily informative of the longer-term future. By incorporating prior knowledge, the Bayesian method of \cite{RLG14} demonstrated the superior forecast accuracy of the long-term projection of life expectancy.

A limitation of the current study is that the comparative analysis among the five methods focuses on errors that aggregate over all age groups for one- to 30-step-ahead mortality forecasts. In future research, it is possible that the analysis of the forecast errors for certain key age groups, such as those above 65, might shed light on the results of more detailed analysis. For a relatively long time series, geometrically decaying weights can be imposed on the computation of functional principal components \citep[see, for example,][]{HS09} for achieving potentially improved forecast accuracy. In addition, the product-ratio and multilevel functional data methods could be applied to model and forecast other demographic components, such as age-specific immigration, migration, and population size by sex or other attributes for national and sub-national populations. Reconciling these forecasts across different levels of a hierarchy is worthwhile to investigate in the future \citep[see an early work by][]{SH16}. 

\begin{singlespace}
\section*{Supplement to: ``Mortality and life expectancy forecasting for a group of populations in developed countries: A multilevel functional data method." by H. L. Shang} 
This supplement contains a PDF divided into four sections.
\begin{description}
\item[Supplement A:] Some theoretical properties of multilevel functional principal component decomposition; 
\item[Supplement B:] Derivation of posterior density of principal component scores and other variance parameters; 
\item[Supplement C:] WinBUGS computational code used for sampling principal component scores and estimating variance parameters from full conditional densities; 
\item[Supplement D:] Additional results for point and interval forecast accuracy of mortality and life expectancy, based on maximum forecast error measures.
\end{description}
\end{singlespace}

\begin{center}
\textbf{Supplement to ``Mortality and life expectancy forecasting for a group of populations in developed countries: A multilevel functional data method} by H. L. Shang
\end{center}
\vspace{.1in}

\begin{center}
\large Supplement A: Some theoretical properties of multilevel functional principal component decomposition
\end{center}

Let $R$ and $U^{j}$ be two stochastic processes defined on a compact set $\mathcal{I}$, with finite variance. The covariance functions of $R$ and $U^{j}$ are defined to be the function $\mathcal{K}: \mathcal{I}\times \mathcal{I}\rightarrow R$, such that
\begin{align*}
\mathcal{K}^R(w,v) &= \text{cov}\{R(w), R(v)\} = \text{E}\left\{[R(w)-\mu(w)]\otimes [R(v)-\mu(v)]\right\},\\
\mathcal{K}^{U^{j}}(w,v) &= \text{cov}\left\{U^{j}(w), U^{j}(v)\right\} = \text{E}\left\{[U^{j}(w)-\mu(w)]\otimes [U^{j}(v)-\mu(v)]\right\},
\end{align*}
where $\otimes$ represents the tensor product and $j$ represents the index of sub-populations. In a finite dimension, the tensor product reduces to matrix multiplication.

Mercer's theorem \citep[][Chapter 4]{Indritz63} provides the following consistent spectrum decomposition,
\begin{align*}
  \mathcal{K}^R(w,v) &= \text{cov}\left\{R(w), R(v)\right\}= \sum^{\infty}_{k=1}\lambda_k \phi_k(w)\phi_k(v),\\
  \mathcal{K}^{U^{j}}(w,v) &= \text{cov}\left\{U^{j}(w), U^{j}(v)\right\}= \sum^{\infty}_{l=1}\lambda^{j}_l\psi_l^{j}(w)\psi_l^{j}(v),
\end{align*}

\noindent where $\lambda_1\geq \lambda_2\geq \dots$ are the ordered population eigenvalues and $\phi_k(\cdot)$ is the $k^{\text{th}}$ orthonormal eigenfunction of $\mathcal{K}^{R}(\cdot,\cdot)$ in the $L^2$ norm. Similarly, $\lambda^{j}_1\geq \lambda^{j}_2\geq\dots$ are the ordered population eigenvalues and $\psi_l^{j}(\cdot)$ is the $l^{\text{th}}$ orthonormal eigenfunction of $\mathcal{K}^{U^{j}}(\cdot,\cdot)$ in the $L^2$ norm.

With Mercer's lemma, stochastic processes $R$ and $U^{j}$ can be expressed by the Karhunen-Lo\`{e}ve expansion \citep{Karhunen46, Loeve46}. In practice, we reduce the dimensionality of functional data by truncating the infinite series to finite dimension, such as the first $K$ number of principal components \citep*{YMW05, HH06,Hossein13}. These can be expressed as:
\begin{align*}
  R_t(x) &= \sum^{\infty}_{k=1}\beta_{t,k}\phi_k(x)\approx  \sum^{K}_{k=1}\beta_{t,k}\phi_k(x),\\
  U_t^{j}(x) &= \sum^{\infty}_{l=1}\gamma_{t,l}^{j}\psi_{l}^{j}(x)\approx \sum^{L}_{l=1}\gamma_{t,l}^{j}\psi_{l}^{j}(x), 
\end{align*}

\noindent where $\beta_{t,k}=\int_{\mathcal{I}} R_t(x)\phi_k(x)dx$, $\gamma_{t,l}^{j}=\int_{\mathcal{I}}U_t^{j}(x)\psi_l^{j}(x)dx$ are the uncorrelated principal component scores with $\text{E}(\beta_{t,k}) = \text{E}\left(\gamma_{t,l}^{j}\right)=0$, $\text{Var}(\beta_{t,k})=\lambda_k<\infty$, $\text{Var}(\gamma_{t,l}^{j})=\lambda_l^{j}<\infty$, $K$ and $L$ represent the retained numbers of principal components, and $\mathcal{I}$ represents the domain of $x$ variable, such as $x\in [0,95+]$ in our context. 

\newpage

\begin{center}
\large Appendix B: Derivation of posterior density of principal component scores
\end{center}

We present derivations for the multilevel functional data model, including its specification and full conditional densities. The full conditionals are also given in \cite{DCC+09}, which provides a foundation for this work. Here, we extend it by adding an additional stochastic variance for the pre-smoothing step. This stochastic variance takes into account the varying uncertainty across observations. 
\[ \left\{ \begin{array}{ll}
         y_t^j(x_i) = f_t^j(x_i)+\delta_t^j(x_i)\epsilon_{t,i}^j  \\
	f_t^j(x_i) = \mu(x_i) + \eta^j(x_i) + \sum^K_{k=1}\beta_{t,k}\phi_k(x_i) + \sum^L_{l=1}\gamma_{t,l}^j\psi_l^j(x_i)+\varepsilon_t^j(x_i) \\
        \beta_{t,k}\sim N\left(0,\lambda_k\right); \gamma_{t,l}^j\sim N\left(0,\lambda_l^{j}\right); \varepsilon_t^j(x_i)\sim N(0,(\sigma^2)^j); \delta_t^j(x_i) \sim N(0, (\kappa_i^2)^j) \\
        \frac{1}{(\sigma^2)^j} \sim \text{Gamma}(\alpha_1,\alpha_2) 
        \end{array} \right. \] 
\begin{enumerate}        
\item The full conditional density of inverse error variance given other parameters is
\begin{align*}
1/\left(\sigma^2\right)^j\big|\text{others} \sim \text{Gamma}\left(\alpha_1^{\text{post}},\alpha_2^{\text{post}}\right), 
\end{align*}
where
\begin{align*}
\alpha_1^{\text{post}} &= \alpha_1 + \frac{1}{2}Jnp \\
\alpha_2^{\text{post}} &= \alpha_2 + \frac{1}{2}\sum^J_{j=1}\sum^n_{t=1}\sum^p_{i=1}\left[\varepsilon_t^j(x_i)\right]^2
\end{align*}
and
\begin{equation*}
\varepsilon_t^j(x_i) = f_t^j(x_i) - \mu(x_i) - \eta^j(x_i) - \sum^K_{k=1}\beta_{t,k}\phi_k(x_i) - \sum^L_{l=1}\gamma_{t,l}^j\psi_l^j(x_i),
\end{equation*}
where $J$ denotes the number of populations, $n$ denotes the sample size, and $p$ denotes the total number of age groups.
\item The full conditional density of principal component scores for the common trend given other parameters is
\begin{equation*}
\beta_{t,k}\big|\text{others}\sim N\left(\mu_{\beta_{t,k}}^{\text{post}}, v_{\beta_{t,k}}^{\text{post}}\right)
\end{equation*}
where
\begin{align*}
\mu_{\beta_{t,k}}^{\text{post}} &= \frac{\lambda_k J \sum^p_{i=1}\phi_k(x_i)^2}{\lambda_k J \sum^p_{i=1}\phi_k(x_i)^2 + (\sigma^2)^j}\cdot \frac{\sum^J_{j=1}\sum^p_{i=1}\phi_k(x_i)\left[\varepsilon_t^j(x_i)+\beta_{t,k}\phi_k(x_i)\right]}{J\sum^p_{i=1}\phi_k(x_i)^2}, \\
v_{\beta_{t,k}}^{\text{post}} &= \frac{\lambda_k (\sigma^2)^j}{\lambda_k J\cdot \sum_{i=1}^p\phi_k(x_i)^2+(\sigma^2)^j},
\end{align*}
where $\lambda_k$ denotes the $k$th eigenvalue of the common covariance function.
\item The full conditional density of principal component scores for the population-specific residual trend given other parameters is
\begin{equation*}
\gamma_{t,l}^j|\text{others} \sim N\left(\mu_{\gamma_{t,l}^j}^{\text{post}}, v_{\gamma_{t,l}^j}^{\text{post}}\right),
\end{equation*}
where
\begin{align*}
\mu_{\gamma_{t,l}^j}^{\text{post}} &= \frac{\lambda_l^j\cdot \sum^{p}_{i=1}\psi_l^j(t_i)^2}{\lambda_l^j\cdot \sum_{i=1}^p\psi_l^j(x_i)^2+(\sigma^2)^j}\cdot \frac{\sum_{i=1}^p\phi_k(x_i)\left[\varepsilon_{t}^j(x_i)+\gamma_{t,l}^j\psi_l^j(x_i)\right]}{\sum_{i=1}^p\psi_l^{j}(x_i)^2}, \\
v_{\gamma_{t,l}^j}^{\text{post}}&=\frac{\lambda_l^j(\sigma^2)^j}{\lambda_l^j\cdot\sum_{i=1}^p\psi_l^j(x_i)^2+(\sigma^2)^j},
\end{align*}
where $\lambda_l^j$ denotes the $l$th eigenvalue of the population-specific covariance function. 

\end{enumerate}

Since the first step involves a nonparametric smoothing with heteroscedastic of unknown form. We can incorporate this nonparametric smoothing step in our Markov chain Monte Carlo (MCMC) iterations. For different ages or age groups, variances are unequal as shown in equation (2.4) of the main manuscript. Following the early work by \citet[][Chapter 6.4]{Koop03}, we consider a linear regression with heterscedastic errors and its Bayesian computation algorithm is documented in \citet[][pp. 127-128]{Koop03}

Let $(\omega_1,\omega_2,\dots,\omega_p)=\left[1/\delta^2(x_1),1/\delta^2(x_2),\dots,1/\delta^2(x_p)\right]$ be the precision parameters for different ages. Consider the following Gamma prior for $\omega_i$:
\begin{equation*}
\pi(\omega_i) = f_{\text{G}}(1,v_{\omega}),\qquad i=1,2,\dots,p,
\end{equation*}
where the prior for $\omega_i$ depends upon a hyperparameter $v_{\omega}$ and assume that each precision $\omega_i$ comes from the same distribution, but can differ from each other in values. 

Each of the conditional posteriors for $\omega_i$ has the form of a Gamma density, given by
\begin{align*}
\pi(\omega_i|v_{\omega}, \text{others}) &= f_{\text{G}}\left(\frac{v_{\omega}+1}{\sum^n_{t=1}\left[y_t(x_i) - f_t(x_i)\right]^2+v_{\omega}}, v_{\omega}+1 \right), \\
\pi(v_{\omega}|\omega_i,\text{other}) &\propto \left(\frac{v_{\omega}}{2}\right)^{p\cdot \frac{v_{\omega}}{2}}\Gamma\left(\frac{v_{\omega}}{2}\right)^{-p}e^{-\eta \cdot v_{\omega}},
\end{align*}
where $\eta = \frac{1}{v_{\omega}}+\frac{1}{2}\sum^p_{i=1}\left[\ln \left(\frac{1}{\omega_i}\right)+\omega_i\right]$, and $\Gamma(\cdot)$ denotes a Gamma function. 

\newpage

\begin{center}
\large Supplement C: WinBUGS code used for estimating variance parameters
\end{center}

Statistical software WinBUGS is used to estimate variances in the principal component scores and error function. From the estimated variances, the principal component scores and error function are simulated from normal distributions with zero mean. Below is a modified version of WinBUGS given by \cite{CG10}, for modeling age- and sex-specific mortality rates.

\lstloadlanguages{R}

\lstset{stringstyle=\rmfamily\normalsize, basicstyle=\rmfamily\normalsize,  mathescape=true,  escapeinside=||,  autogobble}

\begin{lstlisting}
model
{
  for (i in 1:N_subj)
  {
	for (t in 1:N_obs)
	{
		W_1[i,t] ~ dnorm(m_1[i,t], taueps_1)
		W_2[i,t] ~ dnorm(m_2[i,t], taueps_2)
		m_1[i,t] <- X[i,t] + U_1[i,t]
		m_2[i,t] <- X[i,t] + U_2[i,t]

		X[i,t] <- inprod(xi[i,], psi_1[t,])
		U_1[i,t] <- inprod(zi[i,], psi_2[t,])
		U_2[i,t] <- inprod(fi[i,], psi_3[t,])
	}
	for(k in 1:dim_space_b)
	{
		xi[i,k] ~ dnorm(0.0, ll_b[k])
	}
	for(l in 1:dim_space_w)
	{
		zi[i,l]  ~ dnorm(0.0, ll_w[l])
	}
	for(j in 1:dim_space_f)
	{	
		fi[i,j] ~ dnorm(0.0, ll_f[j])
	}
  }	
  for(k in 1:dim_space_b)
  {
	ll_b[k] ~ dgamma(1.0E-3, 1.0E-3)
	lambda_b[k] <- 1/ll_b[k]
  }
  for(l in 1:dim_space_w)
  {
	ll_w[l] ~ dgamma(1.0E-3, 1.0E-3)
	lambda_w[l] <- 1/ll_w[l]
  }
  for(j in 1:dim_space_f)
  {
 	ll_f[j] ~ dgamma(1.0E-3, 1.0E-3)
	lambda_f[j] <- 1/ll_f[j]
  }
  taueps_1 ~ dgamma(1.0E-3, 1.0E-3)
  taueps_2 ~ dgamma(1.0E-3, 1.0E-3)
}
\end{lstlisting}

\noindent The definition of all variables is given below:
\begin{enumerate}
\item $\text{N}\_\text{subj}$ is the number of subjects (sample size)
\item $\text{N}\_\text{obs}$ is the number of observations within subjects
\item $\text{W}\_1$[i,t] and $\text{W}\_2$[i,t] are the functional observations 
   at the aggregated level and sex-specific level, for 
   subject i at time t. Both matrices $\text{W}\_1$[,] and $\text{W}\_2$[,] 
   are $\text{N}\_\text{subj} \times \text{N}\_\text{obs}$, are loaded as data and may 
   contain missing observations
\item $\text{m}\_1$[i,t] and $\text{m}\_2$[i,t] are the smoothed means of 
   $\text{W}\_1$[i,t] and $\text{W}\_2$[i,t], respectively, are unknown 
   and their joint distribution is simulated
\item $\text{X}$[i,t] is the mean process at the aggregated level.
   $\text{X}$[,] is a $\text{N}\_\text{subj}\times \text{N}\_\text{obs}$ dimensional matrix of 
   parameters that are estimated from the model
\item $\text{U}\_1$[i,t] and $\text{U}\_2$[i,t] are the sex-specific mean process 
   at the individual level. $\text{U}\_1$[i,t] and $\text{U}\_2$[i,t] are the 
   $\text{N}\_\text{subj} \times \text{N}\_\text{obs}$ dimensional matrices of parameters 
   that are estimated from the model
\item $\text{psi}\_1$[t,], $\text{psi}\_2$[t,], $\text{psi}\_3$[t,] are eigenfunctions at both 
   the aggregated level and sex-specific level, evaluated 
   at the time t. The matrices $\text{psi}\_1$, $\text{psi}\_2$, $\text{psi}\_3$ are 
   $\text{N}\_\text{obs}\times \text{K}\_1$, $\text{N}\_\text{obs} \times \text{K}\_2$, $\text{N}\_\text{obs} \times \text{K}\_3$, 
   where $\text{K}\_1$ is the number of retained components that 
   explains at least 90\% of total variations in total mortality data, 
   $\text{K}\_2$ and $\text{K}\_3$ are the number of retained components 
   that explains at least 90\% of the remaining 10\% total 
   variations in sex-specific data. The matrices of $\text{psi}\_1$, $\text{psi}\_2$, $\text{psi}\_3$ 
   do not contain any missing value, and are loaded as data
\item $\text{xi}$[i,k] are the scores for the subject i on the 
   kth eigenfunction $\text{psi}\_1$[t,k]
\item $\text{zi}$[i,l] are the scores for the subject i on the lth 
   eigenfunction $\text{psi}\_2$[t,l]
\item $\text{fi}$[i,j] are the scores for the subject i on the jth 
    eigenfunction $\text{psi}\_3$[t,j]
\item $\text{ll}\_{\text{b}}$[k] are the precisions for the distribution of 
    scores $\text{xi}$[i,k]
\item $\text{ll}\_{\text{w}}$[l] are the precisions for the distribution of 
    scores $\text{zi}$[i,l]
\item $\text{ll}\_{\text{f}}$[j] are the precisions for the distribution of 
    scores $\text{fi}$[i,j]
\item $\text{taueps}\_1$ is the precision of the error process due to 
    imperfect observations of $\text{W}\_1$[i,t] around its smooth mean 
    $\text{m}\_1$[i,t]. This is a parameter of the model that is estimated
\item $\text{taueps}\_2$ is the precision of the error process due to 
    imperfect observations of $\text{W}\_2$[i,t] around its smooth mean 
    $\text{m}\_2$[i,t]. This is a parameter of the model that is estimated
\item All precision priors are Gamma priors with mean 1 and 
    variance 1000
\end{enumerate}

\newpage

\begin{center}
\large Supplement D: Additional results for point and interval forecast accuracy of mortality and life expectancy
\end{center}

Apart from the averaged forecast error criteria, we also consider the maximum absolute forecast error (Max AFE), maximum root squared forecast error (Max RSFE), and maximum interval score for measuring the extreme errors across different ages ($x_i$) and years in the forecasting period (year $k$). Averaging across 16 countries, they are defined as
\begin{align*}
\text{Max AFE}(h) &= \frac{1}{16}\sum^{16}_{c=1}\max_{k,i}\left|m_k^c(x_i)-\widehat{m}_k^c(x_i)\right|, \\
\text{Max RSFE}(h) &= \frac{1}{16}\sum^{16}_{c=1}\sqrt{\max_{k,i}\left[m_k^c(x_i)-\widehat{m}_k^c(x_i)\right]^2}, \\
\text{Max interval score}(h) &= \frac{1}{16}\sum^{16}_{c=1}\max_{k,i} S_{\alpha,k}^{c}(x_l,x_u;x_i).
\end{align*}

Tables~\ref{tab:1} to~\ref{tab:2} present the Max AFE, Max RSFE, and Max interval score for comparing point and interval forecast accuracies of the age-specific mortality and life expectancy by method, in the case of two populations. 

\begin{table*}[!htbp]
\centering
\tabcolsep 0.26cm
\caption{Point and interval forecast accuracy of mortality and life expectancy for females and males by method, as measured by the Max AFE, Max RSFE and Max interval score. For mortality, the forecast errors were multiplied by 100, in order to keep two decimal places. The minimal forecast errors are underlined for female and male data given in Section 5, whereas the minimal overall forecast error is highlighted in bold.}\label{tab:1}
\begin{tabular}{@{}lrrrrrrrrr@{}}\toprule
Method & \multicolumn{3}{c}{Max AFE} & \multicolumn{3}{c}{Max RSFE} & \multicolumn{3}{c}{Max interval score} \\
		&	F	& 	M 	& $\frac{\text{F+M}}{2}$ &	F 	& M & $\frac{\text{F+M}}{2}$ &	F 	& M & $\frac{\text{F+M}}{2}$ \\\midrule
\underline{Mortality ($\times 100$)} \\
Lee-Carter 			& 7.96 			& 9.37 			& 8.67 		& 0.71 			& 0.99 			& 0.85 		& 77.78 			& 97.47 			& 87.63		\\
Li-Lee	 			& 8.05 			& 8.00 			& 8.03 		& 0.72 			& 0.75 			& 0.74 		& 46.89 			& 40.47 			& 43.68 	\\
Independent FDM 		& \underline{7.11} 	& 8.05 			& \textBF{7.58} & \underline{0.55} 	& 0.72 			& 0.64 		& 35.13 			& 39.32 			& 37.23 \\
Product-ratio			& 7.52 			& 7.95 			& 7.74 		& 0.64 			& 0.69 			& 0.67 		& 38.20 			& 43.81 			& 41.01 \\
Multilevel FDM (arima) 	& 7.25 			& 7.90 			& \textBF{7.58} & 0.57  			& 0.68 			& \textBF{0.63} & \underline{32.06} 	& \underline{38.11} 	& \textBF{35.09} \\
Multilevel FDM (rwf)		& 7.95 			& \underline{7.85} 	& 7.90 		& 0.70 			& \underline{0.67} 	& 0.69 		& 40.03 			& 38.36 			& 39.20 \\
\\
\underline{e(0)}
\\
Lee-Carter 			& 2.85 			& 3.77 			& 3.31 		& 9.19 			& 16.63 			& 12.91 			& 15.74 			& 62.29 			& 39.02 \\ 
Li-Lee				& 3.54 			& 2.62 			& 3.08 		& 14.23 			& 7.91 			& 11.07 			& 24.75 			& 12.57 			& 18.66 \\
Independent FDM 		& \underline{2.22} 	& 3.69 			& 2.96 		& \underline{6.34} 	& 17.48 			& 11.91 			& 12.39 			& 24.61 			& 18.50 \\
Product-ratio			& 2.98 			& 2.86 			& 2.92 		& 11.38 			& 10.04			& 10.71 			& 18.35 			& 12.93 			& 15.64 \\
Multilevel FDM (arima) 	& 2.31 			& 3.01 			& \textBF{2.66} & 6.66 			& 11.75 			& \textBF{9.21} 		& \underline{10.62} 	& 14.05 			& \textBF{12.34} \\
Multilevel FDM (rwf)		& 3.07 			& \underline{2.45} 	& 2.76 		& 12.02 			& \underline{7.35} 	& 9.69 			& 18.86 			& \underline{9.47} 	& 14.17 \\\bottomrule
\end{tabular}
\end{table*}

\begin{table}[!htbp]
\centering
 \setlength\tabcolsep{7pt}
\caption{Point and interval forecast accuracy of mortality and life expectancy across different states (described in Section 6.1) by method, as measured by the Max AFE, Max RSFE, and maximum interval score. The minimal forecast errors are underlined for each state in Section 6, whereas the minimal overall forecast error is highlighted in bold.}
\begin{tabular}{@{}llrrrrrrr@{}}
  \toprule
	& & VIC & NSW & QLD & TAS & SA & WA & Mean \\
  \hline
Mortality &  \textit{Max AFE}		\\
($\times 100$) & Independent FDM 		& 9.01 & 10.43 & 12.12 & 14.47 & 10.91 & 10.44 & 11.23	\\
			& Product-ratio			& 7.57 & 7.36 & 6.42 & 11.93 & 8.85 & 8.20 & 8.39	\\
			& Multilevel FDM (arima) 	& 6.78 & 6.86 & 7.55 & \underline{10.99} & 8.73 & 8.18 & 8.18	\\
			& Multilevel FDM (rwf)	& \underline{6.13} & \underline{6.01} & \underline{6.14} & 11.03 & \underline{7.86} & \underline{7.78} & \textBF{7.49}	\\
			\\
		& \textit{Max RSFE} 	\\
		& Independent FDM &  0.85 & 1.13 & 1.53 & 2.16 & 1.22 & 1.10 & 1.33	\\
		& Product-ratio		& 0.58 & 0.55 & 0.43 & 1.56 & 0.85 & 0.69 & 0.78	\\
		& Multilevel FDM (arima) 	& 0.47 & 0.48 & 0.59 & \underline{1.35} & 0.83 & 0.69 & 0.73	 	\\
		& Multilevel FDM (rwf)	& \underline{0.38} & \underline{0.37} & \underline{0.41} & 1.37 & \underline{0.69} & \underline{0.65} & \textBF{0.65}		\\
\\
& \textit{Maximum interval score} \\
& Independent FDM 		& 9.71 & 7.12 	& 7.59 & 10.40 & 9.00 & 7.80 & 8.60 \\
& Product-ratio			& 4.17 & 4.25	& 3.87 & 3.47 & 3.69 & 3.98 & 3.90 \\
& Multilevel FDM (arima)  & 4.66 & 4.11 	& 3.58 & 3.51 & 2.92 & 4.29  & 3.84 \\
& Multilevel FDM (rwf) 	& \underline{4.08} & \underline{3.82} 	& \underline{3.17} & \underline{3.05} & \underline{2.45} & \underline{3.45} & \textBF{3.34} \\
\\
e(0) & \textit{Max AFE} \\
	& Independent FDM 		& 5.04 & 4.78 & 6.06 & 5.33 & 4.80 & 5.50 & 5.25		\\
	& Product-ratio			& 4.13 & 4.50 & 4.07 & 3.20 & 3.75 & 4.16 & 3.97		\\
	& Multilevel FDM (arima) 	& 3.97 & 4.38 & 4.22 & 3.97 & 3.72 & 4.08 & 4.06 		\\
	& Multilevel FDM (rwf)	& \underline{3.94} & \underline{4.30} & \underline{3.80} & \underline{2.96} & \underline{3.58} & \underline{3.95} & \textBF{3.75}		\\
	\\
	& \textit{Max RSFE} \\
	& Independent FDM 		& 30.80 & 27.51 & 42.28 & 32.41 & 26.44 & 35.11 & 32.43		\\
	& Product-ratio			& 19.85 & 23.14 & 19.02 & 11.99 & 15.88 & 19.61 & 18.25		\\
	& Multilevel FDM (arima)	& 18.47 & 22.00 & 20.43 & 18.26 & 15.59 & 18.98 & 18.95		\\
	& Multilevel FDM (rwf)	& \underline{18.13} & \underline{21.25} & \underline{16.85} & \underline{10.45} & \underline{14.41} & \underline{17.82} & \textBF{16.48}		\\
\\
& \textit{Maximum interval score} \\
& Independent FDM 		& 31.56 & 37.77 & 48.15 & 39.13 & 29.77 & 27.51 & 35.65 \\
& Product-ratio 			& 33.95 & 37.31 & 23.67 & 32.42 & 29.60 & 33.73 & 31.78 \\
& Multilevel FDM (arima) 	& 31.69 & 35.27 & 28.16 & 33.12 & 28.47 & 31.94 & 31.44 \\
& Multilevel FDM (rwf) 	& \underline{28.10} & \underline{31.34} & \underline{18.18} & \underline{25.98} & \underline{24.35} & \underline{27.93} & \textBF{25.98} \\
  \bottomrule  
\end{tabular}
\end{table}

\begin{table}[!htbp]
\centering
\begin{small}
 \setlength\tabcolsep{9pt}
\caption{Point and interval forecast accuracy of mortality ($\times 100$) across states and sexes (described in Section 6.2) by method, as measured by the Max AFE, Max RSFE, and maximum interval score. The minimal forecast errors are underlined for female and male data and their average, whereas the minimal overall forecast error is highlighted in bold.}
\begin{tabular}{@{}llrrrrrrr@{}}\toprule
Sex & Method &  VIC & NSW & QLD & TAS & SA & WA & Mean \\\midrule
&\multicolumn{6}{l}{\textit{Max AFE}}  & \\
F & Independent FDM 		& 9.26 & 8.90 & 17.32 & 10.07 & 10.44 & 10.48 & 11.08 	\\
	& Product-ratio 			& 7.42 & 7.41 & \underline{7.14} & 14.52 & 8.88 & 8.14 & 8.92	\\
	& Multilevel FDM (arima) 	& 6.45 & 6.20 & 13.94 & 7.64 & \underline{7.65} & 7.81 & 8.28	\\
	& Multilevel FDM (rwf)	& \underline{5.81} & \underline{5.85} & 13.63 & \underline{5.67} & \underline{7.65} & \underline{7.34} & \textBF{7.66}	\\
\\
M & Independent FDM 		& 15.22 & 11.96 & 28.65 & 24.66 & 16.12 & 20.76 & 19.56	\\
	& Product-ratio 			& 13.59 & 11.26 & \underline{12.67} & 27.04 & 13.78 & 19.17 & 16.25	\\
	& Multilevel FDM (arima) 	& 14.39 & 11.54 & 27.27 & \underline{13.62} & 13.79 & 20.03 & 16.77	\\
	& Multilevel FDM (rwf)	& \underline{12.40} & \underline{10.22} & 27.03 & 14.49 & \underline{13.16} & \underline{18.52} & \textBF{15.97}	\\
\\
$\frac{\text{F}+\text{M}}{2}$ & Independent FDM & 12.24 & 10.43 & 22.98 & 17.37 & 13.28 & 15.62 & 15.32 \\
	& Product-ratio 			& 10.50 & 9.34 &  \underline{9.90} & 20.78 & 11.33 & 13.66 & 12.59 \\
	& Multilevel FDM (arima) 	& 10.42 & 8.87 & 20.61 & 10.63 & 10.72 & 13.92 & 12.52	\\
	& Multilevel FDM (rwf)	&  \underline{9.11}  & \underline{8.04} & 20.33 & \underline{10.08} & \underline{10.41} & \underline{12.93} & \textBF{11.82}	\\
\\
& \multicolumn{6}{l}{\textit{Max RSFE}} & \\
F & Independent FDM 	& 0.89 & 0.85 & 3.09 & 1.04 & 1.10 & 1.12 & 1.35	\\
& Product-ratio 			& 0.56 & 0.56 & \underline{0.53} & 2.27 & 0.82 & 0.70 & 0.91	\\
& Multilevel FDM (arima) 	& 0.42 & 0.40 & 2.18 & 0.61 & \underline{0.64} & 0.65 & 0.82	\\
& Multilevel FDM (rwf) 	& \underline{0.35} & \underline{0.37} & 2.10 & \underline{0.35} & \underline{0.64} & \underline{0.59} & \textBF{0.73}	 \\
\\
M & Independent FDM 		& 2.36 & 1.51 & 9.92 & 6.76 & 2.66 & 4.47 & 4.61	\\
	& Product-ratio 			& 1.87 & 1.30 & \underline{1.78} & 9.67 & 2.06 & 3.86 & 3.42	\\
	& Multilevel FDM (arima) 	& 2.11 & 1.38 & 9.67 & \underline{1.95} & 2.02 & 4.16 & 3.55	\\
	& Multilevel FDM (rwf) 	& \underline{1.57} & \underline{1.09} & 9.89 & 2.36 & \underline{1.89} & \underline{3.66} & \textBF{3.41}	 \\
	\\
$\frac{\text{F}+\text{M}}{2}$ & Independent FDM & 1.62 &1.18 & 6.50 & 3.90 & 1.88 & 2.80 & 2.98  \\
				& Product-ratio 			& 1.22 & 0.93 & \underline{1.16} & 5.97 & 1.44 & 2.28 & 2.17 	\\
				& Multilevel FDM (arima) 	& 1.26 & 0.89 & 5.92 & \underline{1.28} & 1.33 & 2.41 & 2.19	\\
				& Multilevel FDM (rwf) 	& \underline{0.96} & \underline{0.73} & 6.00 & 1.36 & \underline{1.26} & \underline{2.12} & \textBF{2.07} 	\\
\\
	& \multicolumn{6}{l}{\textit{Maximum interval score}} \\
F  & Independent FDM		& 9.75 & 4.33 & 7.76 & 8.19 & 6.29 & 7.15 	& 7.24 \\
    & Product-ratio	    		& 4.15 & 4.20 & 4.48 & 3.50 & 3.70 & 3.76	& 3.97 \\
    & Multilevel FDM (arima)	& 3.33 & 3.30 & 3.88 & 3.19 & 2.95 & 2.98	& 3.27 \\
    & Multilevel FDM (rwf)		& \underline{2.99} & \underline{2.98} & \underline{3.72} & \underline{2.53} & \underline{2.85} & \underline{2.87}	& \textBF{2.99} \\
\\
M & Independent FDM		& 11.45 	& 8.21 	& 10.76 	& 11.70 	& 9.82 & 10.70 & 10.44 \\
    & Product-ratio			& \underline{4.71} 	& \underline{4.50} 	& 6.10 	& 3.98 	& \underline{4.62} & \underline{5.35} 	& \textBF{4.88} \\
    & Multilevel FDM (arima)	& 10.45 	& 10.15 	& 10.58 	& 8.60 	& 8.98 & 11.00 	& 9.96 \\
    & Multilevel FDM (rwf)		& 6.84 	& \underline{4.50} 	& \underline{4.68} 	& \underline{3.81} 	& 6.30 & 7.43 	& 5.59 \\
\\
 $\frac{\text{F}+\text{M}}{2}$   & Independent FDM		& 10.60 	& 6.27 & 9.26 & 9.95 & 8.06 & 8.92 & 8.84 \\
    & Product-ratio								& \underline{4.42} 	& 4.35 & 5.29 & 3.74 & \underline{4.16} & \underline{4.56} & 4.42 \\
    & Multilevel FDM (arima)						& 6.89 	& 6.72 & 7.23 & 5.90 & 5.96 & 6.99 & 6.62 \\
    & Multilevel FDM (rwf)							& 4.92 	& \underline{3.74} & \underline{4.20} & \underline{3.17} & 4.58 & 5.15 & \textBF{4.29} \\ 
  \bottomrule
 \end{tabular}
 \end{small}
\end{table}

\begin{table}[!htbp]
\begin{small}
\centering
 \setlength\tabcolsep{11pt}
\caption{Point and interval forecast accuracy of life expectancy across states and sexes (described in Section 6.2) by method, as measured by the Max AFE, Max RSFE, and maximum interval score. The minimal forecast errors are underlined for female and male data and their average, whereas the minimal overall forecast error is highlighted in bold.}\label{tab:2}
\begin{tabular}{@{}llrrrrrrr@{}}\toprule
Sex & Method &  VIC & NSW & QLD & TAS & SA & WA & Mean \\\midrule
&\multicolumn{6}{l}{\textit{Max AFE}}  & \\
F & Independent FDM 		& 5.16 & \underline{3.28} & 5.82 & 5.44 & 4.64 & 5.51 & 4.97	\\
	& Product-ratio 			& 4.13 & 4.46 & 4.06 & 3.63 & 3.93 & 4.03 & 4.04	\\
	& Multilevel FDM (arima) 	& \underline{3.12} & \underline{3.28} & \underline{2.73} & \underline{3.23} & \underline{2.72} & \underline{3.01} & \textBF{3.02}	\\
	& Multilevel FDM (rwf)	& 4.98 & 5.34 & 3.98 & 4.83 & 4.56 & 5.06 & 4.79	\\
\\
M & Independent FDM 		& 5.56 & 5.24 & 6.34 & 6.38 & 5.65 & 6.18 & 5.89	 \\
	& Product-ratio 			& 3.96 & 4.36 & 3.89 & 3.21 & 3.55 & 4.02 & 3.83	\\
	& Multilevel FDM (arima) 	& 5.11 & 5.41 & 5.40 & 5.26 & 4.66 & 5.30 & 5.19	\\
	& Multilevel FDM (rwf)	& \underline{3.18} & \underline{3.45} & \underline{2.57} & \underline{3.01} & \underline{2.94} & \underline{3.03} & \textBF{3.03}	\\
\\
$\frac{\text{F}+\text{M}}{2}$ 	& Independent FDM & 5.36 & 4.26 & 6.08 & 5.91 & 5.14 & 5.84 & 5.43\\
	& Product-ratio 			& \underline{4.04} & 4.41 & 3.97 & \underline{3.42} & \underline{3.74} & \underline{4.03} & 3.94 	\\
	& Multilevel FDM (arima) 	& 4.12 & 4.34 & 4.07 & 4.25 & 3.69 & 4.15 & 4.11	\\
	& Multilevel FDM (rwf)	& 4.08 & \underline{4.39} & \underline{3.27} & 3.92 & 3.75 & 4.04 & \textBF{3.91}	\\
\\
&\multicolumn{6}{l}{\textit{Max RSFE}}  & \\
F & Independent FDM 		& 30.75 & \underline{11.86} & 36.29 & 33.57 & 23.94 & 33.93 & 28.39	\\
	& Product-ratio 			& 19.04 & 22.00 & 18.15 & 14.21 & 16.51 & 17.77 & 17.95	\\
	& Multilevel FDM (arima) 	& \underline{10.72} & 11.87 & \underline{7.78} & \underline{11.26} & \underline{7.78} & \underline{9.74} & \textBF{9.86}	\\
	& Multilevel FDM (rwf)	& 29.88 & 34.01 & 19.49 & 28.14 & 24.54 & 29.94 & 27.67	\\
\\
M & Independent FDM 		& 37.78 & 33.52 & 44.37 & 47.57 & 37.09 & 44.34 & 40.78	\\
	& Product-ratio 			& 19.30 & 22.58 & 17.90 & 11.76 & 14.98 & 19.17 & 17.61	\\
	& Multilevel FDM (arima) 	& 31.55 & 34.65 & 34.52 & 32.76 & 25.62 & 32.61 & 31.95	\\
	& Multilevel FDM (rwf)	& \underline{11.16} & \underline{13.03} & \underline{7.07} & \underline{9.89} & \underline{9.08} & \underline{9.91} & \textBF{10.02}	\\
\\
$\frac{\text{F}+\text{M}}{2}$ & Independent FDM & 34.27 & 22.69 & 40.33 & 40.57 & 30.52 & 39.14 & 34.59 \\
	& Product-ratio 		  	& \underline{19.17} & \underline{22.29} & 18.02 & \underline{12.98} & \underline{15.75} & \underline{18.47} & \textBF{17.78} 	\\
	& Multilevel FDM (arima) 	& 21.14 & 23.26 & 21.15 & 22.01 & 16.70 & 21.18 & 20.91 	\\
	& Multilevel FDM (rwf)	& 20.52 & 23.52 & \underline{13.28} & 19.02 & 16.81 & 19.93 & 18.84	\\
\\	
&\multicolumn{6}{l}{\textit{Maximum interval score}} & \\
F &  Independent FDM 		& 17.59 & 25.84 & 50.67 & 27.77 & 16.03 & 25.19 & 27.18 \\
    & Product-ratio			& 32.12 & 35.33 & 26.88 & 31.40 & 29.39 & 31.15 & 31.04 \\
    & Multilevel FDM (arima)	& 17.63 & 18.90 & 14.29 & 17.62 & 14.03 & 14.77 & 16.20 \\
    & Multilevel FDM (rwf)		& \underline{15.04} & \underline{17.22} & \underline{9.69}  & \underline{12.82} & \underline{12.64} &\underline{13.19} & \textBF{13.43} \\
\\
M & Independent FDM		& 43.97 & 47.23 & 61.70 & 51.64 & 43.99 & 49.21 & 49.62 \\    
    & Product-ratio			& \underline{29.89} & \underline{33.59} & \underline{19.73} & \underline{27.43} & \underline{25.54} & \underline{30.54} & \textBF{27.79} \\
    & Multilevel FDM (arima)	& 44.50 & 46.50 & 42.08 & 44.72 & 40.85 & 45.26 & 43.98 \\
    & Multilevel FDM (rwf)		& 38.28 & 41.93 & 27.04 & 35.80 & 34.05 & 38.96 & 36.01 \\
\\
$\frac{\text{F}+\text{M}}{2}$ 	& Independent FDM 		& 30.78 & 36.53 & 56.19 & 39.71 & 30.01 & 37.20 & 38.40 \\
						& Product-ratio			& 31.00 & 34.46 & 23.30 & 29.41 & 27.46 & 30.85 & 29.41 \\
						& Multilevel FDM (arima)	& 31.06 & 32.70 & 28.19 & 31.17 & 27.44 & 30.01 & 30.09 \\
						& Multilevel FDM (rwf)	& \underline{26.66} & \underline{29.58} & \underline{18.36} & \underline{24.31} & \underline{23.35} & \underline{26.08} & \textBF{24.72} \\
  \bottomrule
 \end{tabular}
 \end{small}
\end{table}

\newpage
\bibliographystyle{ims}
\bibliography{coherent}

\begin{thebibliography}{72}
\expandafter\ifx\csname natexlab\endcsname\relax\def\natexlab#1{#1}\fi
\expandafter\ifx\csname url\endcsname\relax
  \def\url#1{\texttt{#1}}\fi
\expandafter\ifx\csname urlprefix\endcsname\relax\def\urlprefix{URL }\fi
\providecommand{\eprint}[2][]{\url{#2}}

\bibitem[{Akaike(1974)}]{Akaike74}
\textsc{Akaike, H.} (1974).
\newblock A new look at the statistical model identification.
\newblock \textit{IEEE Transactions on Automatic Control}, \textbf{19}
  716--723.

\bibitem[{Alkema et~al.(2011)Alkema, Raftery, Gerland, Clark, Pelletier,
  Buettner and Heilig}]{ARG+11}
\textsc{Alkema, L.}, \textsc{Raftery, A.~E.}, \textsc{Gerland, P.},
  \textsc{Clark, S.~J.}, \textsc{Pelletier, F.}, \textsc{Buettner, T.} and
  \textsc{Heilig, G.~K.} (2011).
\newblock Probabilistic projections of the total fertility rate for all
  countries.
\newblock \textit{Demography}, \textbf{48} 815--839.

\bibitem[{Aue et~al.(2015)Aue, Norinho and H\"{o}rmann}]{ANH15}
\textsc{Aue, A.}, \textsc{Norinho, D.~D.} and \textsc{H\"{o}rmann, S.} (2015).
\newblock On the prediction of stationary functional time series.
\newblock \textit{Journal of the American Statistical Association},
  \textbf{110} 378--392.

\bibitem[{Biatat and Currie(2010)}]{BC10}
\textsc{Biatat, V.~D.} and \textsc{Currie, I.~D.} (2010).
\newblock Joint models for classification and comparison of mortality in
  different countries.
\newblock In \textit{Proceedings of 25th International Workshop on Statistical
  Modelling} (A.~W. Bowman, ed.). Glasgow, 89--94.

\bibitem[{Booth(2006)}]{Booth06}
\textsc{Booth, H.} (2006).
\newblock Demographic forecasting: 1980-2005 in review.
\newblock \textit{International Journal of Forecasting}, \textbf{22} 547--581.

\bibitem[{Booth et~al.(2002)Booth, Maindonald and Smith}]{BMS02}
\textsc{Booth, H.}, \textsc{Maindonald, J.} and \textsc{Smith, L.} (2002).
\newblock {Applying Lee-Carter under conditions of variable mortality decline}.
\newblock \textit{Population Studies}, \textbf{56} 325--336.

\bibitem[{Booth and Tickle(2008)}]{BT08}
\textsc{Booth, H.} and \textsc{Tickle, L.} (2008).
\newblock Mortality modelling and forecasting: A review of methods.
\newblock \textit{Annals of Actuarial Science}, \textbf{3} 3--43.

\bibitem[{Box et~al.(2008)Box, Jenkins and Reinsel}]{BJR08}
\textsc{Box, G. E.~P.}, \textsc{Jenkins, G.~M.} and \textsc{Reinsel, G.~C.}
  (2008).
\newblock \textit{{Time Series Analysis: Forecasting and Control}}.
\newblock 4th ed. John Wiley, Hoboken, New Jersey.

\bibitem[{Cairns et~al.(2011{\natexlab{a}})Cairns, Blake, Dowd, Coughlan,
  Epstein and {Khalaf-Allah}}]{CBD+11}
\textsc{Cairns, A. J.~G.}, \textsc{Blake, D.}, \textsc{Dowd, K.},
  \textsc{Coughlan, G.~D.}, \textsc{Epstein, D.} and \textsc{{Khalaf-Allah},
  M.} (2011{\natexlab{a}}).
\newblock Mortality density forecasts: An analysis of six stochastic mortality
  models.
\newblock \textit{Insurance: Mathematics and Economics}, \textbf{48} 355--367.

\bibitem[{Cairns et~al.(2011{\natexlab{b}})Cairns, Blake, Dowd, Coughlan and
  {Khalaf-Allah}}]{CBD+11b}
\textsc{Cairns, A. J.~G.}, \textsc{Blake, D.}, \textsc{Dowd, K.},
  \textsc{Coughlan, G.~D.} and \textsc{{Khalaf-Allah}, M.}
  (2011{\natexlab{b}}).
\newblock Bayesian stochastic mortality modelling for two populations.
\newblock \textit{ASTIN Bulletin}, \textbf{41} 29--55.

\bibitem[{Chernick(2008)}]{Chernick08}
\textsc{Chernick, M.~R.} (2008).
\newblock \textit{{Bootstrap Methods: A Guide for Practitioners and
  Researchers}}.
\newblock Wiley-Interscience, New Jersey.

\bibitem[{Chiou(2012)}]{Chiou12}
\textsc{Chiou, J.-M.} (2012).
\newblock Dynamical functional prediction and classification, with application
  to traffic flow prediction.
\newblock \textit{The Annals of Applied Statistics}, \textbf{6} 1588--1614.

\bibitem[{Crainiceanu and Goldsmith(2010)}]{CG10}
\textsc{Crainiceanu, C.~M.} and \textsc{Goldsmith, J.~A.} (2010).
\newblock {Bayesian functional data analysis using WinBUGS}.
\newblock \textit{Journal of Statistical Software}, \textbf{32}.

\bibitem[{Crainiceanu et~al.(2009)Crainiceanu, Staicu and Di}]{CSD09}
\textsc{Crainiceanu, C.~M.}, \textsc{Staicu, A.-M.} and \textsc{Di, C.-Z.}
  (2009).
\newblock {Generalized multilevel functional regression}.
\newblock \textit{Journal of the American Statistical Association},
  \textbf{104} 1550--1561.

\bibitem[{{Cuesta-Albertos} and {Febrero-Bande}(2010)}]{CF10}
\textsc{{Cuesta-Albertos}, J.~A.} and \textsc{{Febrero-Bande}, M.} (2010).
\newblock {A simple multiway ANOVA for functional data}.
\newblock \textit{Test}, \textbf{19} 537--557.

\bibitem[{Currie et~al.(2004)Currie, Durban and Eilers}]{CDE04}
\textsc{Currie, I.~D.}, \textsc{Durban, M.} and \textsc{Eilers, P. H.~C.}
  (2004).
\newblock {Smoothing and forecasting mortality rates}.
\newblock \textit{Statistical Modelling}, \textbf{4} 279--298.

\bibitem[{Delwarde et~al.(2006)Delwarde, Denuit, {Guill\'{e}n} and
  {Vidiella-i-Anguera}}]{DDG+06}
\textsc{Delwarde, A.}, \textsc{Denuit, M.}, \textsc{{Guill\'{e}n}, M.} and
  \textsc{{Vidiella-i-Anguera}, A.} (2006).
\newblock {Application of the Poisson log-bilinear projection model to the G5
  mortality experience}.
\newblock \textit{Belgian Actuarial Bulletin}, \textbf{6} 54--68.

\bibitem[{Di et~al.(2009)Di, Crainiceanu, Caffo and Punjabi}]{DCC+09}
\textsc{Di, C.-Z.}, \textsc{Crainiceanu, C.~M.}, \textsc{Caffo, B.~S.} and
  \textsc{Punjabi, N.~M.} (2009).
\newblock Multilevel functional principal component analysis.
\newblock \textit{The Annals of Applied Statistics}, \textbf{3} 458--488.

\bibitem[{Dowd et~al.(2011)Dowd, Cairns, Blake, Coughlan, Epstein and
  {Khalaf-Allah}}]{DCB+11}
\textsc{Dowd, K.}, \textsc{Cairns, A. J.~G.}, \textsc{Blake, D.},
  \textsc{Coughlan, G.~D.}, \textsc{Epstein, D.} and \textsc{{Khalaf-Allah},
  M.} (2011).
\newblock A gravity model of mortality rates for two related populations.
\newblock \textit{North American Actuarial Journal}, \textbf{15} 334--356.

\bibitem[{Girosi and King(2008)}]{GK08}
\textsc{Girosi, F.} and \textsc{King, G.} (2008).
\newblock \textit{{Demographic Forecasting}}.
\newblock Princeton University Press, Princeton.

\bibitem[{Gneiting and Katzfuss(2014)}]{GK14}
\textsc{Gneiting, T.} and \textsc{Katzfuss, M.} (2014).
\newblock Probabilistic forecasting.
\newblock \textit{Annual Review of Statistics and Its Applications}, \textbf{1}
  125--151.

\bibitem[{Gneiting and Raftery(2007)}]{GR07}
\textsc{Gneiting, T.} and \textsc{Raftery, A.~E.} (2007).
\newblock Strictly proper scoring rules, prediction and estimation.
\newblock \textit{Journal of the American Statistical Association},
  \textbf{102} 359--378.

\bibitem[{Greven et~al.(2010)Greven, Crainiceanu, Caffo and Reich}]{GCC+10}
\textsc{Greven, S.}, \textsc{Crainiceanu, C.}, \textsc{Caffo, B.} and
  \textsc{Reich, D.} (2010).
\newblock Longitudinal functional principal component analysis.
\newblock \textit{Electronic Journal of Statistics}, \textbf{4} 1022--1054.

\bibitem[{Hall and Hosseini-Nasab(2006)}]{HH06}
\textsc{Hall, P.} and \textsc{Hosseini-Nasab, M.} (2006).
\newblock On properties of functional principal components analysis.
\newblock \textit{Journal of the Royal Statistical Society (Series B)},
  \textbf{68} 109--126.

\bibitem[{Hall and Vial(2006)}]{HV06}
\textsc{Hall, P.} and \textsc{Vial, C.} (2006).
\newblock {Assessing the finite dimensionality of functional data}.
\newblock \textit{Journal of the Royal Statistical Society (Series B)},
  \textbf{68} 689--705.

\bibitem[{He and Ng(1999)}]{HN99}
\textsc{He, X.} and \textsc{Ng, P.} (1999).
\newblock {COBS: Qualitatively constrained smoothing via linear programming}.
\newblock \textit{Computational Statistics}, \textbf{14} 315--337.

\bibitem[{Hoff(2009)}]{Hoff09}
\textsc{Hoff, P.~D.} (2009).
\newblock \textit{{A First Course in Bayesian Statistical Methods}}.
\newblock Springer, New York.

\bibitem[{Hosseini-Nasab(2013)}]{Hossein13}
\textsc{Hosseini-Nasab, M.} (2013).
\newblock {Cross-validation approximation in functional linear regression}.
\newblock \textit{Journal of the Statistical Computation and Simulation},
  \textbf{83} 1429--1439.

\bibitem[{{Human Mortality Database}(2015)}]{HMD13}
\textsc{{Human Mortality Database}} (2015).
\newblock \textit{{University of California, Berkeley (USA), and Max Planck
  Institute for Demographic Research (Germany)}}.
\newblock Accessed at 8 March 2013. URL: \url{http://www.mortality.org}.

\bibitem[{Hyndman(2010)}]{Hyndman10}
\textsc{Hyndman, R.~J.} (2010).
\newblock \textit{addb: Australian Demographic Data Bank}.
\newblock R package version 3.223. URL:
  \url{http://robjhyndman.com/software/addb/}.

\bibitem[{Hyndman et~al.(2011)Hyndman, Ahmed, Athanasopoulos and
  Shang}]{HAA+11}
\textsc{Hyndman, R.~J.}, \textsc{Ahmed, R.~A.}, \textsc{Athanasopoulos, G.} and
  \textsc{Shang, H.~L.} (2011).
\newblock Optimal combination forecasts for hierarchical time series.
\newblock \textit{Computational Statistics \& Data Analysis}, \textbf{55}
  2579--2589.

\bibitem[{Hyndman et~al.(2013)Hyndman, Booth and Yasmeen}]{HBY13}
\textsc{Hyndman, R.~J.}, \textsc{Booth, H.} and \textsc{Yasmeen, F.} (2013).
\newblock {Coherent mortality forecasting: the product-ratio method with
  functional time series models}.
\newblock \textit{Demography}, \textbf{50} 261--283.

\bibitem[{Hyndman and Khandakar(2008)}]{HK08}
\textsc{Hyndman, R.~J.} and \textsc{Khandakar, Y.} (2008).
\newblock {Automatic time series forecasting: the forecast package for R}.
\newblock \textit{Journal of Statistical Software}, \textbf{27}.

\bibitem[{Hyndman and Shang(2009)}]{HS09}
\textsc{Hyndman, R.~J.} and \textsc{Shang, H.~L.} (2009).
\newblock {Forecasting functional time series (with discussion)}.
\newblock \textit{Journal of the Korean Statistical Society}, \textbf{38}
  199--221.

\bibitem[{Hyndman and Ullah(2007)}]{HU07}
\textsc{Hyndman, R.~J.} and \textsc{Ullah, M.~S.} (2007).
\newblock {Robust forecasting of mortality and fertility rates: A functional
  data approach}.
\newblock \textit{Computational Statistics \& Data Analysis}, \textbf{51}
  4942--4956.

\bibitem[{Indritz(1963)}]{Indritz63}
\textsc{Indritz, J.} (1963).
\newblock \textit{{Methods in Analysis}}.
\newblock {Macmillan \& Collier Macmillan}, New York.

\bibitem[{Janssen et~al.(2013)Janssen, {van Wissen} and Kunst}]{JVK13}
\textsc{Janssen, F.}, \textsc{{van Wissen}, L. J.~G.} and \textsc{Kunst, A.~E.}
  (2013).
\newblock {Including the smoking epidemic in internationally coherent mortality
  projection}.
\newblock \textit{Demography}, \textbf{50} 1341--1362.

\bibitem[{Jarner and Kryger(2011)}]{JK11}
\textsc{Jarner, S.~F.} and \textsc{Kryger, E.~M.} (2011).
\newblock {Modelling adult mortality in small populations: The SAINT model}.
\newblock \textit{Astin Bulletin}, \textbf{41} 377--418.

\bibitem[{Karhunen(1946)}]{Karhunen46}
\textsc{Karhunen, K.} (1946).
\newblock Zur spektraltheorie stochastischer prozesse.
\newblock \textit{Annales Academiae Scientiarum Fennicae}, \textbf{37} 1--37.

\bibitem[{Koop(2003)}]{Koop03}
\textsc{Koop, G.} (2003).
\newblock \textit{{Bayesian Econometrics}}.
\newblock Wiley, Chichester.

\bibitem[{Kwiatkowski et~al.(1992)Kwiatkowski, Phillips, Schmidt and
  Shin}]{KPSS92}
\textsc{Kwiatkowski, D.}, \textsc{Phillips, P. C.~B.}, \textsc{Schmidt, P.} and
  \textsc{Shin, Y.} (1992).
\newblock {Testing the null hypothesis of stationarity against the alternative
  of a unit root: How sure are we that economic time series have a unit root?}
\newblock \textit{Journal of Econometrics}, \textbf{54} 159--178.

\bibitem[{Lee(2000)}]{Lee00}
\textsc{Lee, R.~D.} (2000).
\newblock {The Lee-Carter method for forecasting mortality, with various
  extensions and applications}.
\newblock \textit{{North American Actuarial Journal}}, \textbf{4} 80--92.

\bibitem[{Lee(2006)}]{Lee06}
\textsc{Lee, R.~D.} (2006).
\newblock Mortality forecasts and linear life expectancy trends.
\newblock In \textit{{Perspectives on Mortality Forecasting. Vol. III. The
  Linear Rise in Life Expectancy: History and Prospects}} (T.~Bengtsson, ed.).
  No.~3 in Social Insurance Studies, {Swedish National Social Insurance Board},
  Stockholm, 19--39.

\bibitem[{Lee and Carter(1992)}]{LC92}
\textsc{Lee, R.~D.} and \textsc{Carter, L.~R.} (1992).
\newblock {Modeling and forecasting U.S. mortality}.
\newblock \textit{Journal of the American Statistical Association}, \textbf{87}
  659--671.

\bibitem[{Lee and Miller(2001)}]{LM01}
\textsc{Lee, R.~D.} and \textsc{Miller, T.} (2001).
\newblock {Evaluating the performance of the Lee-Carter method for forecasting
  mortality}.
\newblock \textit{Demography}, \textbf{38} 537--549.

\bibitem[{Li(2013)}]{Li13}
\textsc{Li, J.} (2013).
\newblock {A Poisson common factor model for projecting mortality and life
  expectancy jointly for females and males}.
\newblock \textit{Population Studies}, \textbf{67} 111--126.

\bibitem[{Li and Hardy(2011)}]{LH11}
\textsc{Li, J. S.~H.} and \textsc{Hardy, M.~R.} (2011).
\newblock Measuring basis risk in longevity hedges.
\newblock \textit{North American Actuarial Journal}, \textbf{15} 177--200.

\bibitem[{Li and Lee(2005)}]{LL05}
\textsc{Li, N.} and \textsc{Lee, R.} (2005).
\newblock {Coherent mortality forecasts for a group of population: An extension
  of the Lee-Carter method}.
\newblock \textit{Demography}, \textbf{42} 575--594.

\bibitem[{Li et~al.(2013)Li, Lee and Gerland}]{LLG13}
\textsc{Li, N.}, \textsc{Lee, R.} and \textsc{Gerland, P.} (2013).
\newblock {Extending the Lee-Carter method to model the rotation of age
  patterns of mortality decline for long-term projections}.
\newblock \textit{Demography}, \textbf{50} 2037--2051.

\bibitem[{Lo\`{e}ve(1946)}]{Loeve46}
\textsc{Lo\`{e}ve, M.} (1946).
\newblock Fonctions al\'{e}atoires a decomposition orthogonale exponentielle.
\newblock \textit{La Revue Scientifique}, \textbf{84} 159--162.

\bibitem[{Morris and Carroll(2006)}]{MC06}
\textsc{Morris, J.~S.} and \textsc{Carroll, R.~J.} (2006).
\newblock Wavelet-based functional mixed models.
\newblock \textit{Journal of the Royal Statistical Society. Series B},
  \textbf{68} 179--199.

\bibitem[{Morris et~al.(2003)Morris, Vannucci, Brown and Carroll}]{MVB+03}
\textsc{Morris, J.~S.}, \textsc{Vannucci, M.}, \textsc{Brown, P.~J.} and
  \textsc{Carroll, R.~J.} (2003).
\newblock Wavelet-based nonparametric modeling of hierarchical functions in
  colon carcinogenesis.
\newblock \textit{Journal of the American Statistical Association}, \textbf{98}
  573--583.

\bibitem[{Oeppen and Vaupel(2002)}]{OV02}
\textsc{Oeppen, J.} and \textsc{Vaupel, J.~W.} (2002).
\newblock {Broken limits to life expectancy}.
\newblock \textit{Science}, \textbf{296} 1029--1031.

\bibitem[{Pampel(2005)}]{Pampel05}
\textsc{Pampel, F.~C.} (2005).
\newblock {Forecasting sex differences in mortality from lung cancer in
  high-income nations: the contribution of smoking}.
\newblock \textit{Demographic Research}, \textbf{13} 455--484.

\bibitem[{Preston et~al.(2001)Preston, Heuveline and Guillot}]{PHG01}
\textsc{Preston, S.~H.}, \textsc{Heuveline, P.} and \textsc{Guillot, M.}
  (2001).
\newblock \textit{{Demography: Measuring and Modelling Population Process}}.
\newblock Blackwell, Oxford, UK.

\bibitem[{{R Core Team}(2015)}]{Team13}
\textsc{{R Core Team}} (2015).
\newblock \textit{R: A Language and Environment for Statistical Computing}.
\newblock R Foundation for Statistical Computing, Vienna, Austria.
\newblock URL: \url{http://www.R-project.org/}.

\bibitem[{Raftery et~al.(2013)Raftery, Chunn, Gerland and
  \v{S}ev\v{c}\'{i}kov\'{a}}]{RCG+13}
\textsc{Raftery, A.~E.}, \textsc{Chunn, J.~L.}, \textsc{Gerland, P.} and
  \textsc{\v{S}ev\v{c}\'{i}kov\'{a}, H.} (2013).
\newblock Bayesian probabilistic projections of life expectancy for all
  countries.
\newblock \textit{Demography}, \textbf{50} 777--801.

\bibitem[{Raftery et~al.(2014)Raftery, Lalic and Gerland}]{RLG14}
\textsc{Raftery, A.~E.}, \textsc{Lalic, N.} and \textsc{Gerland, P.} (2014).
\newblock Joint probabilistic projection of female and male life expectancy.
\newblock \textit{Demographic Research}, \textbf{30} 795--822.

\bibitem[{Raftery et~al.(2012)Raftery, Li, \v{S}ev\v{c}\'{i}kov\'{a}, Gerland
  and Heilig}]{RLS+12}
\textsc{Raftery, A.~E.}, \textsc{Li, N.}, \textsc{\v{S}ev\v{c}\'{i}kov\'{a},
  H.}, \textsc{Gerland, P.} and \textsc{Heilig, G.~K.} (2012).
\newblock {Bayesian probabilistic population projection for all countries}.
\newblock \textit{Proceedings of the National Academy of Sciences of the United
  States of America}, \textbf{109} 13915--13921.

\bibitem[{Renshaw and Haberman(2003)}]{RH03}
\textsc{Renshaw, A.~E.} and \textsc{Haberman, S.} (2003).
\newblock {Lee-Carter mortality forecasting with age-specific enhancement}.
\newblock \textit{Insurance: Mathematics and Economics}, \textbf{33} 255--272.

\bibitem[{Rice and Silverman(1991)}]{RS91}
\textsc{Rice, J.} and \textsc{Silverman, B.} (1991).
\newblock Estimating the mean and covariance structure nonparametrically when
  the data are curves.
\newblock \textit{Journal of the Royal Statistical Society. Series B},
  \textbf{53} 233--243.

\bibitem[{{\v{S}}ev\v{c}\'{i}kov\'{a} et~al.(2015){\v{S}}ev\v{c}\'{i}kov\'{a},
  Li, Kantorov\'{a}, Gerland and Raftery}]{SLK+15}
\textsc{{\v{S}}ev\v{c}\'{i}kov\'{a}, H.}, \textsc{Li, N.},
  \textsc{Kantorov\'{a}, V.}, \textsc{Gerland, P.} and \textsc{Raftery, A.~E.}
  (2015).
\newblock {Age-specific mortality and fertility rates for probabilistic
  population projections}.
\newblock Working paper, University of Washington.
\newblock \urlprefix\url{http://arxiv.org/abs/1503.05215}.

\bibitem[{{\v{S}}ev\v{c}\'{i}kov\'{a} and Raftery(2015)}]{SR15}
\textsc{{\v{S}}ev\v{c}\'{i}kov\'{a}, H.} and \textsc{Raftery, A.} (2015).
\newblock \textit{bayesLife: Bayesian Projection of Life Expectancy}.
\newblock R package version 2.2-0,
  \urlprefix\url{http://CRAN.R-project.org/package=bayesLife}.

\bibitem[{Shang(2016)}]{Shang16}
\textsc{Shang, H.~L.} (2016).
\newblock Supplement to ``mortality and life expectancy forecasting for a group
  of populations in developed countries: a multilevel functional data method".

\bibitem[{Shang et~al.(2011)Shang, Booth and Hyndman}]{SBH11}
\textsc{Shang, H.~L.}, \textsc{Booth, H.} and \textsc{Hyndman, R.~J.} (2011).
\newblock Point and interval forecasts of mortality rates and life expectancy:
  A comparison of ten principal component methods.
\newblock \textit{Demographic Research}, \textbf{25} 173--214.

\bibitem[{Shang and Hyndman(2016)}]{SH16}
\textsc{Shang, H.~L.} and \textsc{Hyndman, R.~J.} (2016).
\newblock {Grouped functional time series forecasting: an application to
  age-specific mortality rates}.
\newblock Working paper 04/16, Monash University.
\newblock
  \urlprefix\url{http://business.monash.edu/econometrics-and-business-statistics/research/publications/ebs/wp04-16.pdf}.

\bibitem[{Tickle and Booth(2014)}]{BT14}
\textsc{Tickle, L.} and \textsc{Booth, H.} (2014).
\newblock {The longevity prospects of Australian seniors: An evaluation of
  forecast method and outcome}.
\newblock \textit{Asia-Pacific Journal of Risk and Insurance}, \textbf{8}
  259--292.

\bibitem[{Wi\'{s}niowski et~al.(2015)Wi\'{s}niowski, Smith, Bijak, Raymer and
  Forster}]{WSB+15}
\textsc{Wi\'{s}niowski, A.}, \textsc{Smith, P. W.~F.}, \textsc{Bijak, J.},
  \textsc{Raymer, J.} and \textsc{Forster, J.~J.} (2015).
\newblock {Bayesian population forecasting: Extending the Lee-Carter method}.
\newblock \textit{Demography}, \textbf{52} 1035--1059.

\bibitem[{Woods and Dunstan(2014)}]{WD14}
\textsc{Woods, C.} and \textsc{Dunstan, K.} (2014).
\newblock {Forecasting mortality in New Zealand}.
\newblock Working paper 14-01, Statistics New Zealand.
\newblock
  \urlprefix\url{http://www.stats.govt.nz/methods/research-papers/working-papers-original/forecasting-mortality-14-01.aspx}.

\bibitem[{Yao et~al.(2005)Yao, M\"{u}ller and Wang}]{YMW05}
\textsc{Yao, F.}, \textsc{M\"{u}ller, H.-G.} and \textsc{Wang, J.} (2005).
\newblock Functional data analysis for sparse longitudinal data.
\newblock \textit{Journal of the American Statistical Association},
  \textbf{100} 577--590.

\bibitem[{Zhang(2014)}]{Zhang14}
\textsc{Zhang, J.-T.} (2014).
\newblock \textit{{Analysis of Variance for Functional Data}}.
\newblock Chapman \& Hall, Boca Raton.

\bibitem[{Zivot and Wang(2006)}]{ZW06}
\textsc{Zivot, E.} and \textsc{Wang, J.} (2006).
\newblock \textit{{Modeling Financial Time Series with S-PLUS}}.
\newblock Springer, New York.

\end{thebibliography}

\end{document}